\batchmode
\makeatletter
\def\input@path{{\string"C:/Users/teng-hucheng/SynologyDrive/iPad Pro/1.ED_2D Cooperative transportation (Transactions on mechtronics)/\string"}}
\makeatother
\documentclass[english,conference, twocolumn]{IEEEtran}
\pdfoutput=1
\usepackage[T1]{fontenc}
\usepackage[latin9]{inputenc}
\usepackage{color}
\usepackage{array}
\usepackage{float}
\usepackage{units}
\usepackage{multirow}
\usepackage{amsmath}
\usepackage{amsthm}
\usepackage{amssymb}
\usepackage{graphicx}

\makeatletter

\providecommand{\tabularnewline}{\\}

\theoremstyle{definition}
\newtheorem{assumption}{Assumption}
\theoremstyle{definition}
\newtheorem{problem}{\protect\problemname}
\theoremstyle{definition}
\newtheorem{defn}{\protect\definitionname}
\theoremstyle{remark}
\newtheorem{rem}{\protect\remarkname}
\theoremstyle{plain}
\newtheorem{thm}{\protect\theoremname}

\usepackage{cite}
\IEEEoverridecommandlockouts

\makeatother

\usepackage{babel}
\providecommand{\definitionname}{Definition}
\providecommand{\problemname}{Problem}
\providecommand{\remarkname}{Remark}
\providecommand{\theoremname}{Theorem}

\begin{document}

\title{\textcolor{black}{\bstctlcite{IEEEexample:BSTcontrol}Cooperative
Transportation of UAVs Without Inter-UAV Communication}\thanks{\textcolor{black}{Department of Mechanical Engineering, National Chiao
Tung University, Hsinchu, Taiwan 30010 Email: f29993856@gmail.com,
kb240608@gmail.com, tenghu@g2.nctu.edu.tw}}\textcolor{black}{}\thanks{\textcolor{black}{This research was supported by the Ministry of Science
and Technology, Taiwan (Grant Number 110-2222-E-A49-005-), and partially
supported by Pervasive Artificial Intelligence Research (PAIR) Labs,
Taiwan (Grant Numbers MOST 110-2634-F-009-018-).}}}

\author{\textcolor{black}{Pin-Xian Wu, Cheng-Cheng Yang, and Teng-Hu Cheng}}
\maketitle
\begin{abstract}
\textcolor{black}{A leader-follower system is developed for cooperative
transportation. To the best of our knowledge, this is the first work
that inter-UAV communication is not required and the reference trajectory
of the payload can be modified in real time, so that it can be applied
to a dynamically changing environment. To track the modified reference
trajectory in real time under the communication-free condition, the
leader-follower system is considered as a nonholonomic system in which
a controller is developed for the leader to achieve asymptotic tracking
of the payload. To eliminate the need to install force sensors, UKFs
(unscented Kalman filters) are developed to estimate the forces applied
by the leader and follower. Stability analysis is conducted to prove
the tracking error of the closed-loop system. Simulation results demonstrate
the good performance of the tracking controller. The experiments show
the controllers of the leader and the follower can work in the real
world, but the tracking errors were affected by the disturbance of
airflow in a restricted space.}
\end{abstract}

\begin{IEEEkeywords}
\textcolor{black}{Cooperative Transportation, Force Estimate, Leader-Follower
systems}
\end{IEEEkeywords}

\section{\textcolor{black}{Introduction}}

\textcolor{black}{}
\begin{figure*}[t]
\begin{centering}
\textcolor{black}{\includegraphics[scale=0.4]{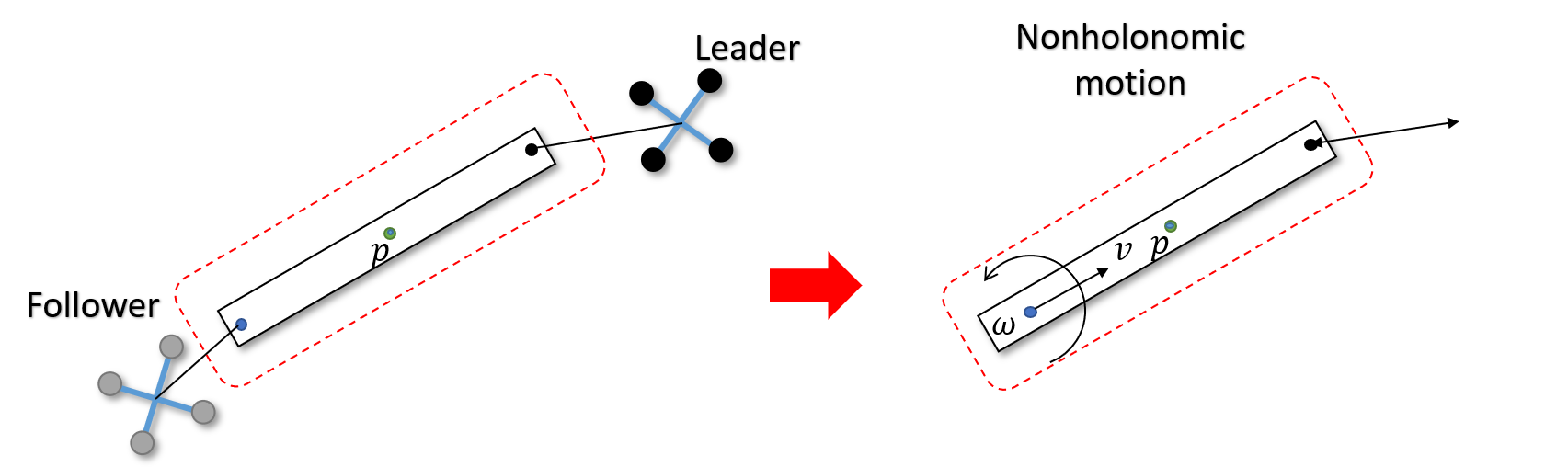}}
\par\end{centering}
\textcolor{black}{\caption{\label{fig:Transportation process}The leader-follower system cooperatively
transporting a payload. The follower and leader are controlled such
that the motion of the payload behaves like a nonholonomic system,
where the end connected to the follower is not allowed to undergo
transverse motion. The controllers are designed in Section \ref{sec:Controller}.}
}
\end{figure*}

\subsection{\textcolor{black}{Background}}

\textcolor{black}{Cooperative tasks executed by a group of robots
have been receiving considerable attention in the field of multi-robot
systems. Cooperative tasks have usually been accomplished by collaborative
ground robots in previous research (cf. \cite{Yamaguchi2010,Yufka2010,Petitti2016,Marino2018,Wang2016a,Wang2016b,Wang2015c,Verginis2018,Tsiamis2015,Tsiamis2015a,Fan2017,PrestonCulbertson2018,Bechlioulis2018,Yang2019}),
but the capability to perform a wide variety of tasks is compromised
by motion constraints (e.g., rugged terrain). Since UAVs are agile
robots that move in higher dimensional space, they have been adopted
to solve a diverse range of challenging problems \cite{Sreenath2013,Lee2015b,Lee2013,Aghdam2016,Lim2017,Liu2018a,Wang2019,Chen2019}.
Achieving cooperative tasks is another important functionality for
a team of UAVs. To achieve such objectives, UAVs usually communicate
(or only intermittently \cite{Yang2020}) state information to their
neighboring UAVs to achieve interaction control. However, communication
is prohibited or unavailable in some applications, which makes using
feedback control for interactions between collaborative robots very
challenging.}

\textcolor{black}{UAVs have been adopted for use in cooperative transportation
to overcome motion constraints. In \cite{HyeonbeomLee2017}, multiple
aerial manipulators are tasked to carry an unknown payload, but the
manipulator's weight and volume represent costs for the UAVs, such
as higher power consumption and lower carrying capacity. In \cite{Masone2016},
an RCDPR (reconfigurable cable-driven parallel robots) approach analyzes
the relative motion between the quadrotors and the payload. A centralized
tracking controller was developed for computing the inputs for the
quadrotors, which requires a wider bandwidth to communicate with other
quadrotors. In \cite{GiuseppeLoianno2018}, controllers for multiple
UAVs to cooperatively carry a payload are developed so that the thrust
and torque of each vehicle are flexible while guaranteeing the system\textquoteright s
stability; however, that design is based on a centralized architecture.
In \cite{Lee2018}, although communication between quadrotors is not
required, a centralized computer is still needed to compute the input
of each quadrotor given the force and torque required for the payload.
\cite{Tognon2018} and \cite{Michael2009} present control methodologies
to control UAVs so that the desired attitude of the payload can be
achieved, but either the use of force sensors or explicit communication
is required. Similarly, a statically rigid cable-suspended aerial
manipulator was developed in \cite{Sanalitro2020}, and the controller
was designed to address system uncertainties, but it was not distributed.}

\textcolor{black}{In summary, the control methods of the aforementioned
works rely on communication to achieve the control objective, but
that can be too restrictive in a communication-denied environment. }

\textcolor{black}{Some control strategies have been invented that
avoid using inter-UAV communication to achieve cooperative transportation.
In \cite{Wang2018}, a group of quadrotors transports an object without
inter-UAV communication by independently solving a local optimization
problem. Although no peer communication is required, the reference
trajectory must be known by all of the quadrotors. Inter-UAV communication
is eliminated in \cite{Mellinger2013,Wang2016} by equally distributing
the net thrust and torque to each quadrotor, but this approach can
be problematic if the predefined trajectory changes due to discrete
events (e.g., obstacle avoidance). Another communication-free control
approach for cable-based collaborative transportation is proposed
in \cite{Gassner2017}, where the follower can the estimate motion
of the leader by monitoring the position of an artificial tag attached
to the payload. However, detecting the tags using cameras can be problematic
in poor lighting conditions. Moreover, motion planning for the leader-follower
system is not clear since the follower's kinematics is not considered.
In \cite{Tagliabue2016} and \cite{Tagliabue2017}, admittance controllers
along with unscented Kalman filters (UKFs) are implemented so that
the UAVs can perform transportation without relying on communication,
where the follower moves passively based on the estimated force applied
by the leader. However, the follower's kinematics is also not considered
for motion planning. Although the aforementioned studies focused on
avoiding inter-UAV communication, methods for changing the predefined
trajectory in real time remain unclear.}

\textcolor{black}{The present work focuses on cooperative transportation
using a leader-follower UAV system in which the leader and the follower
are connected to the two ends of a payload via cables. The developed
method envisions the fixing of an external sensor (an inertial measurement
unit {[}IMU{]}) to the transported object, and the sensor data are
sent to the leader robot. This approach avoids the need for communication
between the two UAVs, with only explicit communication between the
leader UAV and the sensor being required. The main advantages of this
approach are that inter-UAV communication is not required and the
reference trajectory of the payload can be modified by the leader
in real time, so that it can be applied to a dynamically changing
and communication-denied environment. To this end, the leader-follower
system is controlled to behave as a nonholonomic system in which the
follower's motion does not affect the tracking error of the payload;
that is, the follower is controlled such that the end of the payload
connected to the follower allows only longitudinal motion. On the
other hand, the leader is controlled to ensure tracking performance.
To eliminate the need to install force sensors, UKFs are developed
to estimate the forces applied by the leader and follower. Finally,
a hybrid $A^{*}$ trajectory planner is used to find the optimal trajectory
for the leader-follower system under various constraints. Stability
analysis is conducted to prove the stability of the closed-loop system.
Simulation results demonstrate the good performance of the tracking
controller.}

\subsection{\textcolor{black}{Innovation and Challenges}}

\textcolor{black}{The four novel aspects of this work are as follows:}
\begin{enumerate}
\item \textcolor{black}{In contrast to \cite{HyeonbeomLee2017,Masone2016,GiuseppeLoianno2018,Lee2018,Tognon2018,Michael2009},
inter-UAV communication is not required in the present approach.}
\item \textcolor{black}{In contrast to \cite{Wang2018,Mellinger2013,Wang2016,Gassner2017,Tagliabue2016,Tagliabue2017},
the reference trajectory of the payload can be modified by the leader
in real time, and the new reference trajectory can still be tracked.}
\item \textcolor{black}{In contrast to \cite{Tognon2018,Lee2018}, force
sensors are not required since the forces are estimated using UKFs.}
\item \textcolor{black}{To track the modified reference trajectory in real
time under the communication-free condition, the follower motion is
designed to not affect the tracking error of the payload, and the
leader is controlled to ensure tracking control.}
\end{enumerate}

\subsection{\textcolor{black}{Contributions}}

\textcolor{black}{The main contributions of this work can be summarized
as follows: }
\begin{enumerate}
\item \textcolor{black}{To eliminate the need for inter-UAV communication,
two UKFs are designed to be cascaded in the leader to estimate the
force applied by the follower, where an IMU is attached to one end
of the payload. Therefore, the estimated force can be used in the
leader controller for interactions to achieve cooperative transportation.
The details are provided in Section \ref{subsec:UKF-leader}. Force
sensors are not required since the forces are estimated using UKFs;
that is, the kinematics and dynamics of the payload are modeled so
that the force applied by the follower can be estimated using the
leader UKF. }
\item \textcolor{black}{The reference trajectory of the payload can be modified
by the leader in real time since the leader controller can compensate
the follower force applied to the payload so as to minimize the tracking
error.}
\item \textcolor{black}{To avoid affecting the tracking error of the payload,
the follower is controlled in a way to ensure that the end of the
payload close to the follower undergoes only longitudinal motion,
and not transverse motion, as shown in Fig. \ref{fig:Transportation process}.
In other words, the transverse motion of the payload is controlled
to zero velocity, and the longitudinal direction is controlled by
a force controller so that it can follow the leader and achieve nonholonomic
motion. The position of the payload below the leader is controlled
to ensure asymptotic tracking, where the position is estimated implicitly
based on UKFs.}
\item \textcolor{black}{In \cite{Tagliabue2016,Tagliabue2017}, the trajectory
of the follower robot is generated to comply with external forces
based on the admittance controller. However, this implies that the
trajectory of a point on the payload cannot be modified and must continue
to be tracked. This is important when a payload is transported in
a cluttered environment where collision avoidance is necessary. Enforcing
a point on the payload to follow a trajectory that can be modified
in real time is especially important in the presence of unexpected
moving obstacles. A control strategy is therefore designed to achieve
both agility and compactness. In other words, not only is the reference
trajectory not required by the follower, but also the trajectory of
a point on the payload can be modified in real time and the tracking
task can still be guaranteed. }
\end{enumerate}

\section{\textcolor{black}{Preliminaries and Problem Formulation\label{sec:Preliminaries-and-Problem}}}

\textcolor{black}{To facilitate the subsequent analysis, definitions
of symbols are listed in Table \ref{tab1:Symbols-used-for}, \ref{tab2:Symbols-used-for},
and \ref{tab3:Notations}.}

\textcolor{black}{}
\begin{table}[H]
\textcolor{black}{\caption{\label{tab1:Symbols-used-for}Symbols used for the leader-follower
system.}
}
\centering{}\textcolor{black}{}%
\begin{tabular}{lll}
\hline 
\textcolor{black}{Symbol} &
\textcolor{black}{Description} &
\textcolor{black}{Units}\tabularnewline
\hline 
\textcolor{black}{$l_{1},\,l_{2}\in\mathbb{\mathbb{R}}$} &
\textcolor{black}{Lengths of the cables connected to } &
\textcolor{black}{$[\mathrm{m}]$}\tabularnewline
 &
\textcolor{black}{the leader and follower, respectively} &
\tabularnewline
\textcolor{black}{$m_{L},m_{F}\in\mathbb{\mathbb{R}}$} &
\textcolor{black}{Masses of the leader and follower} &
\textcolor{black}{$[\mathrm{kg}]$}\tabularnewline
\textcolor{black}{$p_{L},\,p_{F}\in\mathbb{R^{\mathbf{3}}}$} &
\textcolor{black}{Positions of the leader and follower} &
\textcolor{black}{$[\mathrm{m}]$}\tabularnewline
\textcolor{black}{$v_{L},\,v_{F}\in\mathbb{R^{\mathbf{3}}}$} &
\textcolor{black}{Velocities of the leader and follower} &
\textcolor{black}{$[\mathrm{\mathrm{\frac{m}{s}}}]$}\tabularnewline
\textcolor{black}{$\epsilon_{L},\,\epsilon_{F}\in\mathbb{R^{\mathbf{3}}}$} &
\textcolor{black}{Attitude errors of the leader and follower} &
\textcolor{black}{$[-]$}\tabularnewline
\textcolor{black}{$\omega_{L},\,\omega_{F}\in\mathbb{R^{\mathbf{3}}}$} &
\textcolor{black}{Angular rates of the leader and follower} &
\textcolor{black}{$[\mathrm{\frac{rad}{s}}]$}\tabularnewline
\textcolor{black}{$f_{\zeta}^{i}\in\mathbb{R}$} &
\textcolor{black}{Thrust of the $i^{th}$ propeller in the} &
\textcolor{black}{$[\mathrm{N}]$}\tabularnewline
 &
\textcolor{black}{{} leader or follower UAV} &
\tabularnewline
\textcolor{black}{$p_{c_{1}},\,p_{c_{2}}\in\mathbb{R^{\mathbf{3}}}$} &
\textcolor{black}{Positions of $c_{1}$ and $c_{2}$ } &
\textcolor{black}{$[\mathrm{m}]$}\tabularnewline
\textcolor{black}{$X^{I},\,Z^{I},\,Y^{I}\in\mathbb{R^{\mathbf{3}}}$} &
\textcolor{black}{The three axes of the inertial frame} &
\textcolor{black}{$[-]$}\tabularnewline
\textcolor{black}{$R_{L},R_{F}\in\mathsf{SO}(3)$} &
\textcolor{black}{The rotation matrices from the UAV } &
\textcolor{black}{$[-]$}\tabularnewline
 &
\textcolor{black}{body-fixed frame to the inertial frame} &
\tabularnewline
\textcolor{black}{$g\in\mathbb{R}$} &
\textcolor{black}{The gravitational constant} &
\textcolor{black}{$[\mathrm{\mathrm{\frac{m}{s^{2}}}}]$}\tabularnewline
\hline 
\end{tabular}
\end{table}
\textcolor{black}{}
\begin{table}[H]
\textcolor{black}{\caption{\label{tab2:Symbols-used-for}Symbols used for the payload.}
}
\centering{}\textcolor{black}{}%
\begin{tabular}{lll}
\hline 
\textcolor{black}{Symbol} &
\textcolor{black}{Description} &
\textcolor{black}{Units}\tabularnewline
\hline 
\textcolor{black}{$\theta\in\mathbb{R}$} &
\textcolor{black}{Yaw angle of the payload} &
\textcolor{black}{$[\mathrm{rad}]$}\tabularnewline
\textcolor{black}{$\omega\in\mathbb{R}$} &
\textcolor{black}{Yaw rate of the payload} &
\textcolor{black}{$[\mathrm{\frac{rad}{s}}]$}\tabularnewline
\textcolor{black}{$p_{p}\in\mathbb{R^{\mathbf{3}}}$} &
\textcolor{black}{Position of the payload CoG} &
\textcolor{black}{$[\mathrm{m}]$}\tabularnewline
\textcolor{black}{$v_{c_{2}}\in\mathbb{R^{\mathbf{3}}}$} &
\textcolor{black}{Velocity of the $c_{2}$ point } &
\textcolor{black}{$[\mathrm{\mathrm{\frac{m}{s}}}]$}\tabularnewline
\textcolor{black}{$R\in\mathsf{SO}(3)$} &
\textcolor{black}{The rotation matrix from the payload's } &
\textcolor{black}{$[-]$}\tabularnewline
 &
\textcolor{black}{body-fixed frame to the inertial frame } &
\tabularnewline
\textcolor{black}{$v_{p}^{I}\in\mathbb{R^{\mathbf{3}}}$} &
\textcolor{black}{Velocity of $p_{p}$ in the inertial frame} &
\textcolor{black}{$[\mathrm{\mathrm{\frac{m}{s}}}]$}\tabularnewline
\textcolor{black}{$v_{p}^{B}\in\mathbb{R^{\mathbf{3}}}$} &
\textcolor{black}{Velocity of $p_{p}$ in the body-fixed frame} &
\textcolor{black}{$[\mathrm{\mathrm{\frac{m}{s}}}]$}\tabularnewline
\textcolor{black}{$v\in\mathbb{R}$} &
\textcolor{black}{The component of $v_{p}^{B}$ along $x^{B}$ } &
\textcolor{black}{$[\mathrm{\mathrm{\frac{m}{s}}}]$}\tabularnewline
\textcolor{black}{$F_{F}\in\mathbb{R^{\mathbf{3}}}$} &
\textcolor{black}{External force applied by the follower} &
\textcolor{black}{$[\mathrm{N}]$}\tabularnewline
\textcolor{black}{$F_{L}\in\mathbb{R^{\mathbf{3}}}$} &
\textcolor{black}{External force applied by the leader} &
\textcolor{black}{$[\mathrm{N}]$}\tabularnewline
\textcolor{black}{$F_{F_{x}},\,F_{F_{y}}\in\mathbb{R}$} &
\textcolor{black}{The components of $F_{F}$ along } &
\textcolor{black}{$[\mathrm{N}]$}\tabularnewline
 &
\textcolor{black}{{} $x_{B}$ and $y_{B}$} &
\tabularnewline
\textcolor{black}{$F_{L_{x}},\,F_{L_{y}}\in\mathbb{R}$} &
\textcolor{black}{The components of $F_{L}$ along } &
\textcolor{black}{$[\mathrm{N}]$}\tabularnewline
 &
\textcolor{black}{{} $x_{B}$ and $y_{B}$} &
\tabularnewline
\textcolor{black}{$r_{c_{2}/p}\in\mathbb{R^{\mathbf{3}}}$} &
\textcolor{black}{Position vector from $p$ to $c_{2}$ } &
\textcolor{black}{$[\mathrm{m}]$}\tabularnewline
\textcolor{black}{$L\in\mathbb{R}$} &
\textcolor{black}{Length of the payload} &
\textcolor{black}{$[\mathrm{m}]$}\tabularnewline
\textcolor{black}{$I_{zz}\in\mathbb{R}$} &
\textcolor{black}{Moment of inertia of the payload } &
\textcolor{black}{$[\mathrm{kg\cdot m^{2}}]$}\tabularnewline
 &
\textcolor{black}{along $z^{B}$} &
\tabularnewline
\textcolor{black}{$\theta_{r}\in\mathbb{\mathbb{R}}$} &
\textcolor{black}{Reference yaw angle for the payload} &
\textcolor{black}{$[\mathrm{rad}]$}\tabularnewline
\textcolor{black}{$v_{r}\in\mathbb{\mathbb{R}}$} &
\textcolor{black}{Reference velocity for the payload} &
\textcolor{black}{$[\mathrm{\frac{m}{s}}]$}\tabularnewline
\textcolor{black}{$x_{r},\,y_{r}\in\mathbb{R}$} &
\textcolor{black}{Reference trajectory of the payload} &
\textcolor{black}{$[\mathrm{m}]$}\tabularnewline
\textcolor{black}{$m_{p}\in\mathbb{R}$} &
\textcolor{black}{Mass of the payload} &
\textcolor{black}{$[\mathrm{kg}]$}\tabularnewline
\textcolor{black}{$x^{B},\,z^{B},\,y^{B}\in\mathbb{R^{\mathbf{3}}}$} &
\textcolor{black}{The three axes of the body-fixed } &
\textcolor{black}{$[-]$}\tabularnewline
 &
\textcolor{black}{frame of the payload} &
\tabularnewline
\hline 
\end{tabular}
\end{table}

\textcolor{black}{}
\begin{table}[H]
\textcolor{black}{\caption{\label{tab3:Notations}Notations}
}
\centering{}\textcolor{black}{}%
\begin{tabular}{ll}
\hline 
\textcolor{black}{Symbol} &
\textcolor{black}{Description}\tabularnewline
\hline 
\textcolor{black}{$(\cdot)^{m}$} &
\textcolor{black}{Measurement of the state}\tabularnewline
\textcolor{black}{$\hat{(\cdot)}$} &
\textcolor{black}{Estimate of the state}\tabularnewline
\textcolor{black}{$(\cdot)_{p}$} &
\textcolor{black}{State of the payload}\tabularnewline
\textcolor{black}{$(\cdot)_{\zeta}$} &
\textcolor{black}{State of $\ensuremath{\zeta=\{L,F\}},$leader or
follower UAV}\tabularnewline
\textcolor{black}{$(\cdot)_{e}$} &
\textcolor{black}{Error signal}\tabularnewline
\textcolor{black}{$\phi,\,\psi,\,\theta$} &
\textcolor{black}{Roll, pitch, yaw angles}\tabularnewline
\textcolor{black}{$e_{1},\,e_{2},\,e_{3}$} &
\textcolor{black}{Orthonormal bases}\tabularnewline
\hline 
\end{tabular}
\end{table}

\subsection{\textcolor{black}{Preliminaries\label{subsec:Preliminaries}}}

\begin{figure}[H]
\begin{centering}
\textcolor{black}{\includegraphics[width=0.95\columnwidth]{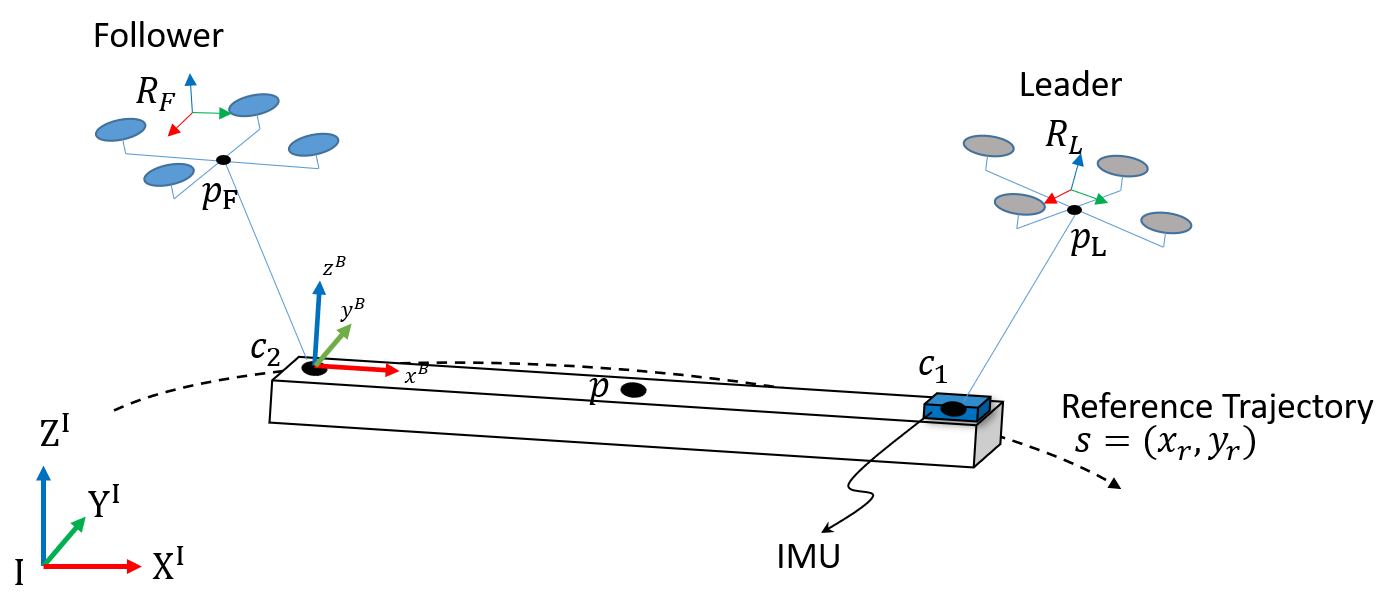}}
\par\end{centering}
\textcolor{black}{\caption{\label{fig2:nonholonomic 2D-motion-in x-y plane}The reference trajectory
of the leader-follower system is designed for point $c_{2}$ to track.
Point $p$ is the center of gravity (CoG) of the payload, and points
$p_{L}$ and $p_{F}$ are the positions of the leader and follower
UAVs, respectively. The attitudes of the leader UAV, follower UAV,
and the payload are $R_{L},$ $R_{F},$ and $R,$ respectively. The
IMU is fixed to the payload at $c_{1},$ and the measurement data
are sent to the leader via a wire.}
}
\end{figure}

\textcolor{black}{This section describes how the leader-follower UAV
system is controlled to behave as a nonholonomic system. As depicted
in Fig. \ref{fig:Transportation process}, the leader pulls the payload
and determines the system velocity while the follower is controlled
to ensure the payload to mimicking nonholonomic motion. As depicted
in Fig. \ref{fig2:nonholonomic 2D-motion-in x-y plane}, the reference
trajectory of $c_{2}$ of the payload can be denoted by $s:\left[0,\,\infty\right)\rightarrow\mathbb{R}^{2},$
defined as $s\left(t\right)\triangleq\left[x_{r}\left(t\right),\,y_{r}\left(t\right)\right]^{T}.$
$l_{1},\,l_{2}\in\mathbb{R}$ are the lengths of the cables connected
to the leader and the follower, respectively, and both cables remain
tight during transportation. Spherical joints are located at points
$c_{1}$ and $c_{2}$ on the payload and are connected to the leader
and the follower via cables, respectively, and $L\in\mathbb{R}$ is
the length of the payload. $l_{1}$, $l_{2},$ and $L$ are known
constants. An IMU is fixed to the payload at $c_{1},$ and the measurement
data are sent to the leader via a wire.}

\textcolor{black}{}

\begin{assumption}
\textcolor{black}{\label{Assumption: rigid body}The payload is a
rigid body. The string can only transmit a tension force (i.e., not
a compression force), and the forces at the two ends must be equal,
opposite, and collinear.}
\end{assumption}
\textcolor{black}{}
\begin{assumption}
\textcolor{black}{\label{Assumption 2D rotatation}The UAVs take off
vertically from the ground until the payload is lifted to a desired
height, and then the payload is transported horizontally (in 2D motion).
This means that vectors $z^{B}$ and $z^{I}$ are parallel, and so
the rotation angles of the payload $\left[\phi,\ \psi,\ \theta\right]$
can be simplified to $\left[0,\ 0,\ \theta\right],$ which is measured
by the IMU attached to the payload at $c_{1}.$ Also, the dimensions
of the IMU are negligible compared to those of the payload. }
\end{assumption}
\textcolor{black}{Given that the payload is a rigid body (as depicted
in Fig. \ref{fig2:nonholonomic 2D-motion-in x-y plane}), the velocities
of point $p$ defined in the inertial frame and in the body-fixed
frame are $v_{p}^{I}$ and $v_{p}^{B},$ respectively, and their relationship
can be expressed as
\begin{equation}
v_{p}^{I}=Rv_{p}^{B},\label{eq:v_p relative velocity}
\end{equation}
where $R$ denotes the rotation matrix representing the attitude of
the payload that satisfies Assumption \ref{Assumption 2D rotatation},
and is defined as:}

\textcolor{black}{
\begin{equation}
R=\left[\begin{array}{ccc}
\cos\theta & -\sin\theta & 0\\
\sin\theta & \cos\theta & 0\\
0 & 0 & 1
\end{array}\right],\label{eq:rotation matrix}
\end{equation}
where $\theta\in\mathbb{R}$ is defined as the angle between $x^{B}$
and the $X^{I}$ axis as depicted in Fig. \ref{fig2:nonholonomic 2D-motion-in x-y plane}. }
\begin{assumption}
\textcolor{black}{\label{Assumption 3:nonholonomic constraint}$c_{2}$
is connected to the follower UAV by a cable and transverse motion
is not allowed at $c_{2}$ (i.e., $y^{B}$ direction), which behaves
like nonholonomic motion; that is, the velocity of $c_{2}$ expressed
in the body-fixed frame of the payload is 
\[
v_{c_{2}}\triangleq\left[\begin{array}{ccc}
v_{c_{2,x}},\, & v_{c_{2,y}},\, & 0\end{array}\right]^{T}\thickapprox\left[\begin{array}{ccc}
v,\, & 0,\, & 0\end{array}\right]^{T}.
\]
}
\end{assumption}
\textcolor{black}{}
\textcolor{black}{Note that Assumption \ref{Assumption 3:nonholonomic constraint}
is made to facilitate the subsequent kinematics analysis, but it does
not affect the stability analysis. Furthermore, the design of the
follower controller in Section \ref{subsec:Follower-Scheme} is designed
to ensure $v_{c_{2,y}}\rightarrow0$ in order to satisfy Assumption
\ref{Assumption 3:nonholonomic constraint}.}

\subsection{\textcolor{black}{Tracking Error of Point $c_{2}$}}

\textcolor{black}{In general applications, reference trajectories
are usually provided in the inertial frame, and therefore the reference
position and orientation of point $c_{2}$ are defined in the inertial
frame as 
\begin{align}
p_{r} & =\left[\begin{array}{ccc}
x_{r}, & y_{r}, & z_{r}\end{array}\right]^{T}\in\mathbb{R}^{\mathbf{3}}\label{eq:ref trajectory}\\
\theta_{r} & =\tan^{-1}\left(\frac{\dot{y}_{r}}{\dot{x}_{r}}\right),\label{eq:ref orientation}
\end{align}
where $z_{r}$ is a constant. The positions of the UAVs and the payload
in the z$^{I}$ direction are assumed to be constant. Since control
inputs (i.e., force applied by UAVs) are defined in the body-fixed
frame, tracking errors $p_{e}\triangleq\left[\begin{array}{ccc}
x_{e}, & y_{e}, & 0\end{array}\right]^{T}\in\mathbb{R}^{\mathbf{3}}$ and $\theta_{e}\in\mathbb{R}$ are defined in the body-fixed frame
as
\begin{align}
p_{e} & \triangleq R^{-1}\left(p_{r}-p_{c_{2}}\right)\label{eq:eq:nonholonomic err transform}\\
\theta_{e} & \triangleq\theta_{r}-\theta\label{eq:theta_e}
\end{align}
where $p_{c_{2}}\triangleq\left[\begin{array}{ccc}
p_{c_{2,x}},\, & p_{c_{2,y}},\, & z_{r}\end{array}\right]^{T}\in\mathbb{R}^{\mathbf{3}}$ is defined as the position of $c_{2}$ on the payload, and its height
is the same as $z_{r}$ defined in (\ref{eq:ref trajectory}). }
\begin{assumption}
\textcolor{black}{\label{Assumption:smooth trajectory}The reference
trajectory is bounded and differentiable; that is, $p_{r},\,\dot{p}_{r},\,\ddot{p}_{r},\,\theta_{r},\,\dot{\theta}_{r},\,\ddot{\theta}_{r}\in\mathcal{L}_{\infty}.$ }
\end{assumption}

\subsection{\textcolor{black}{Control Objectives}}

\textcolor{black}{The control objective is defined in Problem \ref{def:The-control-objective}.}
\begin{problem}
\textcolor{black}{\label{def:The-control-objective}The control objective
is defined as
\begin{align}
p_{e} & \rightarrow0\text{ and }\text{\ensuremath{\theta_{e}\rightarrow}0}\text{ as }t\rightarrow\infty.\label{eq:control objectives}
\end{align}
}
\end{problem}

\section{\textcolor{black}{Kinematics and Dynamics\label{sec:nonholonomic motion}}}

\subsection{\textcolor{black}{Kinematics of the Tracking Error}}

\textcolor{black}{The derivation of the open-loop dynamics is further
facilitated by Definition \ref{def:skew-symmetric operator}.}
\begin{defn}
\textcolor{black}{\label{def:skew-symmetric operator}Given a vector
$\overrightarrow{\varPsi}\triangleq\left[\varPsi_{1},\,\varPsi_{2},\,\varPsi_{3}\right]^{T}\in\mathbb{R}^{3}$,
its skew-symmetric operator $\left[\cdot\right]_{\times}$ is defined
as 
\begin{equation}
\left[\overrightarrow{\varPsi}\right]_{\times}\triangleq\left[\begin{array}{ccc}
0 & -\varPsi_{3} & \varPsi_{2}\\
\varPsi_{3} & 0 & -\varPsi_{1}\\
-\varPsi_{2} & \varPsi_{1} & 0
\end{array}\right],\label{eq:angular velocity matrix}
\end{equation}
and the following equation holds 
\begin{equation}
\left[\overrightarrow{\varPsi}\right]_{\times}P=\overrightarrow{\varPsi}\times P,\label{eq:matrix and vector conversion}
\end{equation}
where $P\in\mathbb{R}^{3}.$}
\end{defn}
\textcolor{black}{}
\textcolor{black}{Using the definition in (\ref{eq:angular velocity matrix})
with the angular velocity vector of the payload $\overrightarrow{\omega}=\left[0,\,0,\,\omega\right]^{T}\in\mathbb{R}^{3}$
and taking the time derivative of (\ref{eq:eq:nonholonomic err transform})
yields
\begin{align}
\dot{p}_{e} & =\left(\frac{d}{dt}R^{-1}\right)\left(p_{r}-p_{c_{2}}\right)+R^{-1}\left(\dot{p}_{r}-\dot{p}_{c_{2}}\right)\nonumber \\
 & =\left(-\left[\overrightarrow{\omega}\right]_{\times}R^{-1}\right)\left(p_{r}-p_{c_{2}}\right)+R^{-1}\left(\dot{p}_{r}-\dot{p}_{c_{2}}\right)\nonumber \\
 & =-\left[\overrightarrow{\omega}\right]_{\times}p_{e}+R^{-1}\dot{p}_{r}-R^{-1}\dot{p}_{c_{2}},\label{eq:nonholonomic transform-2}
\end{align}
where $\omega\triangleq\dot{\theta}\in\mathbb{R}$ is defined as the
angular velocity of the payload along the $z^{B}$ direction. Combining
the definitions of $p_{e}$ and $\dot{p}_{r}$ with (\ref{eq:matrix and vector conversion}),
(\ref{eq:nonholonomic transform-2}) can be further expressed as 
\begin{align}
\dot{p}_{e} & =-\overrightarrow{\omega}\times p_{e}+R^{-1}\dot{p}_{r}-\left[\begin{array}{c}
v\\
0\\
0
\end{array}\right]\nonumber \\
 & =\left[\begin{array}{c}
\omega y_{e}+\dot{x}_{r}\cos\theta+\dot{y}_{r}\sin\theta-v\\
-\omega x_{e}-\dot{x}_{r}\sin\theta+\dot{y}_{r}\cos\theta\\
0
\end{array}\right].\label{eq:translation err dynamics}
\end{align}
Substituting the relations
\[
\begin{array}{c}
\dot{x}_{r}=v_{r}\cos\theta_{r}\\
\dot{y}_{r}=v_{r}\sin\theta_{r},
\end{array}
\]
into (\ref{eq:translation err dynamics}) and using Sum-Difference
formulas
\begin{align*}
\cos\theta_{e} & =\cos\left(\theta_{r}-\theta\right)\\
 & =\cos\theta_{r}\cos\theta+\sin\theta_{r}\sin\theta\\
\sin\theta_{e} & =\sin\left(\theta_{r}-\theta\right)\\
 & =\sin\theta_{r}\cos\theta-\cos\theta_{r}\sin\theta,
\end{align*}
the two terms in (\ref{eq:translation err dynamics}) can be rewritten
as
\begin{align}
\dot{x}_{r}\cos\theta+\dot{y}_{r}\sin\theta & =v_{r}\cos\theta_{e}\label{eq:vr cos}\\
-\dot{x}_{r}\sin\theta+\dot{y}_{r}\cos\theta & =v_{r}\sin\theta_{e},\label{eq:vr sin}
\end{align}
where $v_{r}\in\mathbb{R}$ is the velocity along the reference trajectory
defined in the body-fixed frame.}

\textcolor{black}{The following expression can be obtained based on
(\ref{eq:translation err dynamics}), (\ref{eq:vr cos}), and (\ref{eq:vr sin}):
\begin{align}
\dot{x}_{e} & =\omega y_{e}+v_{r}\cos\theta_{e}-v\nonumber \\
\dot{y}_{e} & =-\omega x_{e}+v_{r}\sin\theta_{e}\label{eq:open-loop dynamics 1}\\
\dot{\theta}_{e} & =\omega_{r}-\omega,\nonumber 
\end{align}
where $\omega_{r}\in\mathbb{R}$ represents the reference angular
velocity and is defined as the time derivative of $\theta_{r}$ defined
in (\ref{eq:ref orientation}). To facilitate the subsequent analysis,
two auxiliary signals defined as $v_{d},\,\omega_{d}\in\mathbb{R}$
are added to and subtracted from the first and third equations of
(\ref{eq:open-loop dynamics 1}), respectively, as
\begin{align}
\dot{x}_{e} & =\omega y_{e}+v_{r}\cos\theta_{e}+\eta_{1}-v_{d}\nonumber \\
\dot{y}_{e} & =-\omega x_{e}+v_{r}\sin\theta_{e}\label{eq:open-loop dynamics 2}\\
\dot{\theta}_{e} & =\omega_{r}+\eta_{2}-\omega_{d},\nonumber 
\end{align}
where $\eta_{1},\,\eta_{2}\in\mathbb{R}$ are defined as
\begin{align}
\eta_{1} & =v_{d}-v\label{eq:eta1}\\
\eta_{2} & =\omega_{d}-\omega.\label{eq:eta2}
\end{align}
}
\begin{rem}
\textcolor{black}{$v_{d}$ and $\omega_{d}$ are considered the kinematics
controller and are designed in Section \ref{subsec:Leader-Controller},
where the control objectives are to achieve $x_{e}\rightarrow0,$
 $y_{e}\rightarrow0,$ and $\theta_{e}\rightarrow0.$}
\end{rem}

\subsection{\textcolor{black}{Open-Loop Dynamics of the Payload}}

\textcolor{black}{Since the motion of the payload is constrained to
be nonholonomic, its dynamics needs to be modeled so that the controllers
for the leader and follower can be designed. As depicted in Fig. \ref{fig2:nonholonomic 2D-motion-in x-y plane},
the external forces applied to the payload are the tensions in the
cables connected to the leader and the follower, respectively, and
are defined as
\begin{align}
F_{L} & =RF_{L}^{B}\label{eq:FL}\\
F_{F} & =RF_{F}^{B},\label{FF}
\end{align}
where $R$ is the rotation matrix defined in (\ref{eq:rotation matrix}),
$F_{L}^{B}=[F_{L_{x}}^{B},\,F_{L_{y}}^{B},\,\frac{1}{2}mg]^{T}$,
$F_{F}^{B}=[F_{F_{x}}^{B},\,F_{F_{y}}^{B},\,\frac{1}{2}mg]^{T}$,
and $F_{L_{x}}^{B},F_{F_{x}}^{B}\in\mathbb{R}$ are defined as the
forces along the $x^{B}$ direction, and $F_{L_{y}}^{B},F_{F_{y}}^{B}\in\mathbb{R}$
are defined as the forces along the $y^{B}$ direction. Specifically,
the kinematics between points $c_{2}$ and $p$ on the payload need
to be considered to facilitate the design of the tracking controller
and is defined as
\begin{equation}
Rr_{\nicefrac{c_{2}}{p}}=p_{c_{2}}-p_{p},\label{eq:p to c2 position}
\end{equation}
where $p_{p}\in\mathbb{R}^{3}$ is the position vector of point $p.$
Taking the time derivative on both sides of (\ref{eq:p to c2 position})
yields
\begin{equation}
\dot{R}r_{\nicefrac{c_{2}}{p}}+R\dot{r}_{\nicefrac{c_{2}}{p}}=\dot{p}_{c_{2}}-\dot{p}_{p},\label{eq:p to c2 position-2}
\end{equation}
which can be rewritten as
\begin{equation}
\overrightarrow{\omega}\times\left(Rr_{\nicefrac{c_{2}}{p}}\right)=R\left[\begin{array}{c}
v\\
0\\
0
\end{array}\right]-\dot{p}_{p}\label{eq:p to c2 position-3}
\end{equation}
due to $\dot{r}_{\nicefrac{c_{2}}{p}}=0$ for the payload being a
rigid body (Assumption \ref{Assumption: rigid body}), where $r_{\nicefrac{c_{2}}{p}}\triangleq[-\frac{L}{2},\,0,\,0]^{T},$
and $\dot{p}_{c_{2}}=Rv_{c_{2}}$ based on Assumption \ref{Assumption 3:nonholonomic constraint}.
Taking the time derivative on both sides of (\ref{eq:p to c2 position-3})
yields
\begin{align}
\dot{\overrightarrow{\omega}}\times\left(Rr_{\nicefrac{c_{2}}{p}}\right)+\overrightarrow{\omega}\times\Bigl(\overrightarrow{\omega}\times\left(Rr_{\nicefrac{c_{2}}{p}}\right)\Bigr)= & \overrightarrow{\omega}\times\left(R\left[\begin{array}{c}
v\\
0\\
0
\end{array}\right]\right)\nonumber \\
 & +R\left[\begin{array}{c}
\dot{v}\\
0\\
0
\end{array}\right]-\ddot{p}_{p}.\label{eq:p to c2 position-4}
\end{align}
Multiplying both sides of (\ref{eq:p to c2 position-4}) by $R^{-1}$
obtains 
\begin{align}
\left[\begin{array}{c}
\dot{v}\\
0\\
0
\end{array}\right]= & -\overrightarrow{\omega}\times\left[\begin{array}{c}
v\\
0\\
0
\end{array}\right]+\overrightarrow{\omega}\times\left(\overrightarrow{\omega}\times r_{\nicefrac{c_{2}}{p}}\right)\nonumber \\
 & +R^{-1}\dot{v}_{p}+\dot{\overrightarrow{\omega}}\times r_{c_{2}/p},\label{eq:vc2_dot}
\end{align}
where $v_{p}\in\mathbb{R^{\mathbf{3}}}$ is defined as $\dot{p}_{p}$,
and $v\in\mathbb{R}$ is the velocity of $c_{2}$ defined in Assumption
\ref{Assumption 3:nonholonomic constraint}. Considering the dynamics
of the payload and the tensions in the two cables as its external
forces yields:
\begin{equation}
R^{-1}\dot{v}_{p}+\left[\begin{array}{c}
0\\
0\\
g
\end{array}\right]=\frac{1}{m_{p}}R^{-1}\left(F_{L}+F_{F}\right).\label{eq:payload equation of motion}
\end{equation}
By substituting (\ref{eq:payload equation of motion}) into (\ref{eq:vc2_dot})
and using (\ref{eq:FL}) and (\ref{FF}), the first row of the equation
can be considered as the open-loop dynamics:
\begin{align}
\dot{v} & =\frac{1}{m_{p}}\left(F_{L_{x}}^{B}+F_{F_{x}}^{B}\right)-r_{\nicefrac{c_{2}}{p,x}}\omega^{2}.\label{eq:open-loop 1}
\end{align}
Given the rotational dynamics of the payload
\[
r_{c_{2}/p}\times F_{F}^{B}+r_{c_{1}/p}\times F_{L}^{B}=I_{zz}\dot{\omega},
\]
where $r_{\nicefrac{c_{1}}{p}}\triangleq[\frac{L}{2},\,0,\,0]^{T},$
the following equation can be obtained
\begin{align}
\dot{\omega} & =\frac{L}{2I_{zz}}(F_{L_{y}}^{B}-F_{F_{y}}^{B}),\label{eq:open-loop 2}
\end{align}
where $I_{zz}$ denotes the moment of inertia of the payload along
the $Z$ axis passing through its center of gravity, $p_{p}$.}
\begin{rem}
\textcolor{black}{The open-loop dynamics described by (\ref{eq:open-loop 1})
and (\ref{eq:open-loop 2}) characterize the relation between the
payload and the two agents, and $F_{L_{x}}^{B},$ $F_{L_{y}}^{B},$
$F_{F_{x}}^{B},$ and $F_{F_{y}}^{B}$ are the control inputs designed
in Section \ref{sec:Controller} to ensure $v\rightarrow v_{d}$ and
$\omega\rightarrow\omega_{d}$ (i.e., $\eta_{1}\rightarrow0$ and
$\eta_{2}\rightarrow0$).}
\end{rem}

\section{\textcolor{black}{Estimation using UKFs\label{sec:UKF-estimate}}}

\textcolor{black}{This section describes how three UKFs are applied
to estimate the signals that are required for feedback control in
the leader and follower controllers. The design of the UKFs is based
on a previous study \cite{EricA.Wan2000}, and it is superior to an
extended Kalman filter for several reasons: }
\begin{enumerate}
\item \textcolor{black}{It is accurate to two terms of the Taylor expansion.}
\item \textcolor{black}{It has a higher efficiency since it does not require
sufficient differentiability of the state dynamics. }
\item \textcolor{black}{It provides a derivative-free way to estimate the
state parameters of nonlinear systems by introducing an \textquotedblleft unscented
transformation.''}
\end{enumerate}
\textcolor{black}{In this study, one of the UKFs runs on the follower,
and the other two run on the leader. The differences in the UKFs implemented
on the follower and leader are as follows: }
\begin{enumerate}
\item \textcolor{black}{Follower: the velocity of follower $v_{F}$ and
the force applied by the follower to payload $F_{F}$ are estimated
based on the kinematics and measurements of the follower (i.e., position
and acceleration as measured by onboard sensors), and the estimation
are used later in feedback control for the follower.}
\item \textcolor{black}{Leader: in the first UKF, the states that are the
same as the follower are estimated (i.e., $v_{L}$, $F_{L}$), and
they are the required feedback signals for implementing the leader
controller.}
\item \textcolor{black}{Leader: in the second UKF, $p_{c_{1}},$ $v_{c_{1}},$
and $p_{c_{2}}$ are estimated based on the kinematics of the payload
and are used later in feedback control for the leader.}
\end{enumerate}

\subsection{\textcolor{black}{UKF Algorithm}}

\textcolor{black}{The kinematics and dynamics model derived in the
following two subsections can be considered as process models and
can be discretized using the forward Euler method as \cite{Tagliabue2016}
\begin{align}
x_{k+1} & =f\left(x_{k},u_{k}\right)+w_{k}\label{eq:UKF Algorithm process model}
\end{align}
where $x_{k}$ and $u_{k}$ are the state vector and thrust input
in the current step, respectively, $x_{k+1}$ is the state vector
in the next step, and $w_{k}$ is the process noise. State $x_{k}$
is the signal to be estimated by the UKF. Because the model for the
external force is unknown, it is assumed that the external force can
be updated as follows:
\begin{equation}
F_{k+1}^{ext}=F_{k}^{ext}.\label{eq:UKF Algorithm force process}
\end{equation}
}

\subsection{\textcolor{black}{The UKF on the Follower\label{subsec:UKF follower}}}

\begin{figure}[H]
\begin{centering}
\textcolor{black}{\includegraphics[width=0.3\columnwidth]{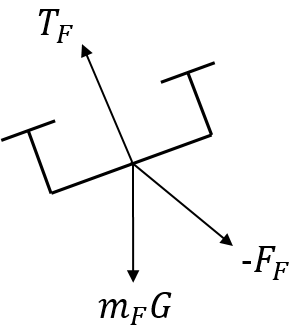}}
\par\end{centering}
\textcolor{black}{\caption{\label{fig:The forces exert on the follower}The forces exerted on
the follower.}
}
\end{figure}

\textcolor{black}{This UKF is used to estimate the follower velocity
and the external force applied by the follower. Assumption \ref{thm:external force of the follower}
is made without any loss of generality.}
\begin{assumption}
\textcolor{black}{\label{thm:external force of the follower}External
forces exerted on the leader and follower, with the exception of the
tension in the cable, are considered as disturbances (e.g., drag,
wind gust), and therefore are not modeled .}
\end{assumption}
\textcolor{black}{}
\textcolor{black}{The free-body diagram of the follower is depicted
in Fig. \ref{fig:The forces exert on the follower}, where the tension
in the cable (i.e., $F_{F}$) that connects to the payload is the
only external force based on Assumption \ref{thm:external force of the follower}.
However, follower vibration caused by unmodeled disturbances or noise
are addressed by the controller developed in Section \ref{sec:Controller}.}

\textcolor{black}{}
\textcolor{black}{The dynamics of the follower is used by the UKF
for state estimation and can be described as follows:
\begin{align}
\dot{p}_{F} & =v_{F}\label{eq:pf}\\
m_{F}\dot{v}_{F} & =R_{F}T_{F}-m_{F}G-F_{F},\label{eq:af}
\end{align}
where $p_{F}\mathbb{\in R}^{3}$ and $m_{F}\in\mathbb{R}$ are the
position and mass of the follower, respectively, $R_{F}\in SO(3)$
is the rotation matrix from the body-fixed frame of the follower to
the inertial frame and is obtained from the flight controller, $T_{F}=[0,\,0,\,T_{F_{z}}]^{T}$
is the thrust, $v_{F}\in\mathbb{R}^{3}$ is the follower velocity
expressed in the inertial frame, $G=[0,\,0,\,g]^{T}$, where $g\in\mathbb{R}$
is the gravity, and $F_{F}\in\mathbb{R}^{3}$ is the external force
to be estimated. The state of the UKF is defined as
\begin{eqnarray}
\hat{x}_{F} & = & \left[\begin{array}{ccccc}
\hat{p}_{F}^{T} & \hat{v}_{F}^{T} & \hat{F}_{F}^{T} & \hat{\epsilon}_{F}^{T} & \hat{\omega}_{F}^{T}\end{array}\right]^{T},\label{eq:xF}
\end{eqnarray}
where $\hat{\left(\cdot\right)}$ denotes the state estimate, $\omega_{F}^{T}\in\mathbb{R}^{3}$
is the angular velocity, and $\epsilon_{F}^{T}\in\mathbb{R^{\mathbf{3}}}$
is the error quaternion. The measurement is defined as
\begin{eqnarray}
y_{F} & = & \left[\begin{array}{cccc}
p_{_{F}}^{m}{}^{T}, & v_{_{F}}^{m}{}^{T}, & \epsilon_{_{F}}^{m}{}^{T}, & \omega_{_{F}}^{m}{}^{T}\end{array}\right]^{T},\label{eq:yF}
\end{eqnarray}
where the superscript $\left(\cdot\right)^{m}$ is used to indicate
a measured value, $p_{_{F}}^{m}$ is the follower's position as measured
by an onboard positioning sensor (e.g., GPS, motion capture systems,
or a visual odometry system), $v_{_{F}}^{m}$ is obtained from the
sensor fusion in the flight controller, and $\omega_{_{F}}^{m}$ is
also  obtained from the IMU in the follower flight controller. To
avoid a singularity of the state covariance, \cite{Crassidis2003}
presents an unscented quaternion estimation based on error quaternion
$\epsilon_{F}^{T}\in\mathbb{R^{\mathbf{3}}}$ instead of the attitude
quaternion. However, the attitude quaternion is considered the feedback
state in a general control system. \cite{Tagliabue2017} uses modified
Rodrigues parameters to convert the error quaternion into the attitude
quaternion.}
\begin{rem}
\textcolor{black}{$\hat{v}_{F}$ and $\hat{F}_{F}$ defined in (\ref{eq:xF})
are used as feedforward and feedback signals in the follower controller.}
\end{rem}

\subsection{\textcolor{black}{Two UKFs on the Leader\label{subsec:UKF-leader}}}

\subsubsection{\textcolor{black}{The First UKF}}

\textcolor{black}{The first UKF is used to estimate the external force
applied by the leader and the leader velocity. Since the dynamics
model of the leader is same as that of the follower, the same state
and measurement as defined in Section \ref{subsec:UKF follower} can
be used to derive the leader state; that is, by replacing the subscript
$F$ by $L$ in (\ref{eq:pf})-(\ref{eq:yF}) and applying the UKF,
estimates $v_{L}$ and $F_{L}$ can be obtained.}

\subsubsection{\textcolor{black}{The Second UKF\label{subsec:The-Second-UKF}}}

\textcolor{black}{In the second UKF, the velocity of $c_{2}$ and
the force applied by follower $F_{F}$ are estimated based on the
kinematics model of the payload and measurements. }

\textcolor{black}{}
\begin{figure}[H]
\begin{centering}
\textcolor{black}{\includegraphics[width=1\columnwidth]{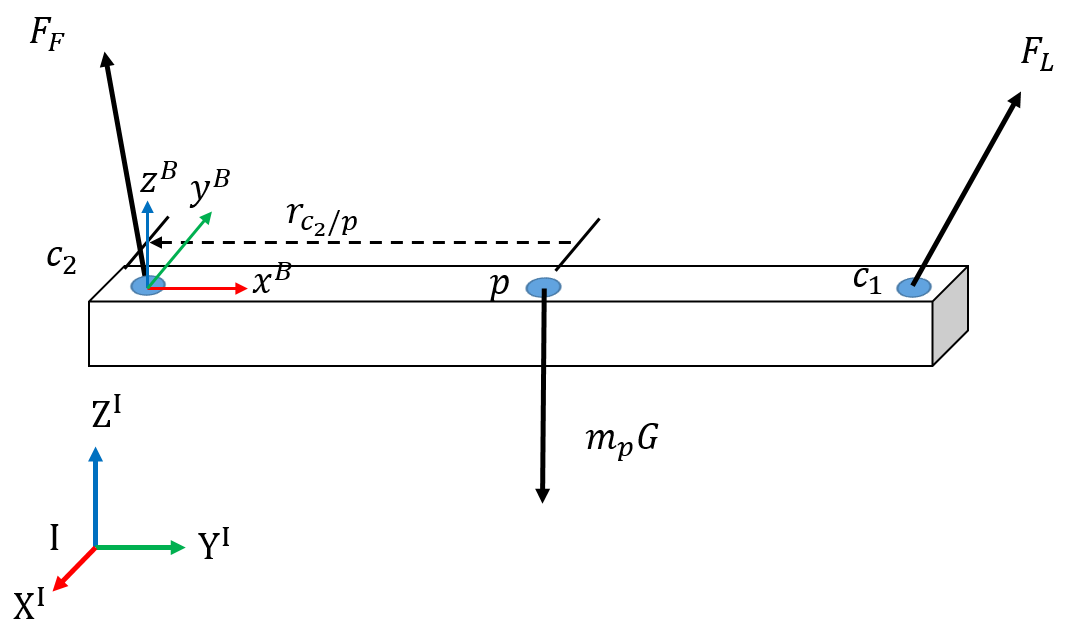}}
\par\end{centering}
\textcolor{black}{\caption{\label{fig:Free-body-diagram-of-payload}Definition of coordinate
frames and the free-body diagram of the payload, where point $p$
is the CoG, and $B$ and $I$ denote the body-fixed and inertial frames,
respectively.}
}
\end{figure}
\textcolor{black}{The dynamics of the payload as depicted in Fig.
\ref{fig:Free-body-diagram-of-payload} is required for state estimation
and is described as
\begin{align}
v_{c_{1}} & =\dot{p}_{c_{1}}\nonumber \\
a_{c_{1}} & =\dot{v}_{c_{1}}\nonumber \\
a_{p} & =a_{c_{1}}+\dot{\overrightarrow{\omega}}\times(Rr_{\nicefrac{p}{c_{1}}})+\overrightarrow{\omega}\times(\overrightarrow{\omega}\times(Rr_{\nicefrac{p}{c_{1}}}))\label{eq:a_p}\\
F_{F} & =m_{p}a_{p}-F_{L}+m_{p}G,\label{eq:F_Lz}
\end{align}
where $r_{\nicefrac{p}{c_{1}}}\triangleq[-\frac{L}{2},\,0,\,0]^{T},$
$F_{L}$ is obtained from the first leader UKF, $a_{p}\in\mathbb{R}^{3}$
is the acceleration of point $p,$ and $a_{c_{1}}\in\mathbb{R}^{3}$
denotes the acceleration of point $c_{1}$ and is measured from the
IMU attached to the payload as described in Section \ref{subsec:Preliminaries}.
Note that $\dot{\overrightarrow{\omega}}$ can be obtained by taking
the time derivative of the measured $\overrightarrow{\omega}.$ }

\textcolor{black}{Based on (\ref{eq:a_p}) and (\ref{eq:F_Lz}), the
state of the UKF is defined as
\begin{equation}
\hat{x}_{L}=\left[\hat{p}_{c_{1}}^{T},\,\hat{v}_{c_{1}}^{T},\,\hat{a}_{c_{1}}^{T},\,\hat{F}_{F}^{T}\right]^{T},\label{eq:x_p}
\end{equation}
and its measurement is
\begin{equation}
y_{L}=\left[p_{c_{1}}^{m}{}^{T},\,a_{c_{1}}^{m}{}^{T}\right]^{T},\label{eq:y_p}
\end{equation}
where $a_{c_{1}}^{m}$ is obtained based on the IMU attached to the
payload as described in Section \ref{subsec:Preliminaries}, and $p_{c_{1}}^{m}$
is obtained based on estimated force $\hat{F}_{L}$ as
\begin{equation}
\hat{p}_{c_{1}}=p_{L}-\frac{\hat{F}_{L}}{\|\hat{F}_{L}\|}l_{1},\label{eq:p_c1 using UKF}
\end{equation}
where $p_{L}\in\mathbb{R}^{3}$ is defined as the leader's position
as shown in Fig. \ref{fig2:nonholonomic 2D-motion-in x-y plane} and
is known from an onboard positioning sensor (e.g., GPS). $\hat{p}_{c_{1}}$
can be used to estimate $p_{c_{2}}$ based on the length and orientation
of the payload:
\begin{align}
\hat{p}_{c_{2}} & =\hat{p}_{c_{1}}+2Rr_{\nicefrac{p}{c_{1}}}.\label{eq:p_c1}
\end{align}
}

\textcolor{black}{Since $v$ is the required signal to obtain $\eta_{1}$
defined in (\ref{eq:eta1}), the following relation obtained based
on Assumption \ref{Assumption: rigid body} can be utilized:
\begin{equation}
\hat{v}=\hat{v}_{c_{1}}^{B}\cdot[1,\,0,\,0]^{T},\label{eq:v_c2}
\end{equation}
where the velocities of $c_{1}$ and $c_{2}$ along direction $x^{B}$
are equal.}
\begin{rem}
\textcolor{black}{Since inter-UAV communication is not used, (\ref{eq:p_c1})
provides a way for the leader to obtain the position of $c_{2},$
which is required in the leader controller to achieve trajectory tracking.
Specifically, $\hat{p}_{c_{1}}$ defined in (\ref{eq:p_c1 using UKF})
is obtained based on $\hat{F}_{L},$ which is estimated using the
leader UKF, and therefore communication is not required. }
\end{rem}

\section{\textcolor{black}{Controller Design\label{sec:Controller}}}

\textcolor{black}{To ensure control robustness, a backstepping controller
for the leader is developed to ensure that the payload achieves trajectory
tracking. In contrast, a switching controller along with triggering
conditions is developed for the follower in order to avoid vibration
on the follower, since the follower controller is sensitive to the
estimate error obtained from the UKFs.}

\subsection{\textcolor{black}{\label{subsec:Leader-Controller}Leader Controller}}

\begin{figure}[H]
\centering{}\textcolor{black}{\includegraphics[width=0.9\columnwidth]{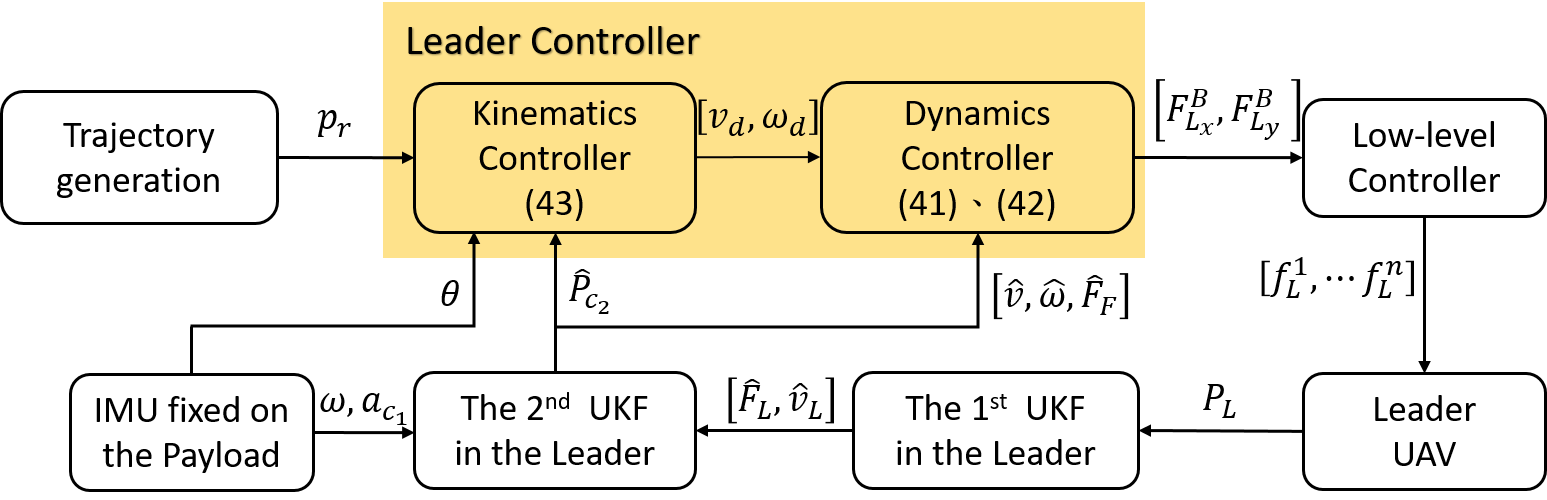}\caption{\label{fig:leader controller}Control block diagram of the leader
UAV controller. The reference trajectory for the controller is generated
by a quadratic programming (QP) motion planner. The leader controller
consists of a kinematics controller and a dynamics controller as defined
in (\ref{eq:leader controller1}), (\ref{eq:leader controller 2}),
and (\ref{eq:control input of kinematics}). The signals for controller
implementation are estimated from the first and second UKFs on the
leader.}
}
\end{figure}

\textcolor{black}{This section describes a backstepping controller
designed for the leader controller. Fig. \ref{fig:leader controller}
shows a block diagram of the controller. The backstepping controller
consists of two controllers, the kinematics controller and the dynamics
controller, which are cascaded by signals $v_{d}$ and $\omega_{d}$.
Specifically, $v_{d}$ and $\omega_{d}$ are considered as the kinematics
controller to ensure that the payload undergoes nonholonomic motion,
and also act as reference signals for the dynamics controller to achieve
the objective
\[
\eta_{1}\rightarrow0\text{ and \ensuremath{\eta_{2}\rightarrow0}.}
\]
}

\textcolor{black}{}

\begin{rem}
\textcolor{black}{In Fig. \ref{fig:leader controller}, the estimates
from the two UKFs are either feedback or feedforward signals, and
the estimate errors are first considered to be zero in order to facilitate
the development of the controller design. The system stability affected
by the estimate error is analyzed in Section \ref{subsec:Robustness of the leader controller}. }
\end{rem}
\textcolor{black}{Leader controller $F_{L_{x}}^{B}$ and $F_{L_{y}}^{B}$
are designed as
\begin{align}
F_{L_{x}}^{B}= & m_{p}\left(\dot{v}_{d}+x_{e}+k_{v}\eta_{1}\right)\label{eq:leader controller1}\\
 & +m_{p}\Bigl(r_{\nicefrac{c_{2}}{p,x}}\omega^{2}\Bigr)\nonumber \\
F_{L_{y}}^{B}= & \frac{2I_{zz}}{L}\left(\dot{\omega}_{d}+k_{\omega}\eta_{2}+\frac{\sin\theta_{e}}{k_{2}}\right)+F_{F_{y}}^{B},\label{eq:leader controller 2}
\end{align}
where $F_{F_{y}}$ can be estimated from the leader UKF defined in
Section \ref{sec:UKF-estimate}, and the kinematics controller defined
by $v_{d}$ and $\omega_{d}$ defined in (\ref{eq:open-loop dynamics 2})
are designed as
\begin{equation}
\left[\begin{array}{c}
v_{d}\\
\omega_{d}
\end{array}\right]=\left[\begin{array}{c}
v_{r}\cos(\theta_{e})+k_{1}x_{e}\\
\omega_{r}+v_{r}k_{2}y_{e}+k_{3}\sin\theta_{e}
\end{array}\right].\label{eq:control input of kinematics}
\end{equation}
}
\begin{rem}
\textcolor{black}{Although the reference trajectory is defined for
$c_{2},$ its tracking errors $x_{e},$ $y_{e},$ and $\theta_{e}$
can be obtained by the leader to implement the controller defined
in (\ref{eq:leader controller1}) and (\ref{eq:leader controller 2});
that is, the reference trajectory of $c_{2}$ is known to the leader
and $p_{c_{2}}$ can be obtained by the leader based on its position
(\ref{eq:p_c1}), which does not require communication with the follower.
Moreover, the trajectory can be modified by the leader online, which
is another advantage over other approaches.}
\end{rem}

\subsection{\textcolor{black}{Follower Controller\label{subsec:Follower-Scheme}}}

\subsubsection{\textcolor{black}{Control Objectives}}

\textcolor{black}{For the payload to mimic nonholonomic motion, the
control objective for the follower can be defined as 
\begin{equation}
F_{Fx}^{B}\rightarrow0,\ v_{c_{2},y}\rightarrow0,\label{eq:control_objective}
\end{equation}
where $F_{Fx}^{B}$ is the follower force applied to the payload as
defined in (\ref{FF}), and $v_{c_{2},y}$ is defined in Assumption
\ref{Assumption 3:nonholonomic constraint}. When $F_{Fx}^{B}\rightarrow0$
is close to being achieved, the follower will follow the leader in
order to eliminate the internal force in the longitudinal direction.
To avoid Zeno behavior (i.e., high frequency switching behavior) induced
by the error in the force estimate, a triggering mechanism and switching
controllers are developed as described below.}

\subsubsection{\textcolor{black}{Triggering Mechanism\label{subsec:Triggering-Mechanism}}}

\begin{figure}[H]
\begin{centering}
\textcolor{black}{\includegraphics[width=0.9\columnwidth]{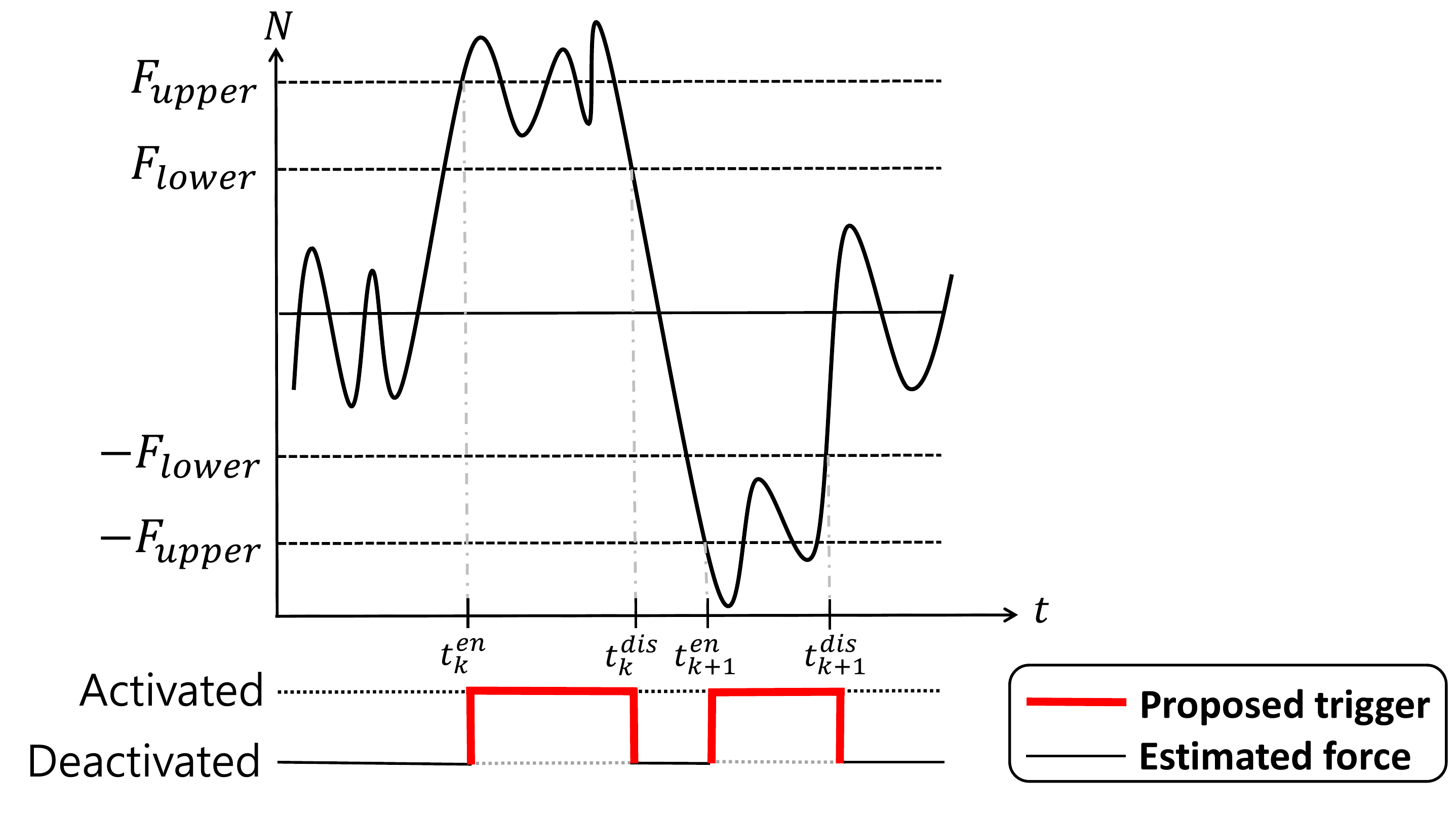}}
\par\end{centering}
\textcolor{black}{\caption{\label{fig:Trigger-time-of}Triggering mechanism developed based on
the estimated external force $\hat{F}_{F_{x}}$. As the figure depicts,
there are two different thresholds for the activated and deactivated
states. The use of thresholds $F_{upper}$ and $F_{lower}$ allows
the sensitivity of the trigger to be adjusted. }
}
\end{figure}

\textcolor{black}{To facilitate the design of triggering conditions,
the time interval sets are defined as:
\begin{align}
t_{k}^{dis} & =\left\{ t\in\left[0\ ,\infty\right)\mid\left|\hat{F}_{F_{x}}\left(t\right)\right|=F_{lower}\right\} \nonumber \\
t_{k}^{en} & =\min\left\{ t_{k-1}^{dis}<t<t_{k}^{dis}\mid\left|\hat{F}_{F_{x}}\left(t\right)\right|=F_{upper}\right\} \nonumber \\
T^{en} & =\{t\in\left[0\ ,\infty\right)\mid t_{k}^{en}\ <t<t_{k}^{dis}\}\label{eq:T^en}\\
T^{dis} & =\left[0\ ,\infty\right)\setminus T^{en},\label{eq:T^dis}
\end{align}
where $\hat{F}_{F_{x}}$ is the first component of $\hat{F}_{F}$,
$k=0,\,1,\,2,\,\cdots,$ $t_{k}^{en}$ and $t_{k}^{dis}$ denote the
sets consisting of discrete time points, respectively, $T^{en}$ and
$T^{dis}$ represent the sets of time intervals when the impedance
controller is activated and deactivated, respectively, and $F_{lower},\:F_{upper}\in\mathbb{R}$
are constant thresholds that are selected based on the mass of the
payload. As depicted in Fig. \ref{fig:Trigger-time-of}, given the
same payload, selecting higher values of $F_{lower}$ and $F_{upper}$
results in less-frequent activation of the force controller, leading
to a larger internal force in the payload, with less-smooth motion
also being expected. }

\textcolor{black}{}

\subsubsection{\textcolor{black}{Follower Controller for Longitudinal Motion}}

\textcolor{black}{To achieve the control objective $F_{Fx}^{B}\rightarrow0$,
a force controller is designed based on the triggering condition developed
in Section \ref{subsec:Triggering-Mechanism} as follows:
\begin{eqnarray}
F_{F_{x}}^{B}=\begin{cases}
k_{F_{1}}\hat{v}_{F_{x}}^{B}+\hat{F}_{F_{x}}^{B}, & if\;t\in T^{en}\\
k_{F_{2}}\Bigl(p_{d,x}-p_{F,x}\Bigr), & if\;t\in T^{dis}
\end{cases}\label{eq:follower controller 1}
\end{eqnarray}
where $\hat{v}_{F_{x}}^{B}\in\mathbb{R}$ is the element of the follower
velocity:
\[
\hat{v}_{F}^{B}\triangleq[\hat{v}_{F_{x}}^{B},\,\hat{v}_{F_{y}}^{B},\,0]^{T}
\]
and $\hat{F}_{F_{x}}^{B}$ defined in (\ref{FF}) is estimated by
the UKF developed in the Section \ref{subsec:UKF follower},  $k_{F_{1}},\,k_{F_{2}}\in\mathbb{R}_{>0}$
are the control gains, $p_{F,x}$ is the first component of follower
position $p_{F,x},$ and $p_{d,x}$ is the position along the $X$
axis at the time instant when the triggering condition is switching
to $t\in T^{dis}.$ Since $F_{F_{x}}^{B}$ is close to zero, the estimate
error can cause vibration, which triggers the second condition to
activate the second controller. Since the controller defined in (\ref{eq:follower controller 1})
ensures $F_{Fx}^{B}\rightarrow0,$ the equation of motion for the
$x$-axis is neglected.}
\begin{rem}
\textcolor{black}{$\hat{F}_{F}$ in (\ref{eq:follower controller 1})
is estimated from the follower UKF, while $\hat{F}_{F}$ in Fig. \ref{fig:leader controller}
is estimated from the leader UKF.}
\end{rem}

\subsubsection{\textcolor{black}{Follower Controller for Transverse Motion}}

\begin{figure*}[t]
\begin{centering}
\textcolor{black}{\includegraphics[width=0.55\paperwidth]{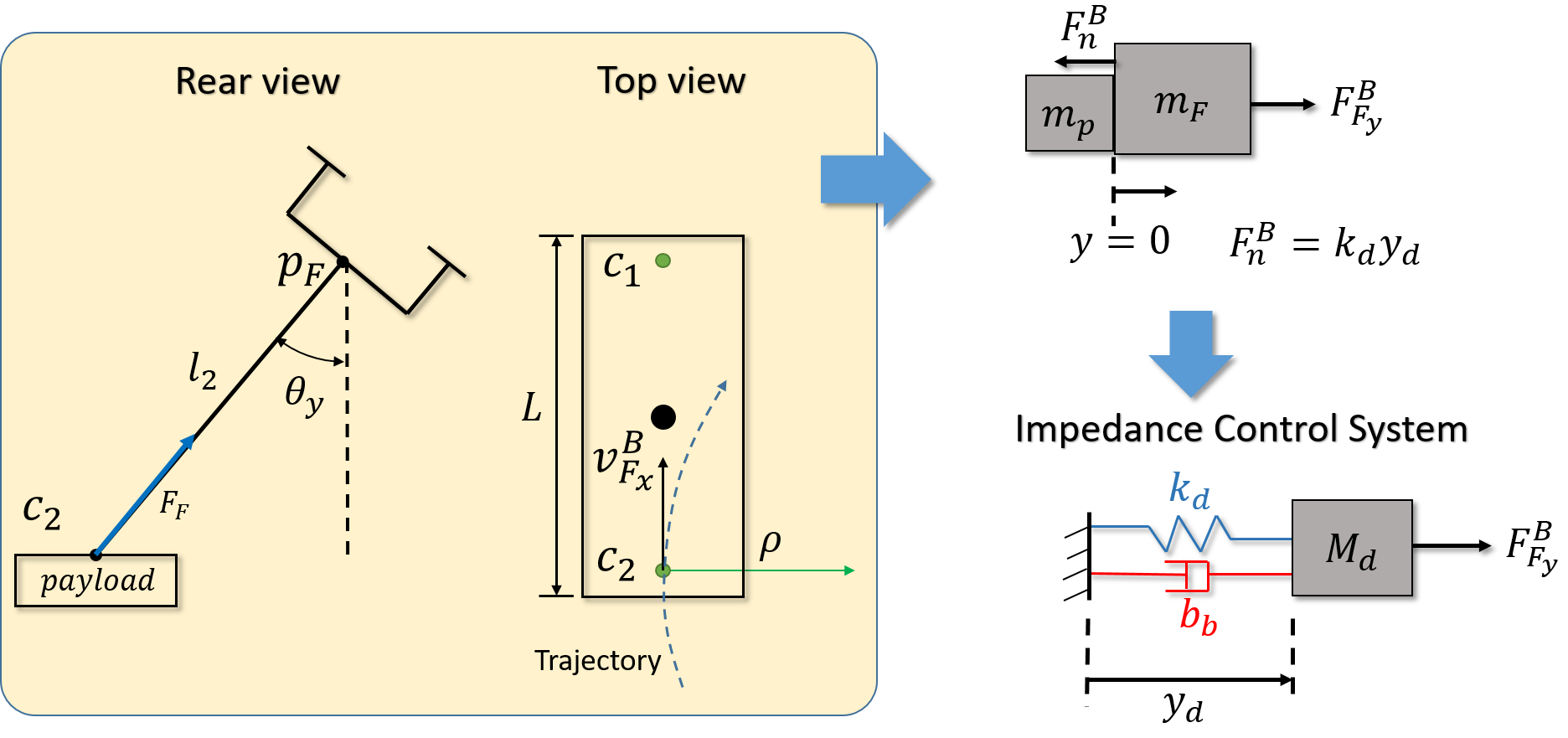}\caption{\label{fig:impedance control}The free-body diagram and kinematics
of the payload in the transverse direction, and conversion into an
impedance control system.}
}
\par\end{centering}
\end{figure*}

\textcolor{black}{The design of this part is based on an impedance
controller as depicted in Fig. \ref{fig:impedance control}, where
the relation between the payload and the follower in the transverse
direction can be first modeled as
\begin{equation}
m_{F}\ddot{y}=F_{F_{y}}^{B}-F_{n}^{B},\label{eq:follower transversal model}
\end{equation}
where $F_{n}^{B}=\frac{m_{p}}{2}a_{n}$ represents the required centripetal
force for the payload to follow a curved trajectory, where $a_{n}=\hat{v}_{c_{2,x}}^{2}/\rho\in\mathbb{R}$
represents the centripetal acceleration, with $\rho\in\mathbb{R}$
being the radius of curvature that is obtained based on the latest
trajectory approximated by a polynomial based on QP optimization. }

\textcolor{black}{}
\textcolor{black}{The dynamics in (\ref{eq:follower transversal model})
can then be converted into an impedance control system in order to
minimize the oscillation between the two so as to achieve nonholonomic
motion that satisfies Assumption \ref{Assumption 3:nonholonomic constraint}.
Force controller $F_{F_{y}}^{B}$ is designed as 
\begin{equation}
F_{F_{y}}^{B}=F_{n}^{B}-\frac{m_{F}}{M_{d}}(b_{d}\dot{e}+k_{d}e),\label{eq:follower controller 2}
\end{equation}
where $e\triangleq y-y_{d}\in\mathbb{R}$ represents the tracking
error, $\dot{y}_{d}$ and $\ddot{y}_{d}$ are zero, $M_{d},\,b_{d},\,k_{d}\in\mathbb{R}$
denote the desired inertia, damping, and stiffness, respectively,
with $k_{d}$ selected to be $k_{d}=\frac{m_{p}g}{2l_{2}\cos\theta_{y,d}}$
so that the virtual spring force can provide centripetal force $k_{d}y_{d}=F_{n}^{B},$
where $\hat{v}_{F_{x}}$ is used as an alternative of $v$ since $v_{F_{x}}$
is controlled to approach $v$ based on the controller defined in
(\ref{eq:follower controller 1}), $y=l_{2}\sin\theta_{y},$ where
$\mathbf{\theta}_{y}$ satisfies $\cos\theta_{y}=\frac{\nicefrac{m_{p}g}{2}}{\hat{F}_{F}}$,
$y_{d}=l_{2}\sin\theta_{y,d}$ (where $\theta_{y,d}$ satisfies $\tan\theta_{y,d}=\frac{a_{n}}{g}$).
The controller described by (\ref{eq:follower controller 2}) is designed
to achieve $v_{c_{2},y}\rightarrow0$, but the convergence of $v_{c_{2},y}$
to zero is subject to the accuracy of estimate $\hat{F}_{F_{y}}.$}

\textcolor{black}{Substituting (\ref{eq:follower controller 2}) into
(\ref{eq:follower transversal model}) yields the closed-loop dynamics:
\[
M_{d}\ddot{e}+b_{d}\dot{e}+k_{d}e=0,
\]
 where the tracking error goes to zero (i.e., $y\rightarrow y_{d}$)
provided that the gains for the second-order linear system are selected
properly.}
\begin{rem}
\textcolor{black}{The follower controller does not require the reference
trajectory for controller implementation. In addition, follower controller
$F_{F}$ can be estimated from the leader's UKF and compensated by
leader controller $F_{L_{y}}^{B}$ defined in (\ref{eq:leader controller 2}).}
\end{rem}

\subsection{\textcolor{black}{Low-Level Controller for the Leader and Follower}}

\textcolor{black}{Similar to (\ref{eq:af}), the dynamics and the
net thrust of the rotors on the UAVs can be given as}

\textcolor{black}{
\begin{equation}
m_{\zeta}\dot{v}_{\zeta}=T_{\zeta}R_{\zeta}e_{3}-F_{\zeta}-m_{\zeta}ge_{3},\label{eq:geometric controller}
\end{equation}
where $e_{3}=\left[\begin{array}{ccc}
0, & 0, & 1\end{array}\right]^{T},$ $\zeta\in\left\{ L,\,F\right\} $ indicates the leader or follower,
$R_{\zeta}\in SO(3)$ denotes the attitude of the UAVs, $F_{\zeta}$
is defined in (\ref{eq:FL}) or (\ref{FF}), and $T_{\zeta}\in\mathbb{R}$
represents the net thrust generated by the rotors on the UAV. Based
on \cite{Lee2010}, the geometric tracking controller for the leader
and follower UAVs is designed as
\begin{align}
T_{\zeta} & =\left(F_{\zeta}+m_{\zeta}ge_{3}+m_{\zeta}\dot{v}_{\zeta,d}\right)\cdot R_{\zeta}e_{3},\label{eq:geometric controller thrust}\\
M_{\zeta} & =k_{e,\zeta}e_{R_{\zeta}}+k_{\Omega,\zeta}e_{\Omega_{\zeta}}+\Omega_{\zeta}\times J_{\zeta}\Omega_{\zeta}\nonumber \\
 & -J_{\zeta}(\left[\Omega_{\zeta}\right]_{\times}R_{\zeta}^{T}R_{\zeta,d}\Omega_{\zeta,d}-R_{\zeta}^{T}R_{\zeta,d}\dot{\Omega}_{\zeta,d}),\label{eq:geometric controller moment}
\end{align}
where $F_{\zeta}$ are force controllers defined in (\ref{eq:leader controller1}),
(\ref{eq:leader controller 2}), (\ref{eq:follower controller 1}),
and (\ref{eq:follower controller 2}), $\dot{v}_{\zeta,d}\in\mathbb{R}^{3}$
represents the reference acceleration, $k_{e,\zeta},\,k_{\Omega,\zeta}\in\mathbb{R}$
are positive control gains, $\Omega_{\zeta}\in\mathbb{R}^{3}$ is
defined as the angular velocity of the UAV, $R_{\zeta,d},\,\Omega_{\zeta,d}\in\mathbb{R}^{3}$
represent the reference attitude and reference angular velocity of
the UAV, respectively, and attitude error $e_{R_{\zeta}}$ and angular
velocity error $e_{\Omega_{\zeta}}$ of the UAV are defined as
\begin{align}
e_{R_{\zeta}} & =\frac{1}{2}\left(R_{\zeta,d}^{T}R_{\zeta}-R_{\zeta}^{T}R_{\zeta,d}\right)\\
e_{\Omega_{\zeta}} & =\Omega_{\zeta}-R_{\zeta}^{T}R_{\zeta,d}\Omega_{\zeta,d}.
\end{align}
The thrust command for each rotor on the UAV can be obtained based
on the following relation:
\begin{equation}
\left[\begin{array}{c}
T_{\zeta}\\
M_{\zeta,x}\\
M_{\zeta,y}\\
M_{\zeta,z}
\end{array}\right]=\varGamma\left[\begin{array}{c}
f_{\zeta}^{1}\\
f_{\zeta}^{2}\\
\vdots\\
f_{\zeta}^{n}
\end{array}\right],\label{eq:UAV Controller control signal}
\end{equation}
where $\varGamma\in\mathbb{R}^{4\times n}$ is a control allocation
matrix. In (\ref{eq:UAV Controller control signal}), $f_{\zeta}^{i}$
represents the thrust corresponding to the $i^{\text{th}}$ rotors
of the UAV, which is assumed to be proportional to the square of angular
velocity $\omega_{\zeta}^{i},$ and is expressed as 
\begin{equation}
f_{\zeta}^{i}=k_{\zeta}^{i}\left(\omega_{\zeta}^{i}\right)^{2},\ i=1,\,2,\,\ldots,\,n.\label{eq:model of rotors}
\end{equation}
where $n\in\mathbb{R}$ is the number of rotors, and $k_{\zeta}^{i}$
is a positive constant.}

\subsection{\textcolor{black}{System Architecture}}

\textcolor{black}{To make the system architecture clear, the dynamics
and controllers of each subsystem are summarized in Table \ref{tab:Dynamics-and-Controllers}.}

\textcolor{black}{}
\begin{table}[H]
\centering{}\textcolor{black}{\caption{\label{tab:Dynamics-and-Controllers}Dynamics and controllers.}
}%
\begin{tabular}{|c|c|c|}
\hline 
\textcolor{black}{Subsystem} &
\textcolor{black}{Type} &
\textcolor{black}{Equation(s)}\tabularnewline
\hline 
\hline 
\multirow{2}{*}{\textcolor{black}{Payload}} &
\textcolor{black}{Kinematics} &
\textcolor{black}{(\ref{eq:open-loop dynamics 2})}\tabularnewline
\cline{2-3} 
 & \textcolor{black}{Dynamics} &
\textcolor{black}{(\ref{eq:open-loop 1}) and (\ref{eq:open-loop 2})}\tabularnewline
\hline 
\multirow{2}{*}{\textcolor{black}{Leader }} &
\textcolor{black}{Kinematics controller} &
\textcolor{black}{(\ref{eq:control input of kinematics})}\tabularnewline
\cline{2-3} 
 & \textcolor{black}{Dynamics controller ($x$ and $y$ axes)} &
\textcolor{black}{(\ref{eq:leader controller1}) and (\ref{eq:leader controller 2})}\tabularnewline
\hline 
\multirow{2}{*}{\textcolor{black}{Follower}} &
\textcolor{black}{Kinematics controller} &
\textcolor{black}{(\ref{eq:follower transversal model})}\tabularnewline
\cline{2-3} 
 & \textcolor{black}{Dynamics controller ($x$ and $y$ axes)} &
\textcolor{black}{(\ref{eq:follower controller 1}) and (\ref{eq:follower controller 2})}\tabularnewline
\hline 
\textcolor{black}{Flight controller} &
\textcolor{black}{Low-level controller} &
\textcolor{black}{(\ref{eq:geometric controller thrust}) and (\ref{eq:geometric controller moment})}\tabularnewline
\hline 
\end{tabular}
\end{table}

\section{\textcolor{black}{Stability Analysis }}

\subsection{\textcolor{black}{Stability Analysis of the Leader Controller}}

\textcolor{black}{As mentioned in Section \ref{subsec:Follower-Scheme},
the follower is controlled to avoid transverse motion and the tracking
error will be compensated by the leader controller. Therefore, only
the tracking performance of the leader controller is provided in this
section. To this end, a Lyapunov-based approach is developed to analyze
the stability of the closed-loop system. }
\begin{thm}
\textcolor{black}{Given a leader-follower system as depicted in Fig.
\ref{fig2:nonholonomic 2D-motion-in x-y plane}, the leader controller
defined in (\ref{eq:leader controller1}) and (\ref{eq:leader controller 2})
ensures that the system can achieve asymptotic tracking as defined
in (\ref{eq:control objectives}).}
\end{thm}
\begin{IEEEproof}
\textcolor{black}{Let a Lyapunov function be defined as
\begin{equation}
\text{\ensuremath{V_{1}=\frac{1}{2}x_{e}^{2}+\frac{1}{2}y_{e}^{2}+\frac{1-\cos\left(\theta_{e}\right)}{k_{2}},}}\label{eq:Lyapunov of  kinematics}
\end{equation}
where $x_{e}$, $y_{e}$ and $\theta_{e}$ are the tracking errors
defined in (\ref{eq:eq:nonholonomic err transform}) and (\ref{eq:theta_e}),
and $k_{2}$ is a positive constant. Substituting the controller defined
in (\ref{eq:control input of kinematics}) into (\ref{eq:open-loop dynamics 2})
yields the closed-loop error dynamics:
\begin{equation}
\begin{cases}
\dot{x}_{e}=\omega y_{e}-k_{1}x_{e}+\eta_{1}\\
\dot{y}_{e}=-\omega(x_{e})+v_{r}\sin\theta_{e}\\
\dot{\theta}_{e}=-v_{r}k_{2}y_{e}-k_{3}\sin\theta_{e}+\eta_{2}
\end{cases},\label{eq:closed-loop error systems}
\end{equation}
where $k_{1}$ and $k_{3}$ are positive constants. Taking the time
derivative of the Lyapunov function defined in (\ref{eq:Lyapunov of  kinematics})
and using (\ref{eq:closed-loop error systems}) yields
\begin{align}
\dot{V}_{1}= & x_{e}\left(\omega y_{e}-k_{1}x_{e}+\eta_{1}\right)\nonumber \\
 & +y_{e}(-\omega x_{e})+y_{e}v_{r}\sin\theta_{e}\nonumber \\
 & +(-v_{r}k_{2}y_{e}-k_{3}\sin\theta_{e}+\eta_{2})\frac{\sin\theta_{e}}{k_{2}},\label{eq:v1_dot}
\end{align}
and canceling out the cross terms reduces (\ref{eq:v1_dot}) to
\begin{equation}
\dot{V}_{1}=-k_{1}x_{e}^{2}-\frac{k_{3}\left(\sin\theta_{e}\right)^{2}}{k_{2}}+x_{e}\eta_{1}+\frac{\sin\theta_{e}}{k_{2}}\eta_{2}.\label{eq:Lyapunov time derivative}
\end{equation}
A second Lyapunov function that contains $V_{1}$ is now defined as
\begin{equation}
V_{2}=V_{1}+\frac{1}{2}\eta_{1}^{2}+\frac{1}{2}\eta_{2}^{2},\label{eq:Lyapunov payload dynamics}
\end{equation}
where $\eta_{1},\eta_{2}$ are defined in (\ref{eq:eta1}) and (\ref{eq:eta2}).
Taking the time derivative of the Lyapunov function in (\ref{eq:Lyapunov payload dynamics})
yields
\begin{equation}
\dot{V}_{2}=\dot{V}_{1}+\eta_{1}\dot{\eta}_{1}+\eta_{2}\dot{\eta}_{2}.\label{eq:Lyapunov payload dynamics time derivative}
\end{equation}
Taking the time derivative of (\ref{eq:eta1}) and (\ref{eq:eta2})
and substituting the controllers defined in (\ref{eq:leader controller1})
and (\ref{eq:leader controller 2}) into the open-loop dynamics (\ref{eq:open-loop 1})
and (\ref{eq:open-loop 2}) along with the result in (\ref{eq:control_objective})
yields the closed-loop error system as:
\begin{align}
\dot{\eta}_{1} & =-k_{v}\eta_{1}-x_{e}\label{eq:CLES 1}\\
\dot{\eta}_{2} & =-k_{\omega}\eta_{2}-\frac{\sin\theta_{e}}{k_{2}}.\label{eq:CLES 2}
\end{align}
Substituting (\ref{eq:CLES 1}) and (\ref{eq:CLES 2}) into (\ref{eq:Lyapunov payload dynamics time derivative})
and using (\ref{eq:Lyapunov time derivative}) yields
\begin{equation}
\dot{V}_{2}=-k_{v}\eta_{1}^{2}-k_{\omega}\eta_{2}^{2}-k_{1}x_{e}^{2}-\frac{k_{3}\left(\sin\theta_{e}\right)^{2}}{k_{2}},\label{eq:Lyapunov time derivative of V}
\end{equation}
which is a negative semidefinite function provided that $k_{v},$
$k_{\omega},$ $k_{1}$, $k_{2},$ and $k_{3}$ are selected to be
positive constants. Therefore, LaSalle's invariance principle can
be invoked to show
\begin{equation}
\begin{array}{cc}
\begin{array}{c}
x_{e}\rightarrow0\\
y_{e}\rightarrow0\\
\theta_{e}\rightarrow0\\
\eta_{1}\rightarrow0\\
\eta_{2}\rightarrow0
\end{array} & \text{as }t\rightarrow\infty,\end{array}\label{eq:error state}
\end{equation}
which implies that the control objective defined in (\ref{eq:control objectives})
is achieved.}
\end{IEEEproof}

\subsection{\textcolor{black}{Robustness of the Leader Controller\label{subsec:Robustness of the leader controller}}}

\textcolor{black}{To consider the robustness of the leader controller,
the estimate errors from the UKFs are added into the leader controller
defined in (\ref{eq:leader controller1}) and (\ref{eq:leader controller 2})
as 
\begin{align}
F_{L_{x}}= & m_{p}\left(\dot{v}_{d}+x_{e}+k_{v}\eta_{1}+r_{\nicefrac{c_{2}}{p,x}}\omega^{2}\right)+d_{1}\label{eq:leader controller1-with disturbance}\\
F_{L_{y}}= & -\frac{2I_{zz}}{L}\left(\dot{\omega}_{d}+k_{\omega}\eta_{2}+\frac{\sin\theta_{e}}{k_{2}}\right)+F_{F_{y}}+d_{2},\label{eq:leader controller 2--with disturbance}
\end{align}
where $d_{1},\,d_{2}\in\mathbb{R}$ are the additive terms that include
the estimator errors (e.g., $\hat{p}_{c_{2}}-p_{c_{2}},$ $\hat{v}-v,$
$\hat{F}_{F_{y}}-F_{F_{y}}$) from the UKFs and the disturbances.
By substituting (\ref{eq:leader controller1-with disturbance}) and
(\ref{eq:leader controller 2--with disturbance}) into the same Lyapunov
function, $\dot{V}_{2}$ can be upper bounded as
\begin{align}
\dot{V}_{2} & \leq-k_{v}\eta_{1}^{2}-k_{\omega}\eta_{2}^{2}-k_{1}x_{e}^{2}-\frac{k_{3}\left(\sin\theta_{e}\right)^{2}}{k_{2}}+D_{1}\left|\eta_{1}\right|+D_{2}\left|\eta_{2}\right|,\label{eq:Lyapunov func UUB0}
\end{align}
where $D_{1},\,D_{2}\in\mathbb{R}_{>0}$ are positive constant upper
bounds of the estimate errors and disturbances. After completing the
square, (\ref{eq:Lyapunov func UUB0}) can be further upper bounded
as
\begin{align}
\dot{V}_{2} & \leq-k_{v1}\eta_{1}^{2}-k_{\omega1}\eta_{2}^{2}-k_{1}x_{e}^{2}-\frac{k_{3}\left(\sin\theta_{e}\right)^{2}}{k_{2}}+\frac{D_{1}^{2}}{4k_{v2}}+\frac{D_{2}^{2}}{4k_{\omega2}},\label{eq:Lyapunov func UUB}
\end{align}
where $k_{v1},\,k_{v2},\,k_{\omega1},\,k_{\omega2}\in\mathbb{R}_{>0}$
satisfy $k_{v}=k_{v1}+k_{v2}$ and $k_{\omega}=k_{\omega1}+k_{\omega2}$.
(\ref{eq:Lyapunov func UUB0}) implies that tracking error $p_{e}$
and $\theta_{e}$ defined in (\ref{eq:eq:nonholonomic err transform})
and (\ref{eq:theta_e}) are uniformly ultimately bounded, and the
error bounds can be decreased by increasing the control gains $k_{v},\,k_{\omega},$
and $k_{1}.$}

\section{\textcolor{black}{Motion Planning for the Leader-Follower System}}

\textcolor{black}{In general cases, obstacles and physical constraints
are taken into account when planning an optimal trajectory. This section
divides the motion planning for the leader-follower system into two
parts: trajectory planning and trajectory generation. }

\subsection{\textcolor{black}{Trajectory Planning}}

\begin{figure}[H]
\centering{}\textcolor{black}{\includegraphics[width=0.9\columnwidth]{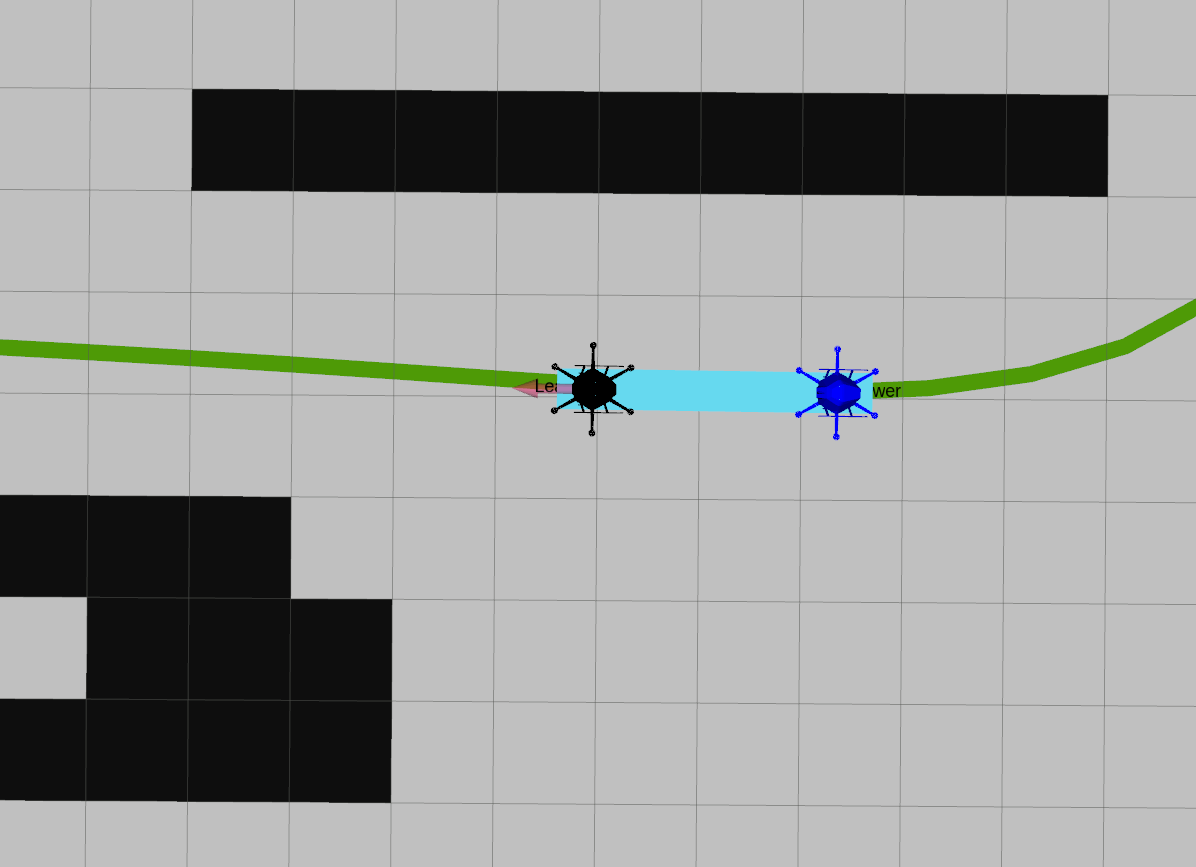}\caption{\label{fig:bitmap}The leader-follower system executing a cooperative
transportation task.}
}
\end{figure}

\textcolor{black}{\cite{Petereit2012} proposed using a hybrid A{*}
planner to find the correct trajectory for a nonholonomic vehicle
in an environment with obstacles from the start to the end point under
various constraints such as nonholonomic motion and the dimensions
of the vehicles. The information about the environment is digitized
to a bitmap before performing the planning; for example, the location
of obstacles in the environment are considered as 1 and are shown
in black in Fig. \ref{fig:bitmap}. The planner output is a sequence
of waypoints. }

\textcolor{black}{}

\subsection{\textcolor{black}{Trajectory Generation\label{subsec:Trajectory-Generation}}}

\textcolor{black}{To generate a desired trajectory for the UAV, \cite{Wang2018}
used QP to find the minimun-snap trajectory passing through all of
the waypoints. Let each trajectory segment presented by a polynomial
be defined as}

\textcolor{black}{
\begin{equation}
s_{i}(t)=\sum_{j=0}^{N}a_{ij}t^{j},\ t_{i-1}\leq t\leq t_{i},\ i\in\{1,\ 2,\ldots,\ M\},\label{eq:single trajectory}
\end{equation}
where $a_{ij}\in\mathbb{R}$ is the $j^{\text{th}}$ order coefficient,
$N\in\mathbb{N}$ represents the order of the polynomial, $M\in\mathbb{N}$
represents the total number of segments, and $t_{i}$ is the time
when $c_{2}$ passes through the $i^{\text{th}}$ waypoint. To generate
a smooth trajectory, $N=7$ is selected to minimize the snap of $s_{i}$
so that Assumption \ref{Assumption:smooth trajectory} can be satisfied
and cost function $J$ is expressed as}

\textcolor{black}{
\begin{align}
J & =\underset{a_{ij}}{\min}\sum_{i=1}^{n}\int_{t_{i-1}}^{t_{i}}\|\frac{d^{4}s_{i}(t)}{dt^{4}}\|^{2}dt\label{eq:cost function}
\end{align}
}

\textcolor{black}{
\[
\text{s.t.}\ AT=B,
\]
where $A$ is a matrix consisting of coefficient $a_{ij}$, $T$ is
a vector consisting of the time intervals defined in (\ref{eq:single trajectory}),
and $B$ is a vector determined by the waypoints. The optimal solution
of (\ref{eq:cost function}) is found using a QP solver \cite{Wang2018}. }

\section{\textcolor{black}{Simulations}}

\textcolor{black}{Two simulations of the leader-follower system were
performed. Simulation 1 evaluated the tracking performance of the
developed controller, including the leader and follower controllers.
Moreover, the estimation performance of the UKFs were also determined
to ensure they provide accurate estimates for control feedback. After
evaluating the control performance in Simulation 1, Simulation 2 was
performed to show that the leader can change the desired trajectory
in real time during flight.}

\subsection{\textcolor{black}{Implementation}}

\textcolor{black}{The simulations were conducted using the ROS and
Gazebo simulator with the UAV model from RotorS provided by \cite{Furrer2016}
along with a customized payload model. All of the ground truth was
provided by the simulator.}

\subsection{\textcolor{black}{Simulation 1: Evaluation of the Tracking Controller\label{subsec:Simulation 1}}}

\textcolor{black}{Simulation 1 was divided into two parts: (1) evaluating
the controller for tracking a desired trajectory, and (2) evaluating
the performance of the UKFs for the UAVs.}

\subsubsection{\textcolor{black}{Evaluation of the Controller}}

\begin{figure}[H]
\centering{}\textcolor{black}{\includegraphics[width=0.95\columnwidth]{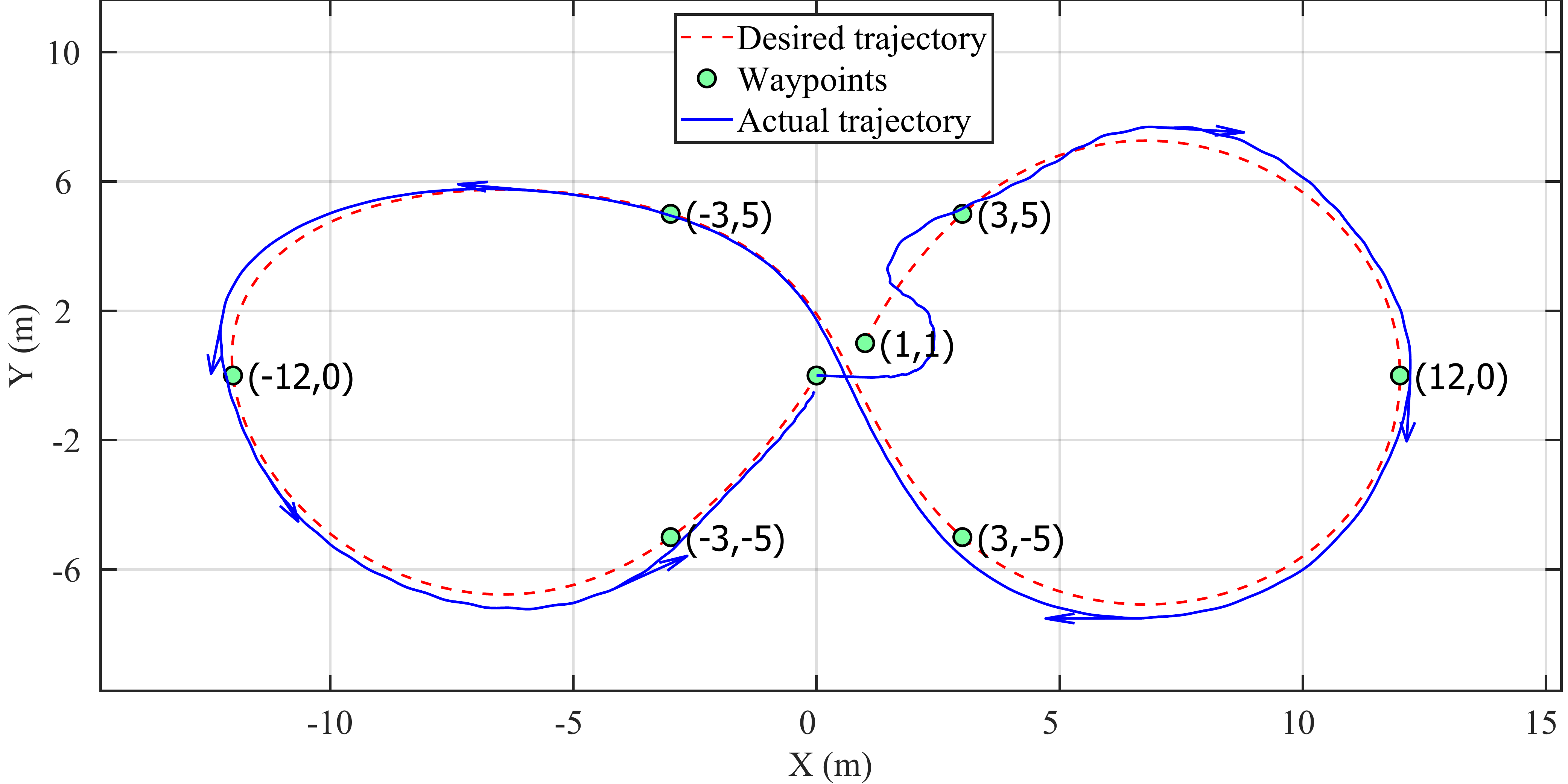}
\caption{\label{fig:figure-8 tracking}Desired trajectory generated from the
QP solver and the actual trajectory of point $c_{2}.$}
}
\end{figure}

\textcolor{black}{Fig. \ref{fig:figure-8 tracking} shows that the
system can track a desired trajectory, which is a high-order polynomial
solved by the QP solver in the leader UAV.}

\textcolor{black}{}
\textcolor{black}{The positions of the waypoints are listed as in
Table \ref{tab:All-segments}.}

\begin{figure}[H]
\centering{}\textcolor{black}{\includegraphics[width=0.95\columnwidth]{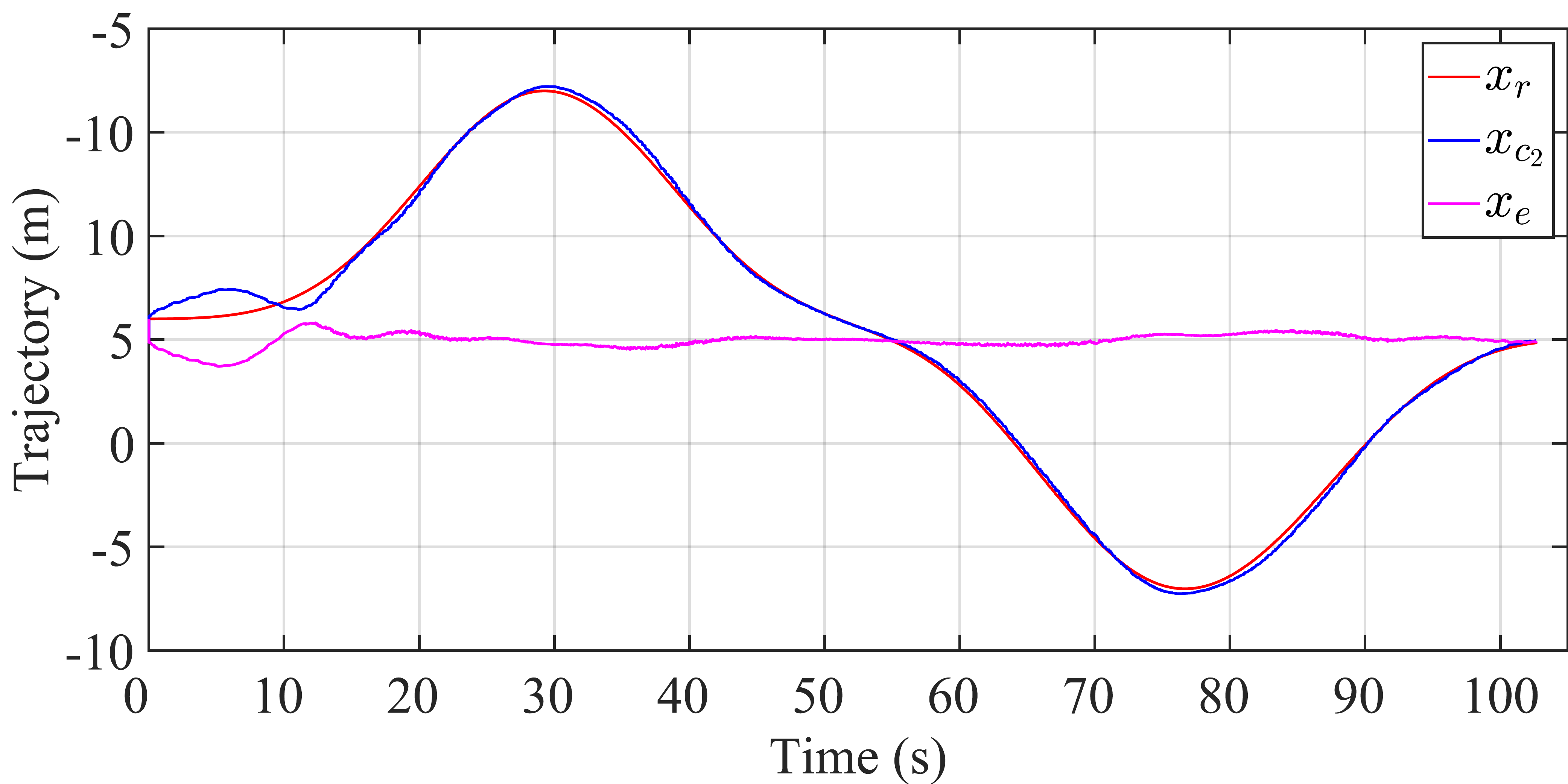}\caption{\label{fig:eval_pos_x}Position and tracking error of point $c_{2}$
along the $X$ axis of the inertial frame.}
}
\end{figure}

\begin{figure}[H]
\begin{centering}
\textcolor{black}{\includegraphics[width=0.95\columnwidth]{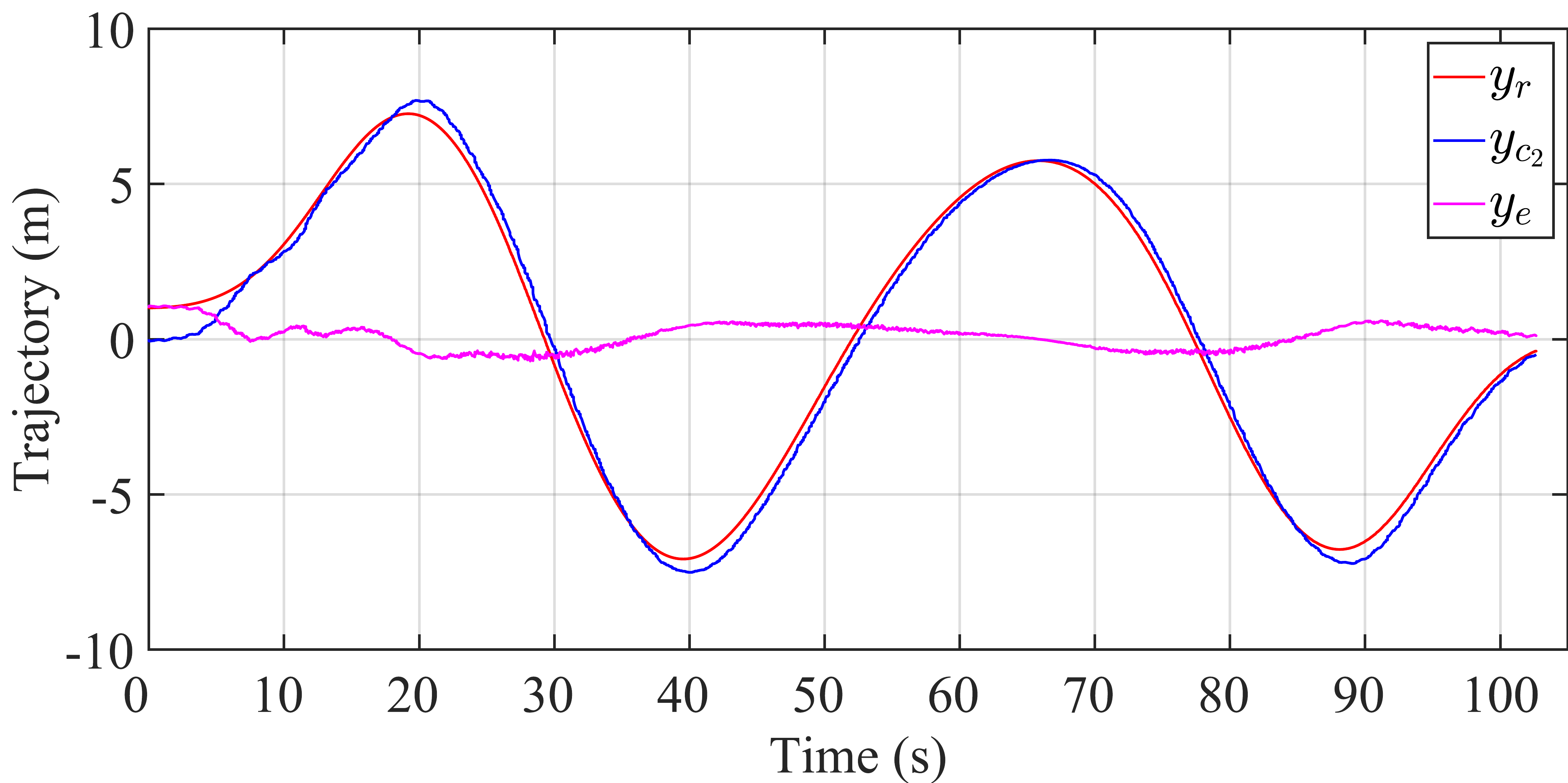}}
\par\end{centering}
\centering{}\textcolor{black}{\caption{\label{fig:eval_pox_y}Position and tracking error of point $c_{2}$
along the $Y$ axis of the inertial frame.}
}
\end{figure}

\textcolor{black}{}
\begin{table}[H]

\begin{centering}
\textcolor{black}{\caption{\label{tab:All-segments}All segments in the trajectory.}
}%
\begin{tabular}{|c|c|c|}
\hline 
\textcolor{black}{Segment} &
\textcolor{black}{Start-End} &
\textcolor{black}{Duration (s)}\tabularnewline
\hline 
\hline 
\textcolor{black}{1} &
\textcolor{black}{$(1,1)-(3,5)$} &
\textcolor{black}{$16.0$}\tabularnewline
\hline 
\textcolor{black}{2} &
\textcolor{black}{$(3,5)-(12,0)$} &
\textcolor{black}{$16.0$}\tabularnewline
\hline 
\textcolor{black}{3} &
\textcolor{black}{$(12,0)-(3,-5)$} &
\textcolor{black}{$16.0$}\tabularnewline
\hline 
\textcolor{black}{4} &
\textcolor{black}{$(3,-5)-(-3,5)$} &
\textcolor{black}{$16.0$}\tabularnewline
\hline 
\textcolor{black}{5} &
\textcolor{black}{$(-3,5)-(-12,0)$} &
\textcolor{black}{$16.0$}\tabularnewline
\hline 
\textcolor{black}{$6$} &
\textcolor{black}{$(-12,0)-(-3,-5)$} &
\textcolor{black}{$16.0$}\tabularnewline
\hline 
\textcolor{black}{7} &
\textcolor{black}{$(-3,-5)-(0,0)$} &
\textcolor{black}{$16.0$}\tabularnewline
\hline 
\end{tabular}
\par\end{centering}
\end{table}
\textcolor{black}{Figs. \ref{fig:eval_pos_x} and \ref{fig:eval_pox_y}
show the tracking errors during the cooperative transportation. $x_{c_{2}}$
and $y_{c_{2}}$ represent the position of $c_{2}$ along the $X$
and $Y$ axes of the inertial frame. Both tracking errors $x_{e}$
and $y_{e}$ decreased over time. Also, the tracking error in the
transverse direction increased as the curvature increased, and decreased
rapidly while traveling along a straight trajectory, which can be
attributed to the efficacy of the impedance controller since the desired
trajectory and the curvature are not available to the follower. The
standard deviations of $x_{e}$ and $y_{e}$ were $0.36$ m and $0.41$m,
respectively.}

\textcolor{black}{}

\begin{figure}[H]
\centering{}\textcolor{black}{\includegraphics[width=0.95\columnwidth]{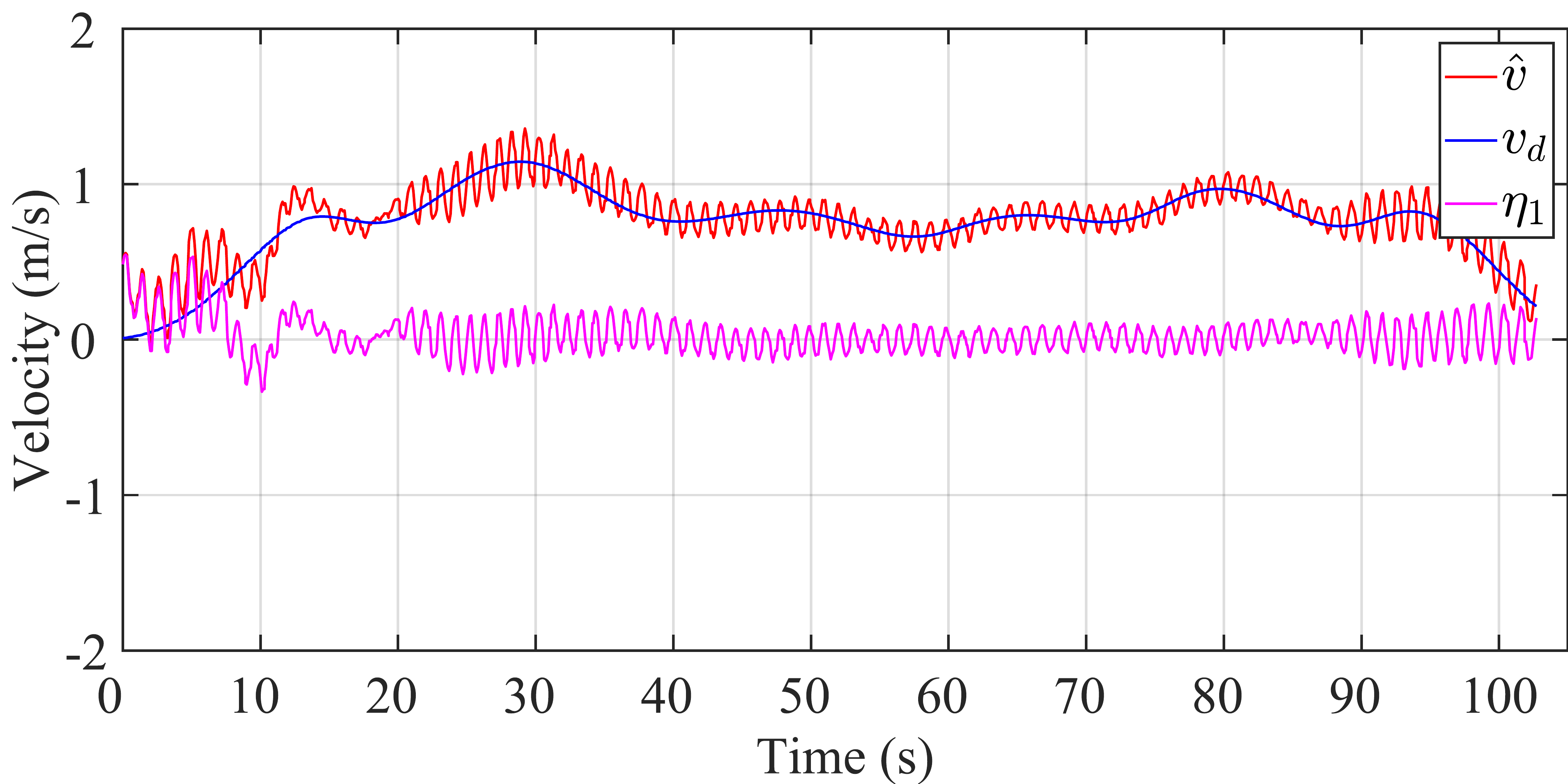}\caption{\label{fig:Evaluation-of-linear-velocity} $\hat{v}$ is estimated
by the UKF in Section \ref{subsec:The-Second-UKF}, and $v_{d}$ is
the desired linear velocity in (\ref{eq:control input of kinematics}).
Signal $\eta_{1}$ indicates the tracking performance of the leader
controller during translational motion.}
}
\end{figure}

\begin{figure}[H]
\centering{}\textcolor{black}{\includegraphics[width=0.95\columnwidth]{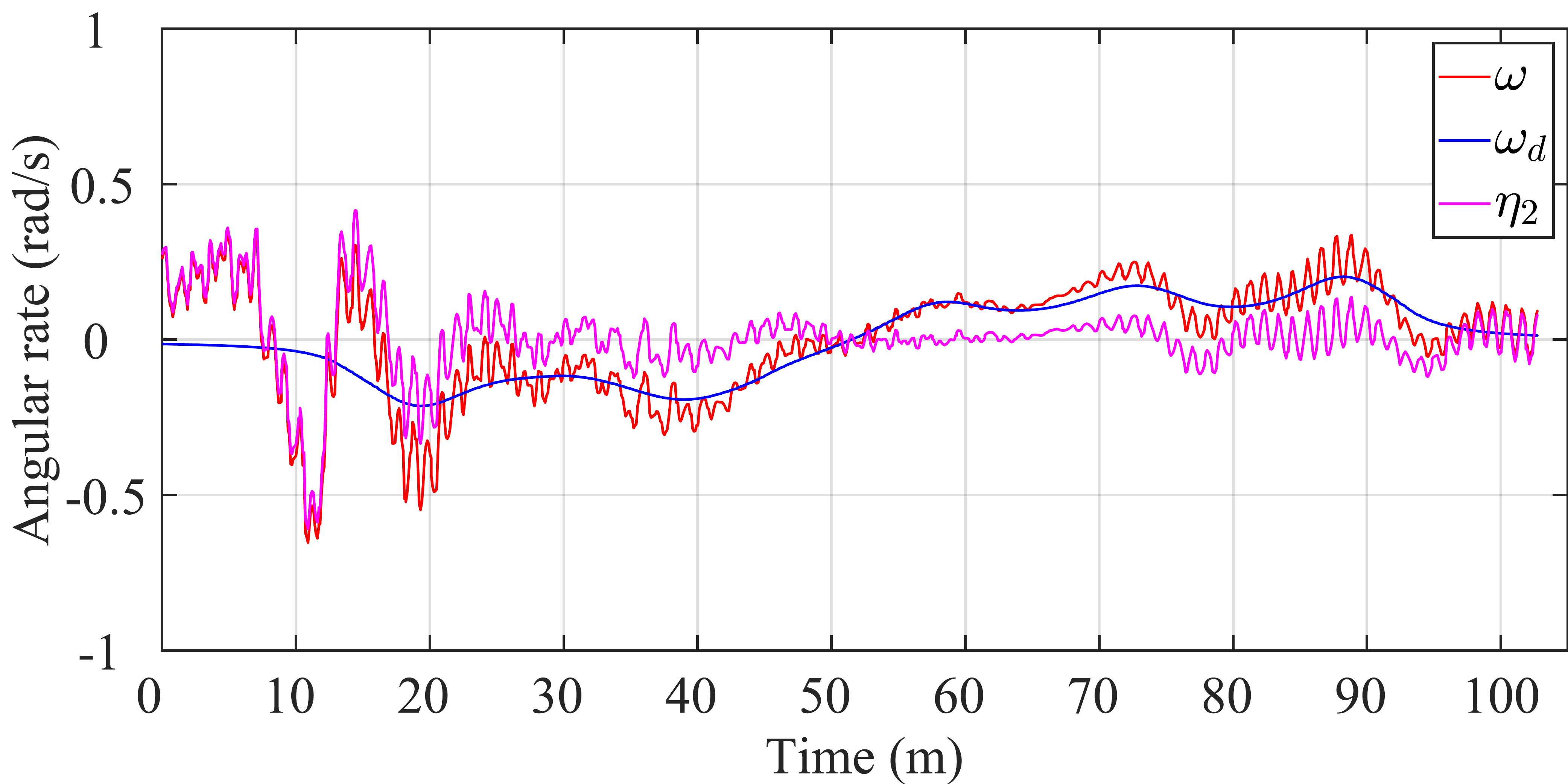}\caption{\label{fig:Evaluation-of-angular-velocity}$\omega$ is measured by
the IMU attached to the payload at $c_{1},$ and $\omega_{d}$ is
the desired angular rate in (\ref{eq:control input of kinematics}).
Signal $\eta_{2}$ indicates the tracking performance of the leader
controller during rotational motion.}
}
\end{figure}

\textcolor{black}{}
\textcolor{black}{Figs. \ref{fig:Evaluation-of-linear-velocity} and
\ref{fig:Evaluation-of-angular-velocity} show the tracking performance
of the kinematics controller defined in (\ref{eq:control input of kinematics}).
Signals $\eta_{1}$ and $\eta_{2}$ are close to zero in the steady
state, and the oscillation is due to the payload swinging along $x^{B}$
during flight and corresponds to position errors that are compensated
by the leader controller.}

\textcolor{black}{}

\begin{figure}[H]
\centering{}\textcolor{black}{\includegraphics[width=0.9\columnwidth]{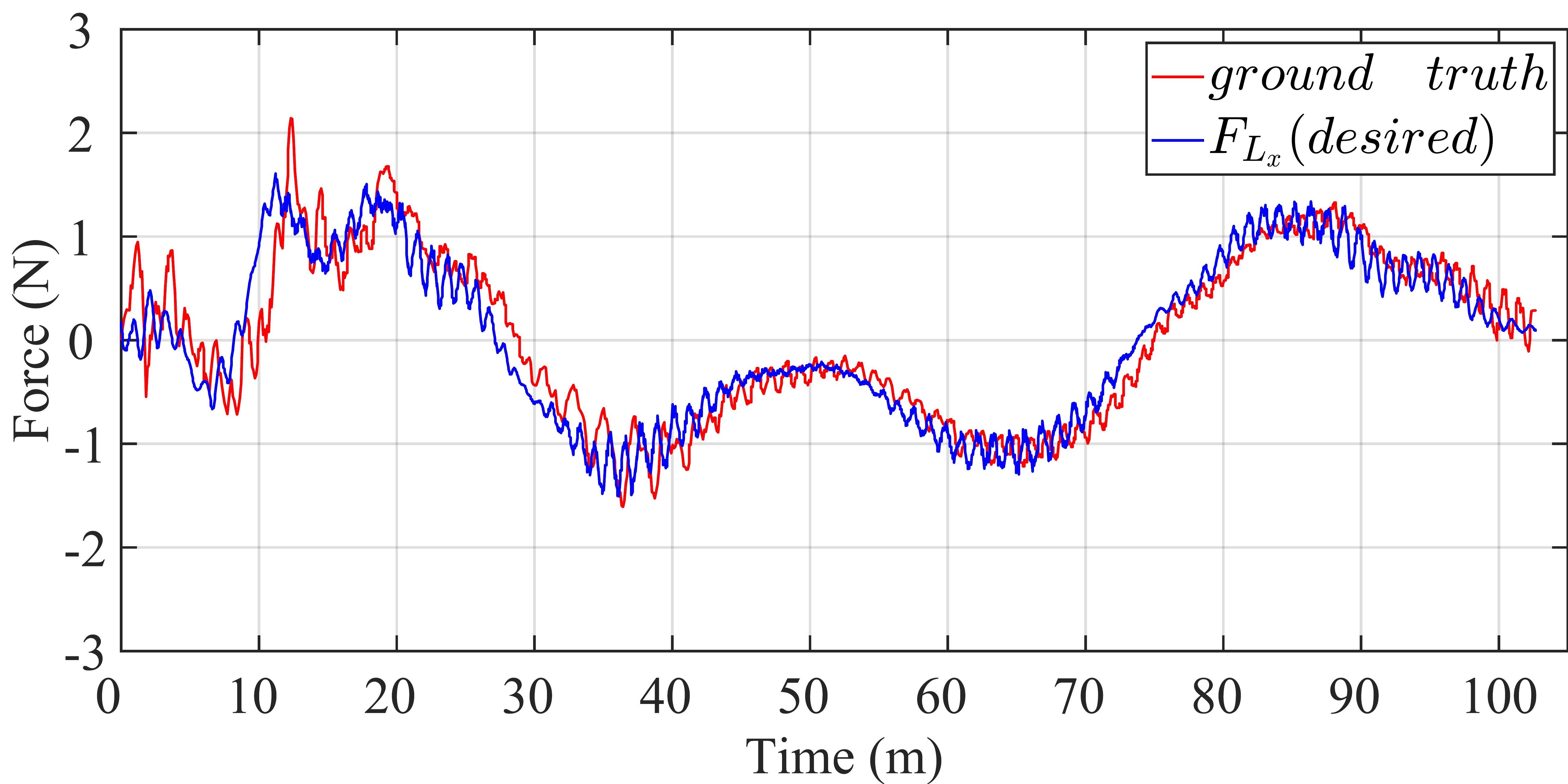}\caption{\label{fig:applied force x}The ground truth of $F_{L_{x}}$estimated
by the virtual force sensor on the payload, and desired force $F_{L_{x}}$
computed by the leader controller, both expressed in the inertial
frame.}
}
\end{figure}

\begin{figure}[H]
\begin{centering}
\textcolor{black}{\includegraphics[width=0.9\columnwidth]{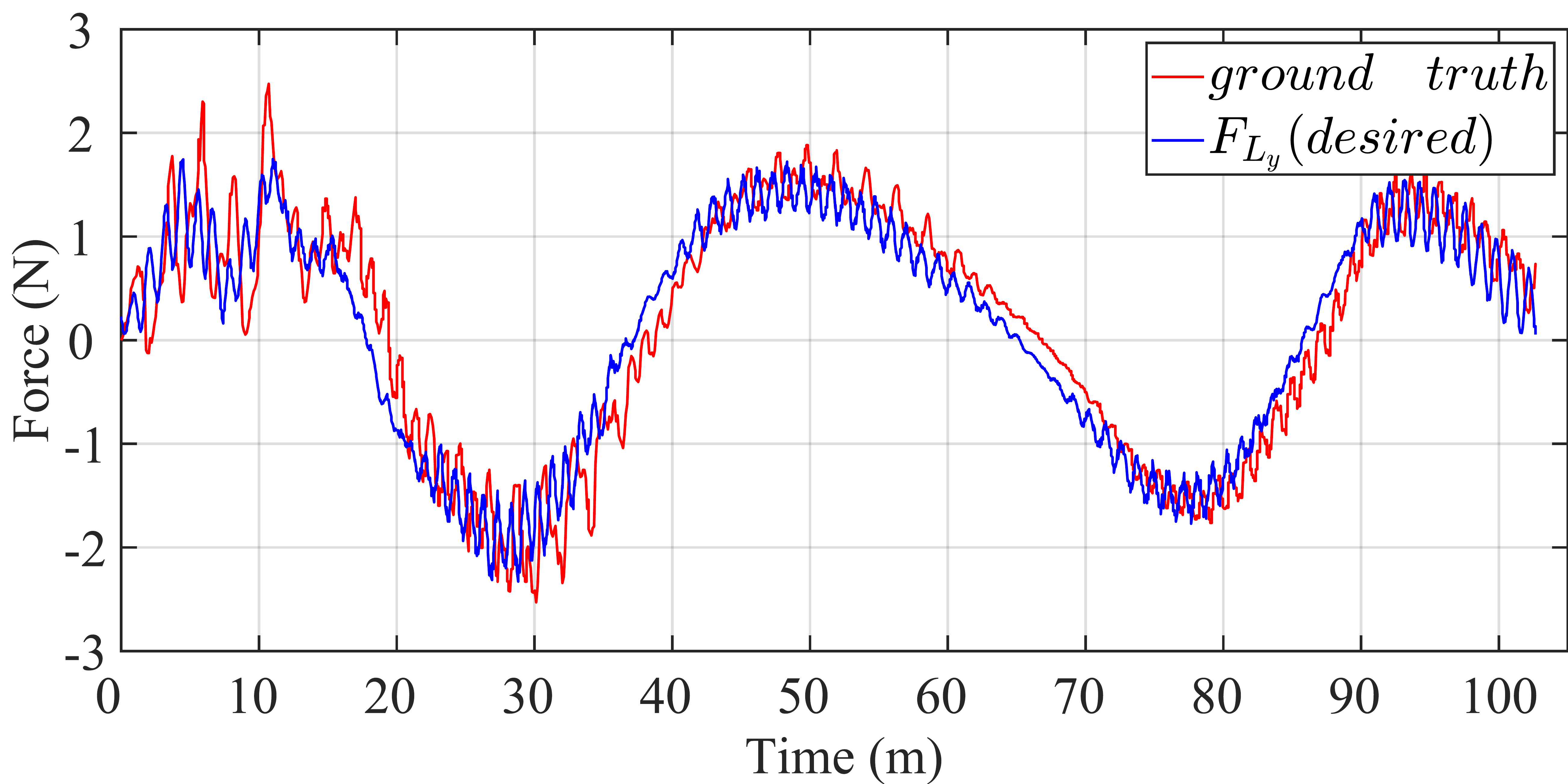}}
\par\end{centering}
\textcolor{black}{\caption{\label{fig:applied force y}The ground truth of $\hat{F}_{L_{y}}$
estimated by the virtual force sensor on the payload, and desired
force $F_{L_{y}}$ computed by the leader controller, both expressed
in the inertial frame.}
}
\end{figure}

\textcolor{black}{}
\textcolor{black}{Figs. \ref{fig:applied force x} and  \ref{fig:applied force y}
show the actual and desired forces applied to point $c_{1}$ by the
leader, which implies that the geometric controller can achieve force
control as described by (\ref{eq:geometric controller}).}

\textcolor{black}{}

\textcolor{black}{}

\begin{figure}[H]
\centering{}\textcolor{black}{\includegraphics[width=0.9\columnwidth]{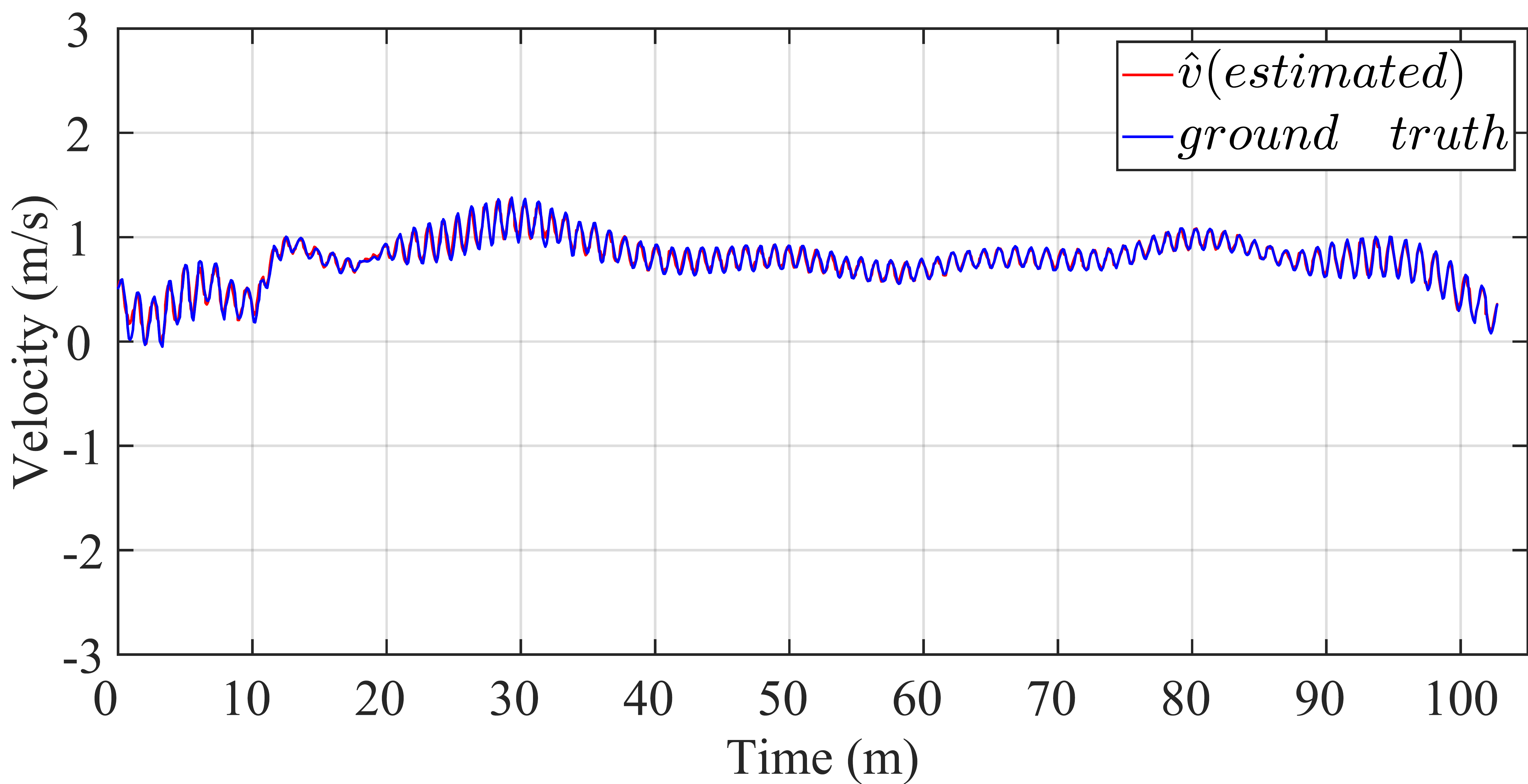}\caption{\label{fig:Evaluation-of-v} $\hat{v}$ estimated by the second UKF
in (\ref{eq:v_c2}) is close to the ground truth, which implies that
the UKFs perform well.}
}
\end{figure}

\begin{figure}[H]
\centering{}\textcolor{black}{\includegraphics[width=0.9\columnwidth]{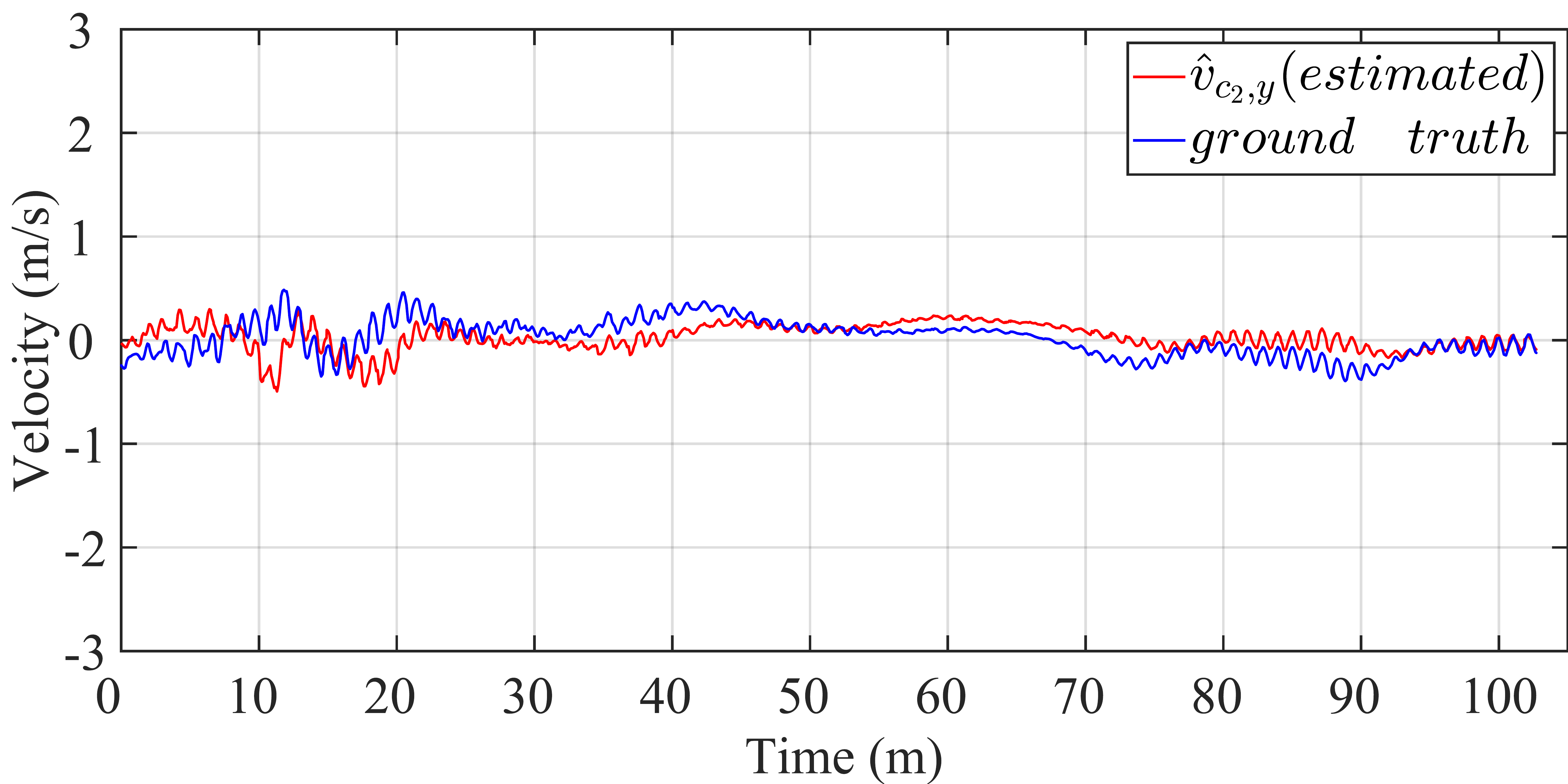}\caption{\label{fig:Evaluation-of-w} Estimated transverse velocity $\hat{v}_{c_{2},y}$
is essentially not zero, which satisfies Assumption \ref{Assumption 3:nonholonomic constraint}.
Nevertheless, the tracking error in the transverse direction can be
compensated by the leader controller as indicated in Fig. \ref{fig:figure-8 tracking}.}
}
\end{figure}

\textcolor{black}{}
\begin{table}
\centering{}\textcolor{black}{\caption{RMSE values of the error signals in the leader controller.}
}%
\begin{tabular}{|c|c|c|}
\hline 
\textcolor{black}{RMSE} &
\textcolor{black}{Value} &
\textcolor{black}{Unit}\tabularnewline
\hline 
\hline 
\textcolor{black}{$x_{e}$} &
\textcolor{black}{$0.36$} &
\textcolor{black}{m}\tabularnewline
\hline 
\textcolor{black}{$y_{e}$} &
\textcolor{black}{$0.42$} &
\textcolor{black}{m}\tabularnewline
\hline 
\textcolor{black}{$\eta_{1}$} &
\textcolor{black}{$0.13$} &
\textcolor{black}{$\text{m}/\text{s}$}\tabularnewline
\hline 
\textcolor{black}{$\eta_{2}$} &
\textcolor{black}{$0.13$} &
\textcolor{black}{$\text{rad}/\text{s}$}\tabularnewline
\hline 
\end{tabular}
\end{table}
\textcolor{black}{To assess the performance of the UKF in Section
\ref{subsec:The-Second-UKF}, Figs. \ref{fig:Evaluation-of-v} and
\ref{fig:Evaluation-of-w} show the difference between the estimation
and the ground truth. Velocity $\hat{v}$ estimated by the leader
UKF is very close to the ground truth.}

\textcolor{black}{}

\textcolor{black}{}

\subsubsection{\textcolor{black}{Evaluation of the UKFs}}

\begin{figure}[H]
\centering{}\textcolor{black}{\includegraphics[width=0.9\columnwidth]{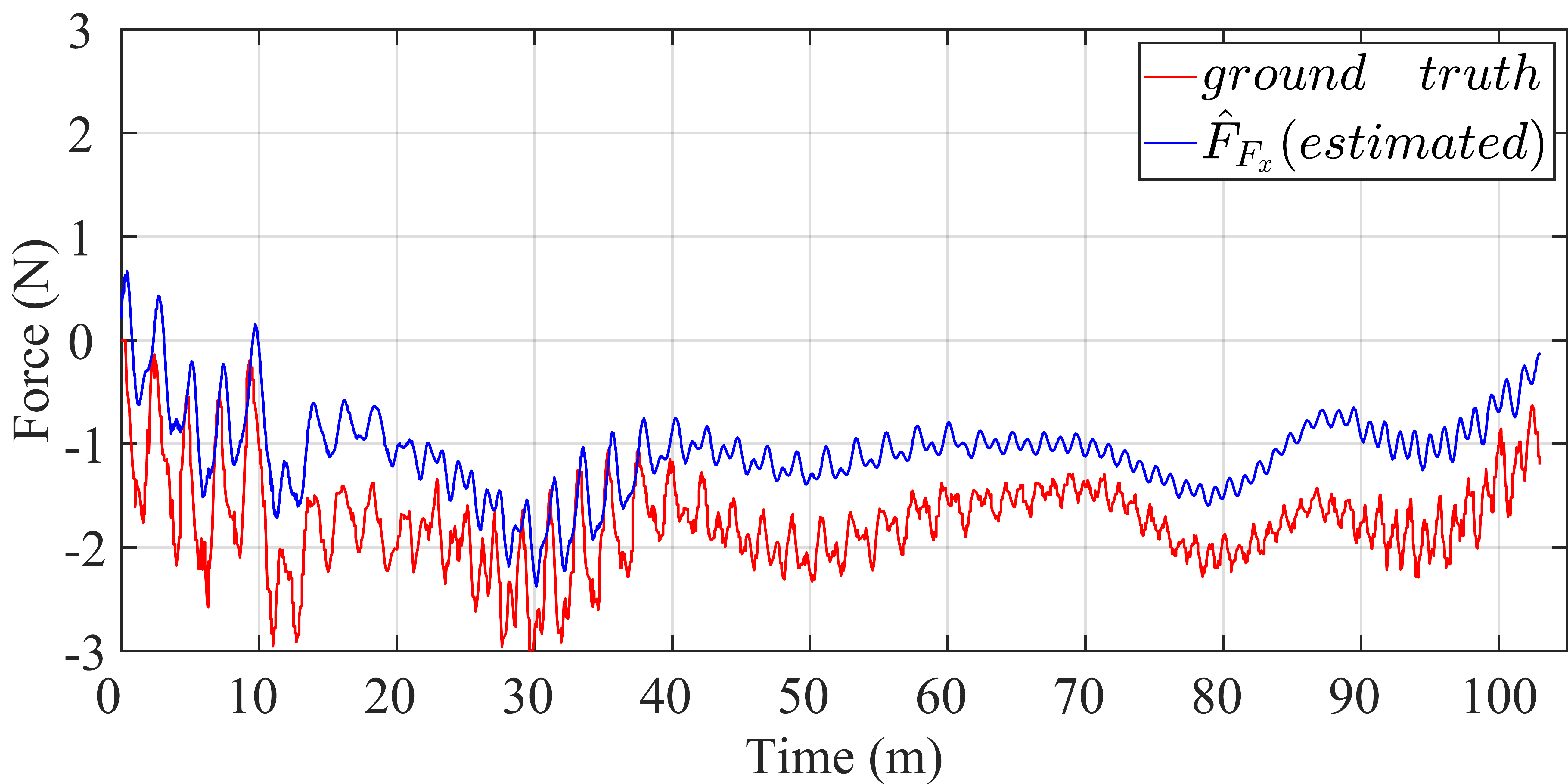}\caption{\label{fig:estimated force x} Estimate $\hat{F}_{F_{x}}$ is not
sufficiently accurate since the rotors defined in (\ref{eq:model of rotors})
is not well modeled. However, the mismatch can be compensated for
by analyzing the effects of $d_{1}$ on system stability using (\ref{eq:leader controller1-with disturbance}).}
}
\end{figure}

\begin{figure}[H]
\centering{}\textcolor{black}{\includegraphics[width=0.9\columnwidth]{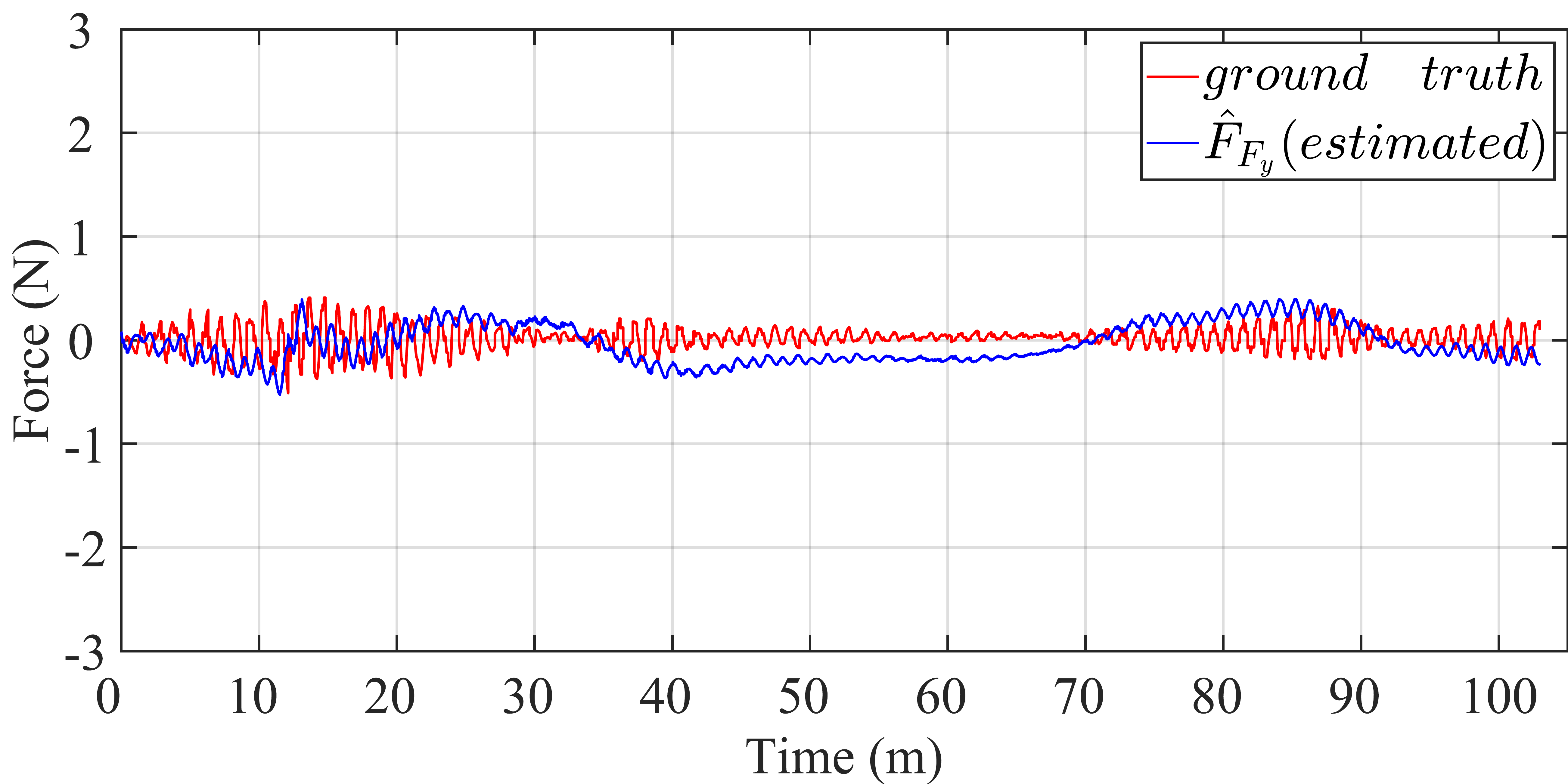}\caption{\label{fig:estimated force y} Estimate $\hat{F}_{F_{y}}$ is not
sufficiently accurate since the rotors defined in (\ref{eq:model of rotors})
is not well modeled; however, the mismatch can be compensated for
by analyzing the effects of $d_{2}$ on system stability using (\ref{eq:leader controller 2--with disturbance}). }
}
\end{figure}

\textcolor{black}{Figs. \ref{fig:estimated force x} and  \ref{fig:estimated force y}
show estimated forces $\hat{F}_{F_{x}}$ and $\hat{F}_{F_{y}}.$}

\textcolor{black}{}

\textcolor{black}{}

\begin{figure}[H]
\centering{}\textcolor{black}{\includegraphics[width=0.9\columnwidth]{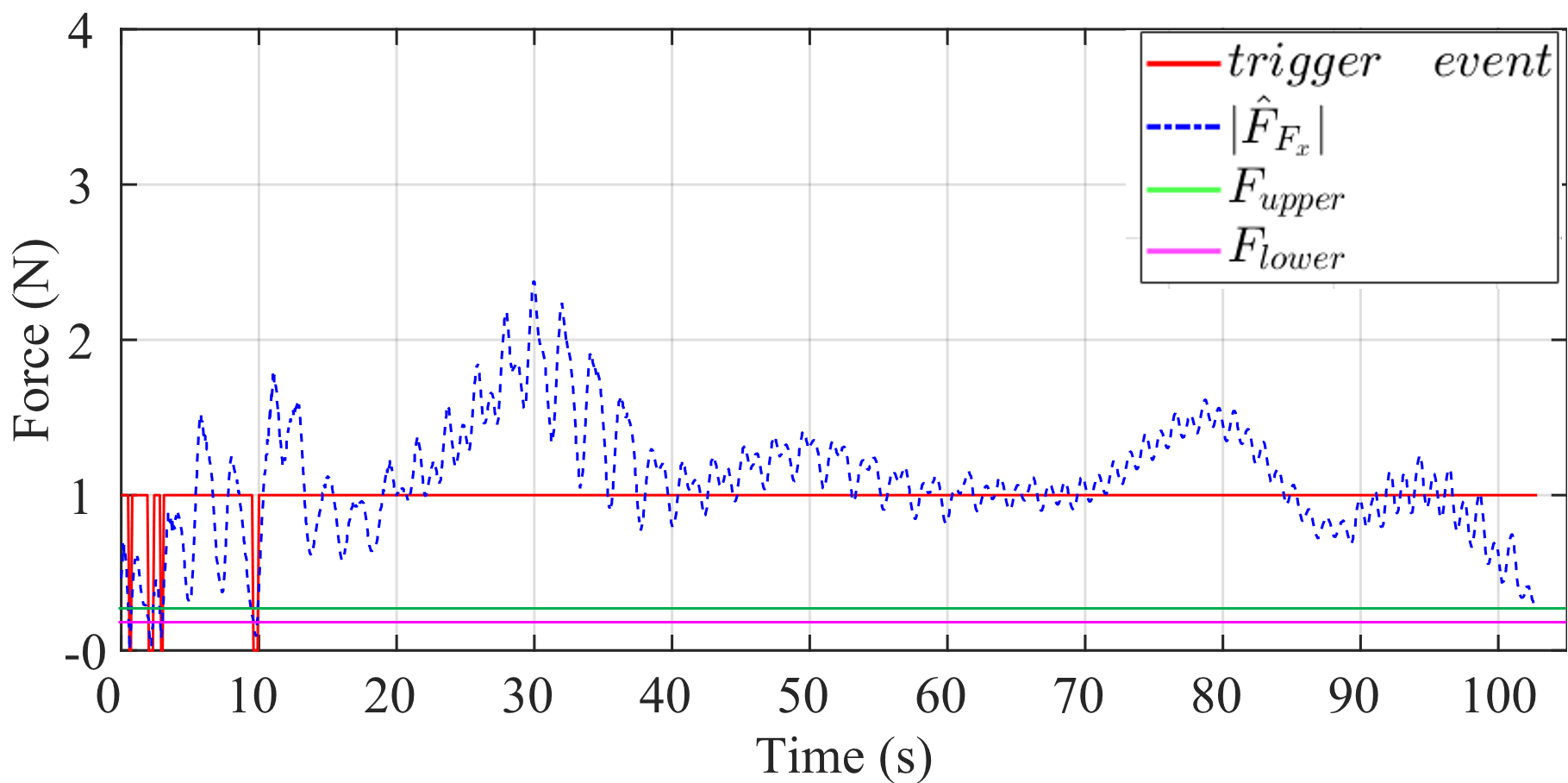}\caption{\label{fig:The-trigger-events}A triggering signal of $1$ implies
$t\in T^{en},$ while it is $0$ otherwise. The trigger switches frequently
at the beginning, while no switching occurs thereafter since the estimated
force is far from zero in the steady state.}
}
\end{figure}

\textcolor{black}{Fig. \ref{fig:The-trigger-events} shows the effect
of the triggering event described in Section \ref{subsec:Triggering-Mechanism}.}

\textcolor{black}{}

\textcolor{black}{}
\begin{table}
\centering{}\textcolor{black}{\caption{RMSE values of the estimates of the first UKF for the follower.}
}%
\begin{tabular}{|c|c|c|}
\hline 
\textcolor{black}{RMSE} &
\textcolor{black}{Value} &
\textcolor{black}{Unit}\tabularnewline
\hline 
\hline 
\textcolor{black}{$\hat{F}_{F_{x}}$} &
\textcolor{black}{0.75} &
\textcolor{black}{N}\tabularnewline
\hline 
\textcolor{black}{$\hat{F}_{F_{y}}$} &
\textcolor{black}{0.25} &
\textcolor{black}{N}\tabularnewline
\hline 
\end{tabular}
\end{table}

\textcolor{black}{}
\begin{table*}[t]
\centering{}\textcolor{black}{\caption{\textcolor{blue}{\label{tab:Simulation-results.}}Simulation results
with different disturbances and noise. The disturbances of UAV $i\in\left\{ L,\,F\right\} $
were obtained from a zero-mean normal distribution with variance $\sigma_{i}$
defined as $\mathcal{N}_{i}(0,\sigma_{i}),$ and the noise was also
obtained from a zero-mean normal distribution.}
}%
\begin{tabular}{|c|c|c|c|c|c|c|c|c|c|c|}
\hline 
\multirow{2}{*}{\textcolor{black}{Simulation}} &
\multicolumn{2}{c|}{\textcolor{black}{Disturbances}} &
\multicolumn{2}{c|}{\textcolor{black}{Noise}} &
\multicolumn{6}{c|}{\textcolor{black}{Errors}}\tabularnewline
\cline{2-11} 
 & \textcolor{black}{Leader} &
\textcolor{black}{Follower} &
\textcolor{black}{Accelerometer} &
\textcolor{black}{Gyro} &
\textcolor{black}{$x_{e}$} &
\textcolor{black}{$y_{e}$} &
\textcolor{black}{$\eta_{1}$} &
\textcolor{black}{$\eta_{2}$} &
\textcolor{black}{$\hat{F}_{F_{x}}$} &
\textcolor{black}{$\hat{F}_{F_{y}}$}\tabularnewline
\hline 
\textcolor{black}{1} &
\textcolor{black}{$\mathcal{N}_{L}(0,0.5)$} &
\textcolor{black}{$\mathcal{N}_{F}(0,0.5)$} &
\multirow{3}{*}{\textcolor{black}{$\mathcal{N}(0,1)$}} &
\multirow{3}{*}{\textcolor{black}{$\mathcal{N}(0,0081)$}} &
\textcolor{black}{0.75} &
\textcolor{black}{0.69} &
\textcolor{black}{0.33} &
\textcolor{black}{0.29} &
\textcolor{black}{0.88} &
\textcolor{black}{0.4}\tabularnewline
\cline{1-3} \cline{6-11} 
\textcolor{black}{2} &
\textcolor{black}{$\mathcal{N}_{L}(0,1.5)$} &
\textcolor{black}{$\mathcal{N}_{F}(0,0.5)$} &
 &  & \textcolor{black}{0.77} &
\textcolor{black}{0.82} &
\textcolor{black}{0.52} &
\textcolor{black}{0.6} &
\textcolor{black}{1.56} &
\textcolor{black}{1.28}\tabularnewline
\cline{1-3} \cline{6-11} 
\textcolor{black}{3} &
\textcolor{black}{$\mathcal{N}_{L}(0,1.5)$} &
\textcolor{black}{$\mathcal{N}_{F}(0,1.5)$} &
 &  & \textcolor{black}{0.9} &
\textcolor{black}{0.82} &
\textcolor{black}{0.55} &
\textcolor{black}{0.56} &
\textcolor{black}{1.3} &
\textcolor{black}{0.7}\tabularnewline
\hline 
\end{tabular}
\end{table*}

\subsection{\textcolor{black}{Simulation 2: Changing the Trajectory of the Payload
During Flight}}

\begin{figure}[H]
\centering{}\textcolor{black}{\includegraphics[width=0.95\columnwidth]{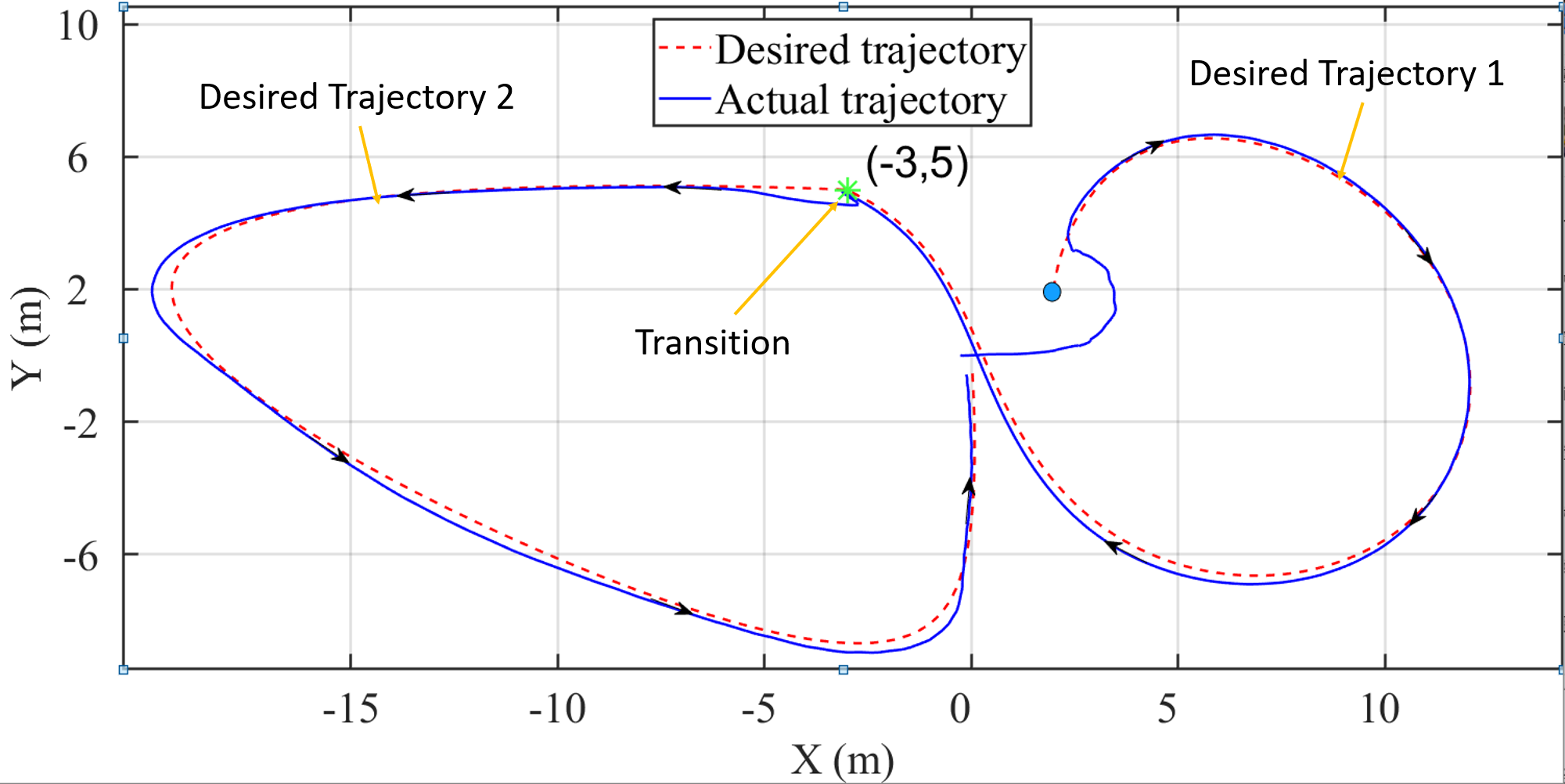}
\caption{\label{fig:Changing the trajectory}Trajectories 1 and 2. The leader-follower
transportation system started at the blue dot from trajectory 1 and
transited to trajectory 2 at the green star in real time.}
}
\end{figure}

\textcolor{black}{The tracking performance of the controllers are
evaluated in Section \ref{subsec:Simulation 1}. Another simulation
was conducted to evaluate the efficacy of the control architecture
that allows changing the desired trajectory in real time. The system
starts from the blue dot in Fig. \ref{fig:Changing the trajectory},
and attempted to follow a double circular trajectory like that shown
in Fig. \ref{fig:figure-8 tracking} until $c_{2}$ point on the payload
reached the green star. At that instant, the leader decided to transit
$c_{2}$ from desired trajectory 1 to desired trajectory 2, and the
follower was not aware of the transition. Fig. \ref{fig:Changing the trajectory}
indicates that the controller can ensure good tracking performance
while traveling along a straight trajectory, and that the tracking
error increases as the curvature increases, which can again be attributed
to the efficacy of the impedance controller since the desired trajectory
and the curvature are not available to the follower.}

\textcolor{black}{}

\subsection{\textcolor{black}{System Parameters}}

\textcolor{black}{The system parameters and control gains used in
the simulations are listed in Table \ref{tab4:The-system-parameters.}
and \ref{tab5:The-control-gains.}, respectively .}

\textcolor{black}{}
\begin{table}[H]
\centering{}\textcolor{black}{\caption{\label{tab4:The-system-parameters.}The system parameters.}
}%
\begin{tabular}{|c|c|c|c|c|c|c|c|}
\hline 
\textcolor{black}{Parameter} &
\textcolor{black}{$m_{p}$} &
\textcolor{black}{$I_{zz}$} &
\textcolor{black}{$L$} &
\textcolor{black}{$l_{1}$} &
\textcolor{black}{$l_{2}$} &
\textcolor{black}{$F_{upper}$} &
\textcolor{black}{$F_{lower}$}\tabularnewline
\hline 
\hline 
\textcolor{black}{Value} &
\textcolor{black}{$0.5$} &
\textcolor{black}{$0.083$} &
\textcolor{black}{$1$} &
\textcolor{black}{$0.18$} &
\textcolor{black}{$0.18$} &
\textcolor{black}{$0.3$} &
\textcolor{black}{$0.2$}\tabularnewline
\hline 
\textcolor{black}{Unit} &
\textcolor{black}{kg} &
\textcolor{black}{$\text{kg}\text{m}^{2}$} &
\textcolor{black}{m} &
\textcolor{black}{m} &
\textcolor{black}{m} &
\textcolor{black}{N} &
\textcolor{black}{N}\tabularnewline
\hline 
\end{tabular}
\end{table}

\textcolor{black}{}
\begin{table}[H]
\begin{centering}
\textcolor{black}{\caption{\label{tab5:The-control-gains.}Control gains.}
}%
\begin{tabular}{|c|c|c|c|c|c|}
\hline 
\textcolor{black}{Gain} &
\textcolor{black}{$k_{1}$} &
\textcolor{black}{$k_{2}$} &
\textcolor{black}{$k_{3}$} &
\textcolor{black}{$k_{v}$} &
\textcolor{black}{$k_{\omega}$}\tabularnewline
\hline 
\hline 
\textcolor{black}{Value} &
\textcolor{black}{3} &
\textcolor{black}{1} &
\textcolor{black}{5} &
\textcolor{black}{5} &
\textcolor{black}{11}\tabularnewline
\hline 
\end{tabular}
\par\end{centering}
\textcolor{black}{$\,$}
\centering{}\textcolor{black}{}%
\begin{tabular}{|c|c|c|c|c|c|}
\hline 
\textcolor{black}{Gain} &
\textcolor{black}{$k_{F_{1}}$} &
\textcolor{black}{$k_{F_{2}}$} &
\textcolor{black}{$k_{d}$} &
\textcolor{black}{$b_{d}$} &
\textcolor{black}{$M_{d}$}\tabularnewline
\hline 
\hline 
\textcolor{black}{Value} &
\textcolor{black}{1} &
\textcolor{black}{3} &
\textcolor{black}{13.6} &
\textcolor{black}{2.0} &
\textcolor{black}{1.5}\tabularnewline
\hline 
\end{tabular}
\end{table}

\subsection{\textcolor{black}{Simulations with Noise and Disturbances}}

\textcolor{black}{To make the simulations more-accurate representations
of experiments, noise was added to the IMU and disturbances of various
magnitudes were added to the UAVs to verify the system stability.
The simulation results are summarized in Table \ref{tab:Simulation-results.}. }

\textcolor{black}{The noise was the same for all three simulations.
The errors increased from simulation 1 to simulation 2, since the
disturbances in the leader UAV increased, and similarly from simulation
2 to simulation 3 as the disturbances in the follower increased. The
simulations indicate that the UAV could successfully complete all
of the cooperative transportation tasks. These simulation findings
verify the control effectiveness of the developed controller. }

\section{\textcolor{black}{Experiments}}

\textcolor{black}{Experiments were performed to evaluate the tracking
performance of the leader and follower controllers. Moreover, the
estimation performance of the UKFs was also determined to ensure they
provide accurate estimates for control feedback.}

\subsection{\textcolor{black}{Implementation}}

\subsubsection{\textcolor{black}{Hardware}}

\begin{figure}[H]
\begin{centering}
\textcolor{black}{\includegraphics[width=0.9\columnwidth]{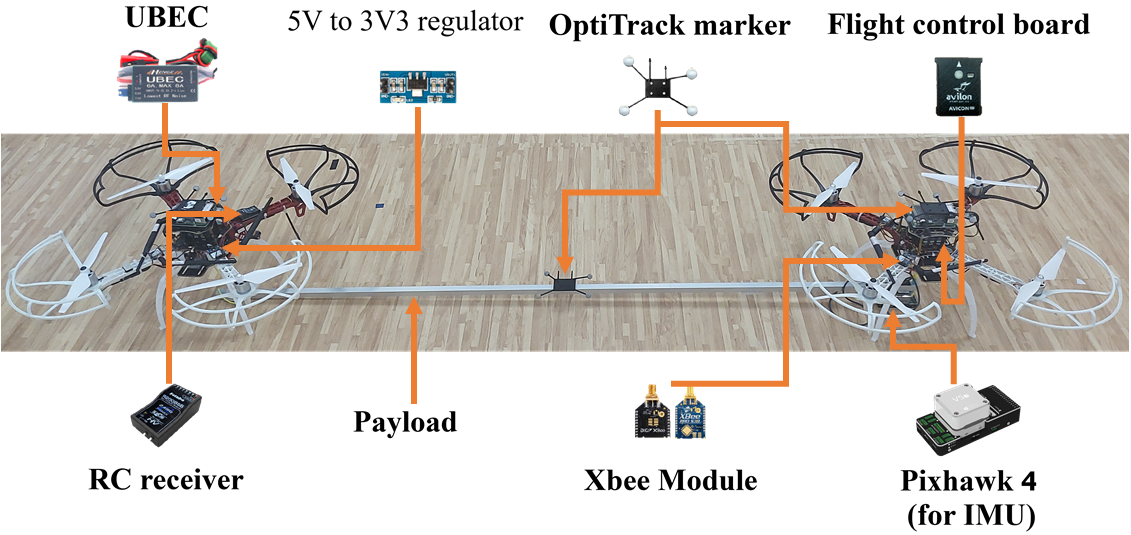}}
\par\end{centering}
\textcolor{black}{\caption{\label{fig:exp-hardware}Hardware architecture.}
}
\end{figure}

\textcolor{black}{A DJI F450 UAV was used in the experiments. This
multirotor was equipped with XBee modules to allow communication between
the multirotor and the ground station for monitoring and collecting
data. Angular velocity and acceleration for obtaining the attitude
of the multirotor were measured using the IMU, and the position of
the multirotor was measured using a motion capture system (Optitrack).
Companion computers (i.e., x5-Z8350 from AAEON}\footnote{\textcolor{black}{https://www.aaeon.com/en/p/up-board-computer-board-for-professional-makers}}\textcolor{black}{)
were implemented on the UAVs for computing the UKFs. The two ends
of the payload were connected to the UAVs with ropes. The configuration
of the multirotor is shown in Fig. \ref{fig:exp-hardware}.}

\textcolor{black}{}

\subsubsection{\textcolor{black}{Software}}

\textcolor{black}{The developed controller and the UKFs were implemented
on an ROS (Robot Operating System) in the companion computers. Since
the controller and UKFs are implemented on the follower and leader,
no inter-agent communication is required. The leader controller was
computed on the UP board, and the follower controller was computed
on the flight control board.}

\subsubsection{\textcolor{black}{System Parameters}}

\textcolor{black}{The system parameters and control gains used in
the experiments are listed in Tables \ref{tab4:Exp-system-parameters}
and \ref{tab5:Exp-control-gains}, respectively. The upper and lower
bounds of the leader controller are listed in Table \ref{tab:leader-controller-bound}.}

\textcolor{black}{}
\begin{table}[H]
\begin{centering}
\textcolor{black}{\caption{\label{tab4:Exp-system-parameters}System parameters.}
}
\par\end{centering}
\centering{}\textcolor{black}{}%
\begin{tabular}{|c|c|c|c|c|c|c|c|}
\hline 
\textcolor{black}{Parameter} &
\textcolor{black}{$m_{p}$} &
\textcolor{black}{$I_{zz}$} &
\textcolor{black}{$L$} &
\textcolor{black}{$l_{1}$} &
\textcolor{black}{$l_{2}$} &
\textcolor{black}{$F_{upper}$} &
\textcolor{black}{$F_{lower}$}\tabularnewline
\hline 
\hline 
\textcolor{black}{Value} &
\textcolor{black}{$0.5$} &
\textcolor{black}{$0.1053$} &
\textcolor{black}{$1.59$} &
\textcolor{black}{$0.2$} &
\textcolor{black}{$0.2$} &
\textcolor{black}{$0.3$} &
\textcolor{black}{$0.25$}\tabularnewline
\hline 
\textcolor{black}{Unit} &
\textcolor{black}{kg} &
\textcolor{black}{$\text{kg}\text{m}^{2}$} &
\textcolor{black}{m} &
\textcolor{black}{m} &
\textcolor{black}{m} &
\textcolor{black}{N} &
\textcolor{black}{N}\tabularnewline
\hline 
\end{tabular}
\end{table}

\textcolor{black}{}
\begin{table}[H]
\begin{centering}
\textcolor{black}{\caption{\label{tab5:Exp-control-gains}Control gains.}
}
\par\end{centering}
\begin{centering}
\textcolor{black}{}%
\begin{tabular}{|c|c|c|c|c|c|}
\hline 
\textcolor{black}{Gain} &
\textcolor{black}{$k_{1}$} &
\textcolor{black}{$k_{2}$} &
\textcolor{black}{$k_{3}$} &
\textcolor{black}{$k_{v}$} &
\textcolor{black}{$k_{\omega}$}\tabularnewline
\hline 
\hline 
\textcolor{black}{Value} &
\textcolor{black}{3} &
\textcolor{black}{0.1} &
\textcolor{black}{1} &
\textcolor{black}{3} &
\textcolor{black}{30}\tabularnewline
\hline 
\end{tabular}
\par\end{centering}
\textcolor{black}{$\,$}
\centering{}\textcolor{black}{}%
\begin{tabular}{|c|c|c|c|c|c|}
\hline 
\textcolor{black}{Gain} &
\textcolor{black}{$k_{F_{1}}$} &
\textcolor{black}{$k_{F_{2}}$} &
\textcolor{black}{$k_{d}$} &
\textcolor{black}{$b_{d}$} &
\textcolor{black}{$M_{d}$}\tabularnewline
\hline 
\hline 
\textcolor{black}{Value} &
\textcolor{black}{0.04} &
\textcolor{black}{4} &
\textcolor{black}{12.26} &
\textcolor{black}{2.0} &
\textcolor{black}{2}\tabularnewline
\hline 
\end{tabular}
\end{table}

\textcolor{black}{}
\begin{table}[H]
\textcolor{black}{\caption{\label{tab:leader-controller-bound}Upper and lower bounds of the
leader controller.}
}
\centering{}\textcolor{black}{}%
\begin{tabular}{|c|c|c|}
\hline 
\multirow{1}{*}{\textcolor{black}{Direction}} &
\textcolor{black}{$\mathit{x}$} &
\textcolor{black}{$\mathit{y}$}\tabularnewline
\hline 
\hline 
\textcolor{black}{Upper bound (N)} &
\textcolor{black}{2.0} &
\textcolor{black}{10.0}\tabularnewline
\hline 
\textcolor{black}{Lower bound (N)} &
\textcolor{black}{-2.6} &
\textcolor{black}{-7.0}\tabularnewline
\hline 
\end{tabular}
\end{table}

\subsection{\textcolor{black}{Evaluation of the Tracking Controller}}

\textcolor{black}{The experiments were divided into two parts: (1)
evaluating the controller by tracking a desired trajectory and (2)
evaluating the performance of the UKFs for the UAVs.}

\subsubsection{\textcolor{black}{Evaluation of the Controller}}

\begin{figure}[H]
\begin{centering}
\textcolor{black}{\includegraphics[width=0.9\columnwidth]{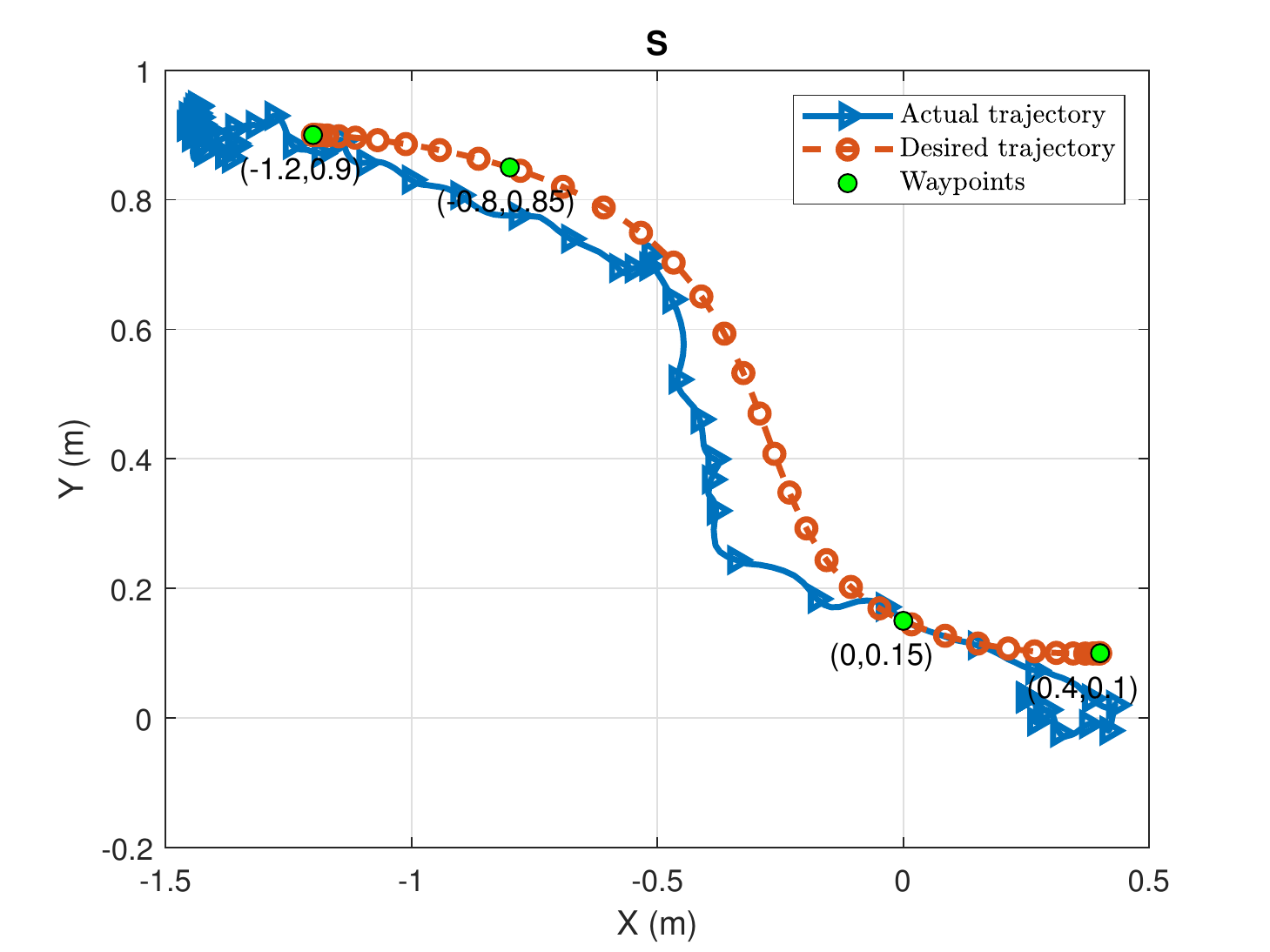}}
\par\end{centering}
\textcolor{black}{\caption{\label{fig:S path plot}Desired trajectory S and the actual trajectory
of point $c_{2}$.}
}
\end{figure}

\begin{figure}[h]
\begin{centering}
\textcolor{black}{\includegraphics[width=0.9\columnwidth]{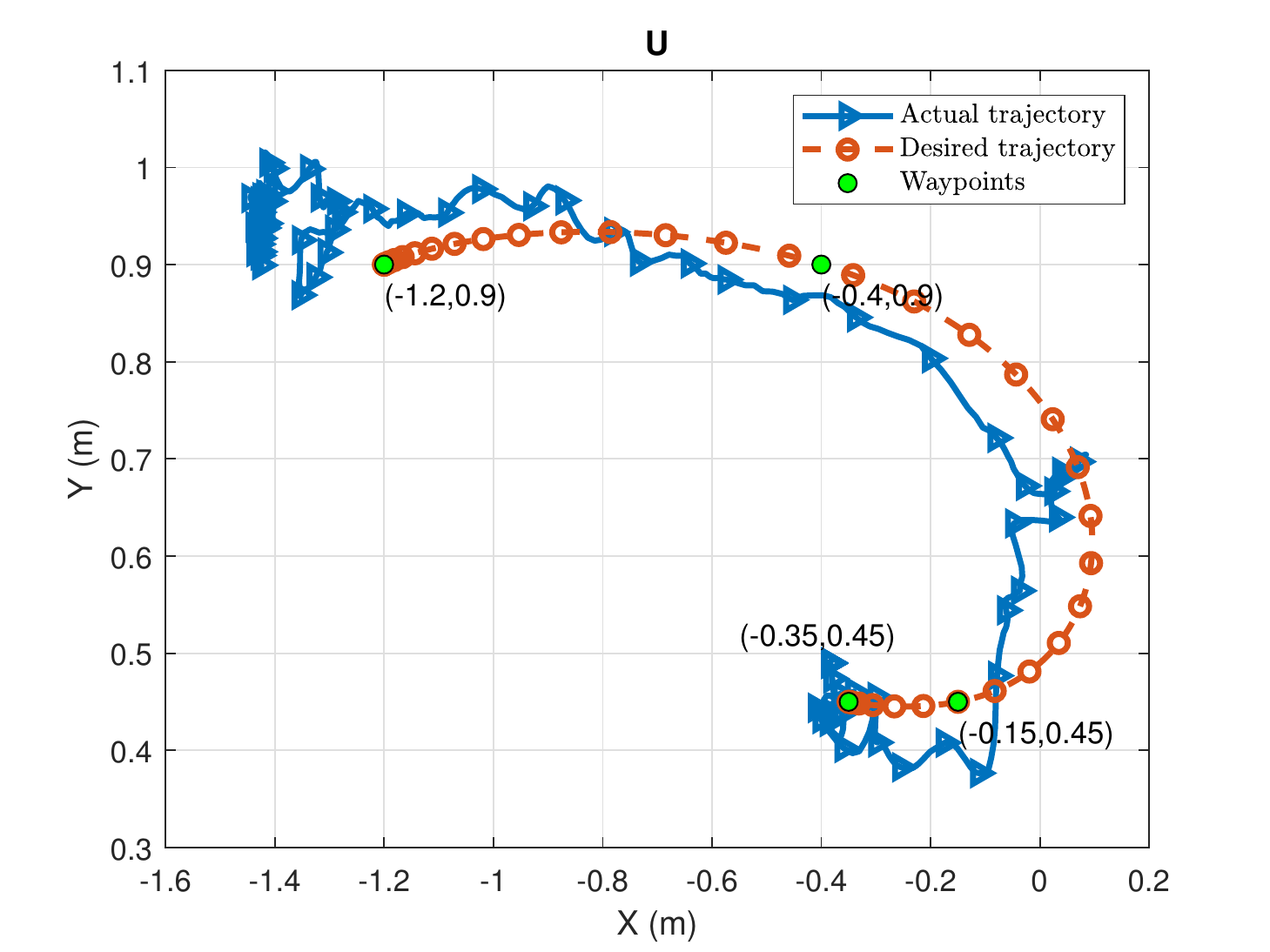}}
\par\end{centering}
\textcolor{black}{\caption{\label{fig:U path plot}Desired trajectory U and the actual trajectory
of point $c_{2}$.}
}

\end{figure}

\textcolor{black}{Figs. \ref{fig:S path plot} and \ref{fig:U path plot}
show that the system can track a desired trajectory, which is a high-order
polynomial solved by the companion computer on the leader UAV.}

\textcolor{black}{}

\textcolor{black}{}
\textcolor{black}{The waypoints of trajectories S and U are listed
in Tables \ref{tab:All-segments-in S} and \ref{tab:All-segments-in U},
respectively. The position RMSE of the trajectory tracking is listed
in Table \ref{tab:RMSE}. The tracking errors were greatly affected
by airflow disturbance due to the experiments being conducted in a
restricted space.}

\textcolor{black}{}
\begin{table}[H]
\textcolor{black}{\caption{\label{tab:All-segments-in S}All segments in trajectory S.}
}
\centering{}\textcolor{black}{}%
\begin{tabular}{|c|c|c|}
\hline 
\textcolor{black}{Segment} &
\textcolor{black}{Start - End} &
\textcolor{black}{Duration (s)}\tabularnewline
\hline 
\hline 
\textcolor{black}{1} &
\textcolor{black}{$(-1.20,0.90)-(-0.80,0.85)$} &
\textcolor{black}{$5.0$}\tabularnewline
\hline 
\textcolor{black}{2} &
\textcolor{black}{$(-0.80,0.85)-(0.00,0.15)$} &
\textcolor{black}{$6.0$}\tabularnewline
\hline 
\textcolor{black}{3} &
\textcolor{black}{$(0.00,0.15)-(0.40,0.10)$} &
\textcolor{black}{$5.0$}\tabularnewline
\hline 
\end{tabular}
\end{table}

\textcolor{black}{}
\begin{table}[H]
\textcolor{black}{\caption{\label{tab:All-segments-in U}All segments in trajectory U.}
}
\centering{}\textcolor{black}{}%
\begin{tabular}{|c|c|c|}
\hline 
\textcolor{black}{Segment} &
\textcolor{black}{Start - End} &
\textcolor{black}{Duration (s)}\tabularnewline
\hline 
\hline 
\textcolor{black}{1} &
\textcolor{black}{$(-1.20,0.90)-(-0.40,0.90)$} &
\textcolor{black}{$7.0$}\tabularnewline
\hline 
\textcolor{black}{2} &
\textcolor{black}{$(-0.40,0.90)-(-0.15,0.45)$} &
\textcolor{black}{$5.0$}\tabularnewline
\hline 
\textcolor{black}{3} &
\textcolor{black}{$(-0.15,0.45)-(-0.35,0.45)$} &
\textcolor{black}{$3.0$}\tabularnewline
\hline 
\end{tabular}
\end{table}

\begin{figure}[H]
\begin{centering}
\textcolor{black}{\includegraphics[width=0.9\columnwidth]{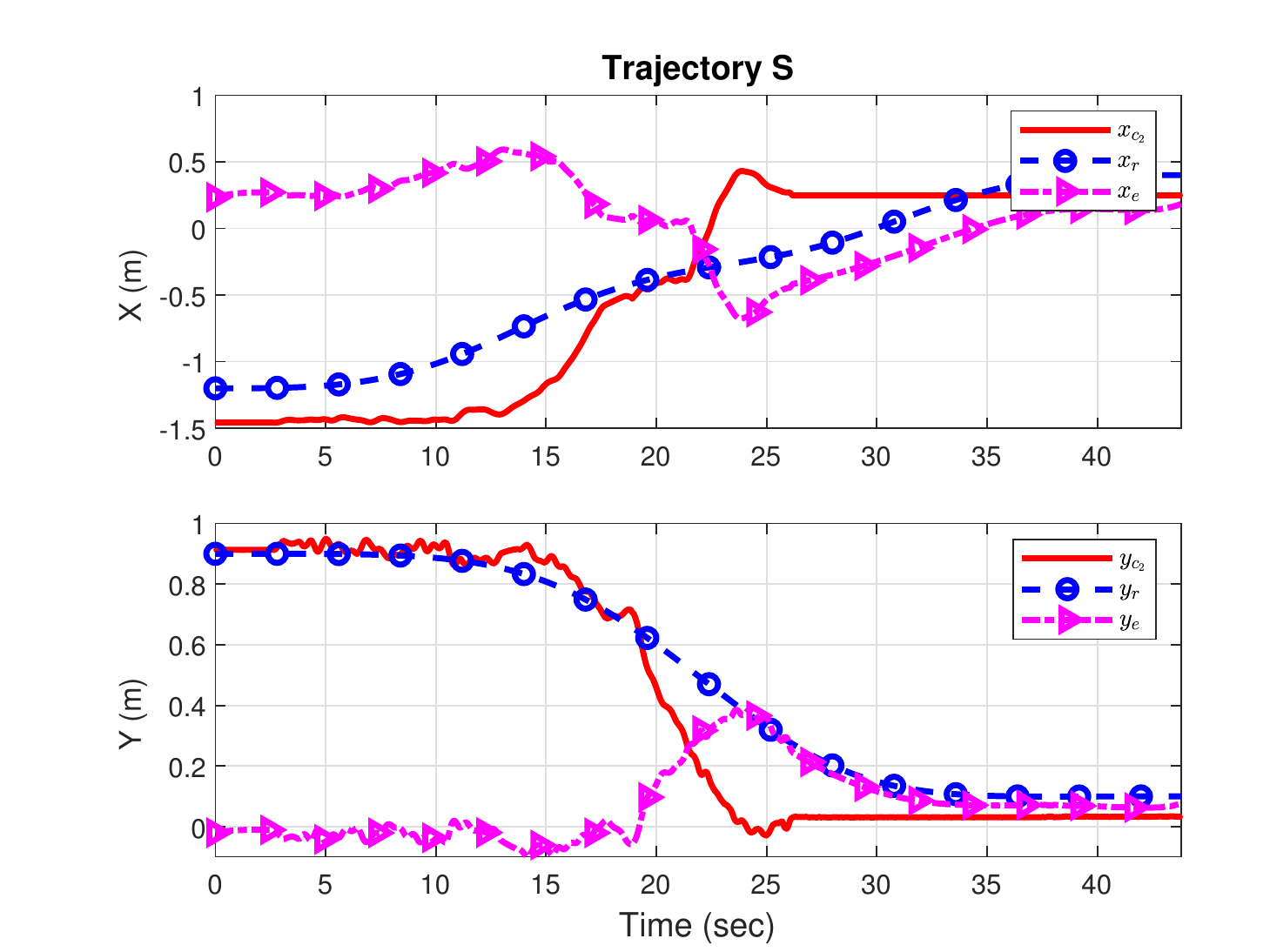}}
\par\end{centering}
\textcolor{black}{\caption{\label{fig:Traj-S-c2}Position and tracking errors of point $c_{2}$
along the $X$ and $Y$ axes in the inertial frame for trajectory
S.}
}

\end{figure}

\begin{figure}[H]
\begin{centering}
\textcolor{black}{\includegraphics[width=0.9\columnwidth]{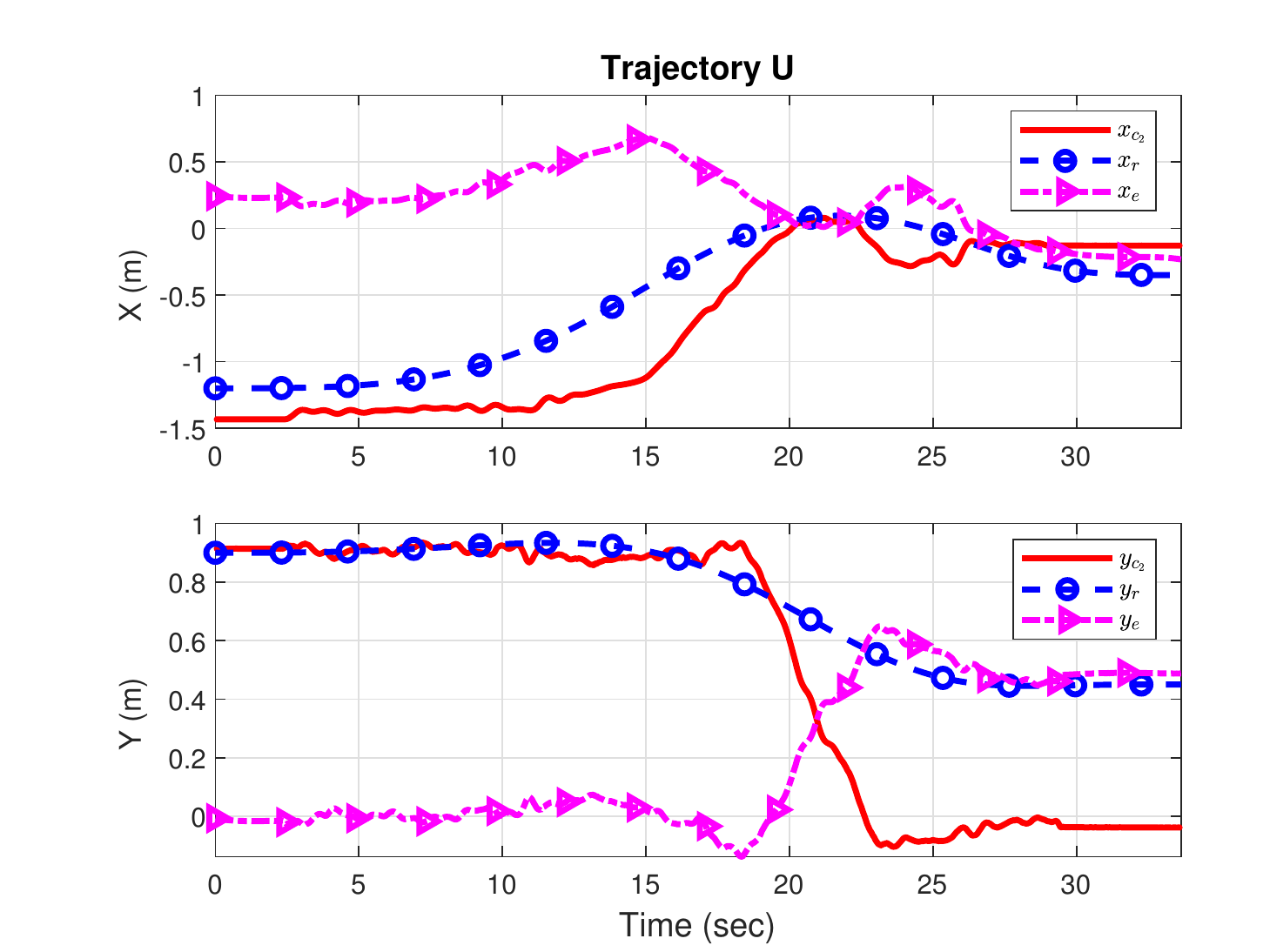}}
\par\end{centering}
\textcolor{black}{\caption{\label{fig:Traj-U-c2}Position and tracking errors of point $c_{2}$
along the $X$ and $Y$ axes in the inertial frame for trajectory
U.}
}
\end{figure}

\textcolor{black}{}
\begin{table}[H]
\textcolor{black}{\caption{\label{tab:RMSE}RMSEs of the two trajectories.}
}
\begin{centering}
\textcolor{black}{}%
\begin{tabular}{|c|c|c|c|c|}
\hline 
\multirow{2}{*}{\textcolor{black}{Trajectory}} &
\multicolumn{2}{c|}{\textcolor{black}{S}} &
\multicolumn{2}{c|}{\textcolor{black}{U}}\tabularnewline
\cline{2-5} 
 & \textcolor{black}{$\mathit{x}$ (m)} &
\textcolor{black}{$\mathit{y}$ (m)} &
\textcolor{black}{$\mathit{x}$ (m)} &
\textcolor{black}{$\mathit{y}$ (m)}\tabularnewline
\hline 
\textcolor{black}{RMSE} &
\textcolor{black}{0.3176} &
\textcolor{black}{0.1366} &
\textcolor{black}{0.2835} &
\textcolor{black}{0.1376}\tabularnewline
\hline 
\textcolor{black}{Geometric RMSE} &
\multicolumn{2}{c|}{\textcolor{black}{0.3457}} &
\multicolumn{2}{c|}{\textcolor{black}{0.3151}}\tabularnewline
\hline 
\end{tabular}
\par\end{centering}
\end{table}
\textcolor{black}{Figs. \ref{fig:Traj-S-c2} and \ref{fig:Traj-U-c2}
show the trajectories and the tracking errors during the cooperative
transportation. $x_{c_{2}}$ and $y_{c_{2}}$ represent the position
of $c_{2}$ along the $X$ and $Y$ axes in the inertial frame. The
tracking error in the transverse direction increased as the curvature
increased, and decreased rapidly while traveling along a straight
trajectory. This behavior can be attributed to the efficacy of the
impedance controller, since the desired trajectory and the curvature
are not available to the follower. Table \ref{tab:The-standard-deviations-of-c2}
lists the standard deviations of $x_{e}$ and $y_{e}$ for the two
trajectories.}

\textcolor{black}{}

\textcolor{black}{}

\begin{figure}[H]
\begin{centering}
\textcolor{black}{\includegraphics[width=0.9\columnwidth]{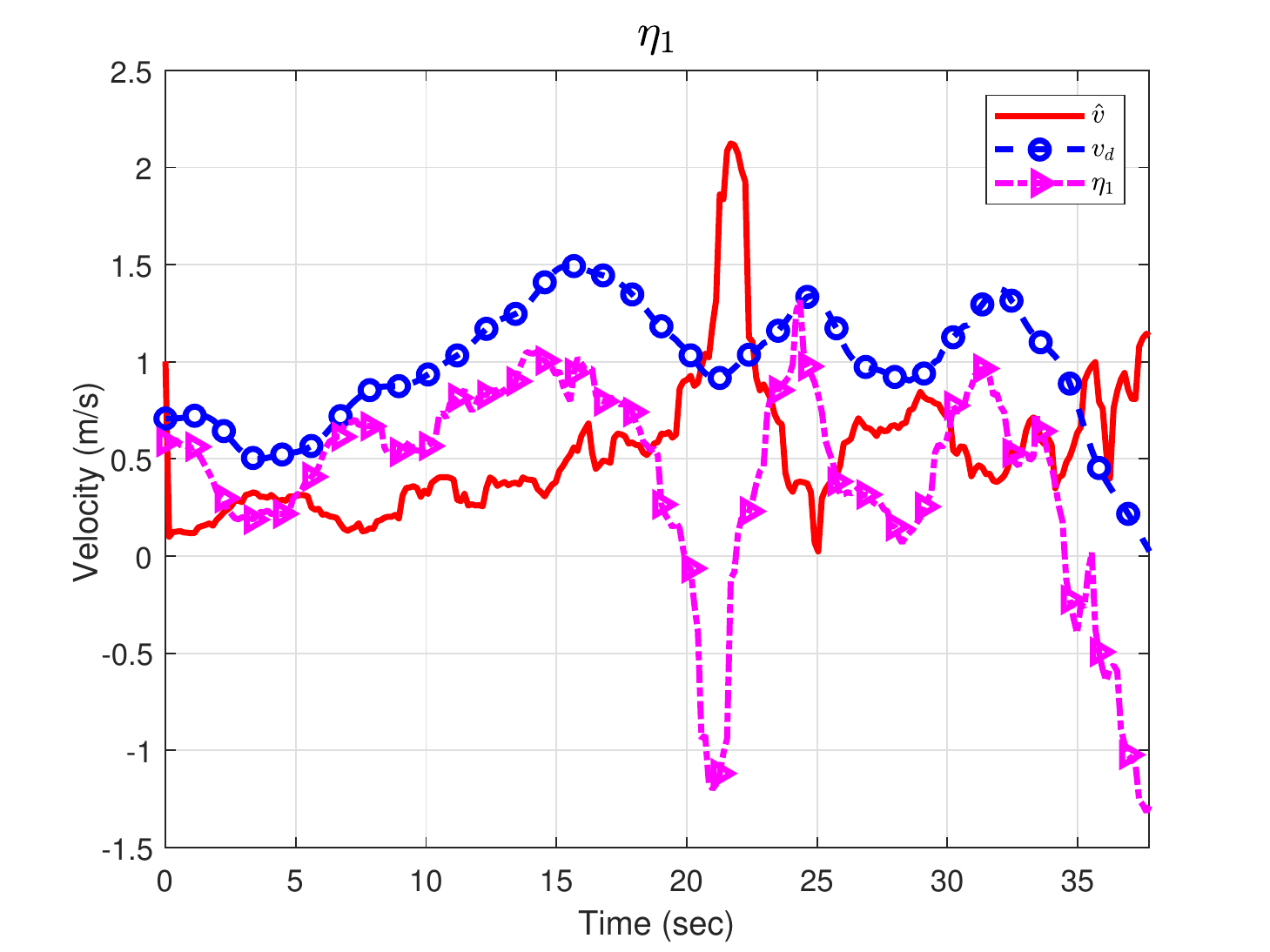}}
\par\end{centering}
\textcolor{black}{\caption{\label{fig:S-eta1}Trajectory S. $\hat{v}$ is estimated by the UKF
described in Section \ref{subsec:The-Second-UKF}, and $v_{d}$ is
the desired velocity according to (\ref{eq:control input of kinematics}).
Signal $\eta_{1}$ indicates the tracking error of the leader controller
during translational motion.}
}
\end{figure}

\begin{figure}[H]
\begin{centering}
\textcolor{black}{\includegraphics[width=0.9\columnwidth]{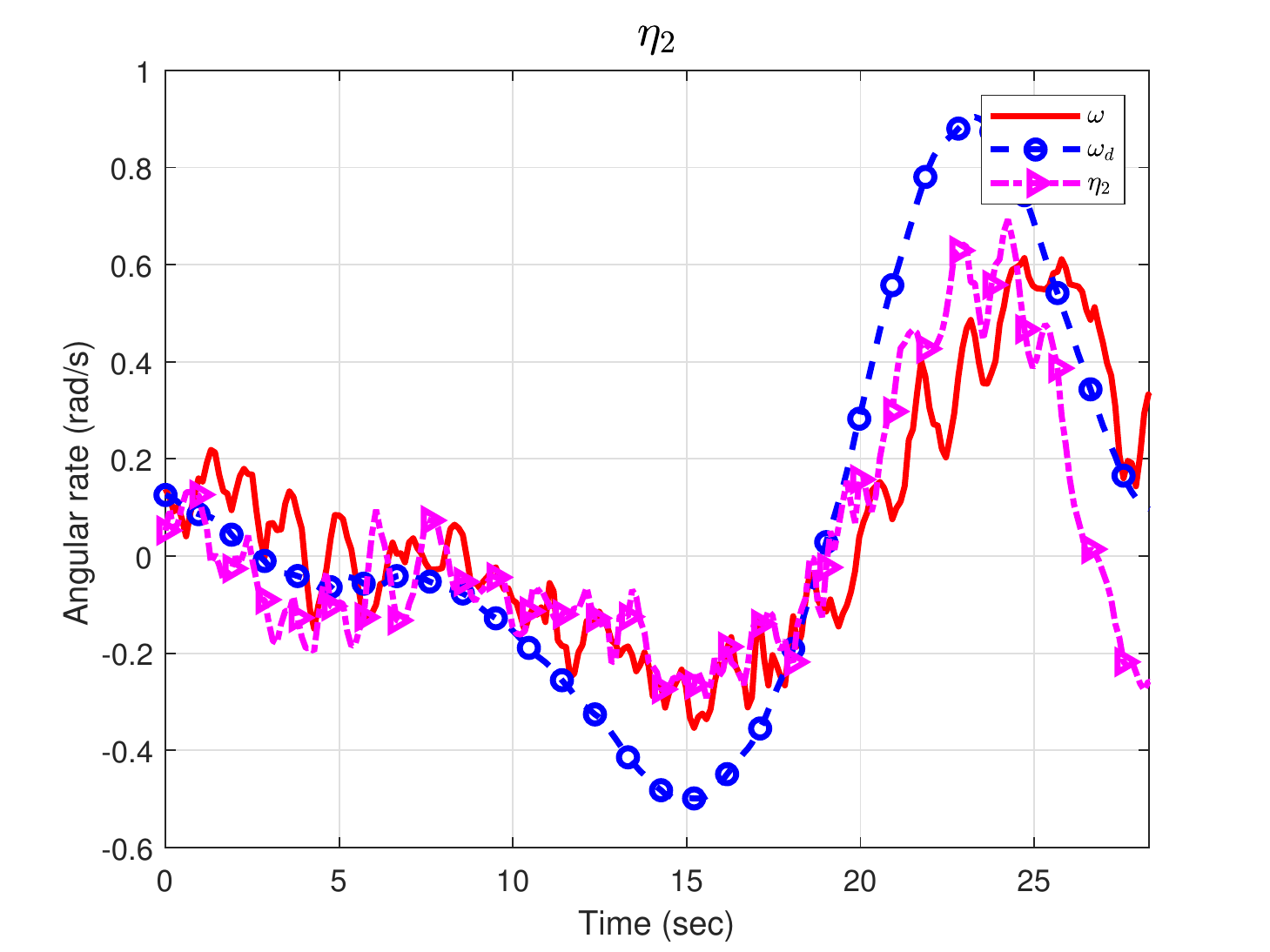}}
\par\end{centering}
\textcolor{black}{\caption{\label{fig:S-eta2}Trajectory S. $\omega$ is measured by the IMU
attached to the payload at $c_{1},$ and $\omega_{d}$ is the desired
angular rate according to (\ref{eq:control input of kinematics}).
Signal $\eta_{2}$ indicates the tracking error of the leader controller
during rotational motion.}
}
\end{figure}

\begin{figure}[h]
\begin{centering}
\textcolor{black}{\includegraphics[width=0.9\columnwidth]{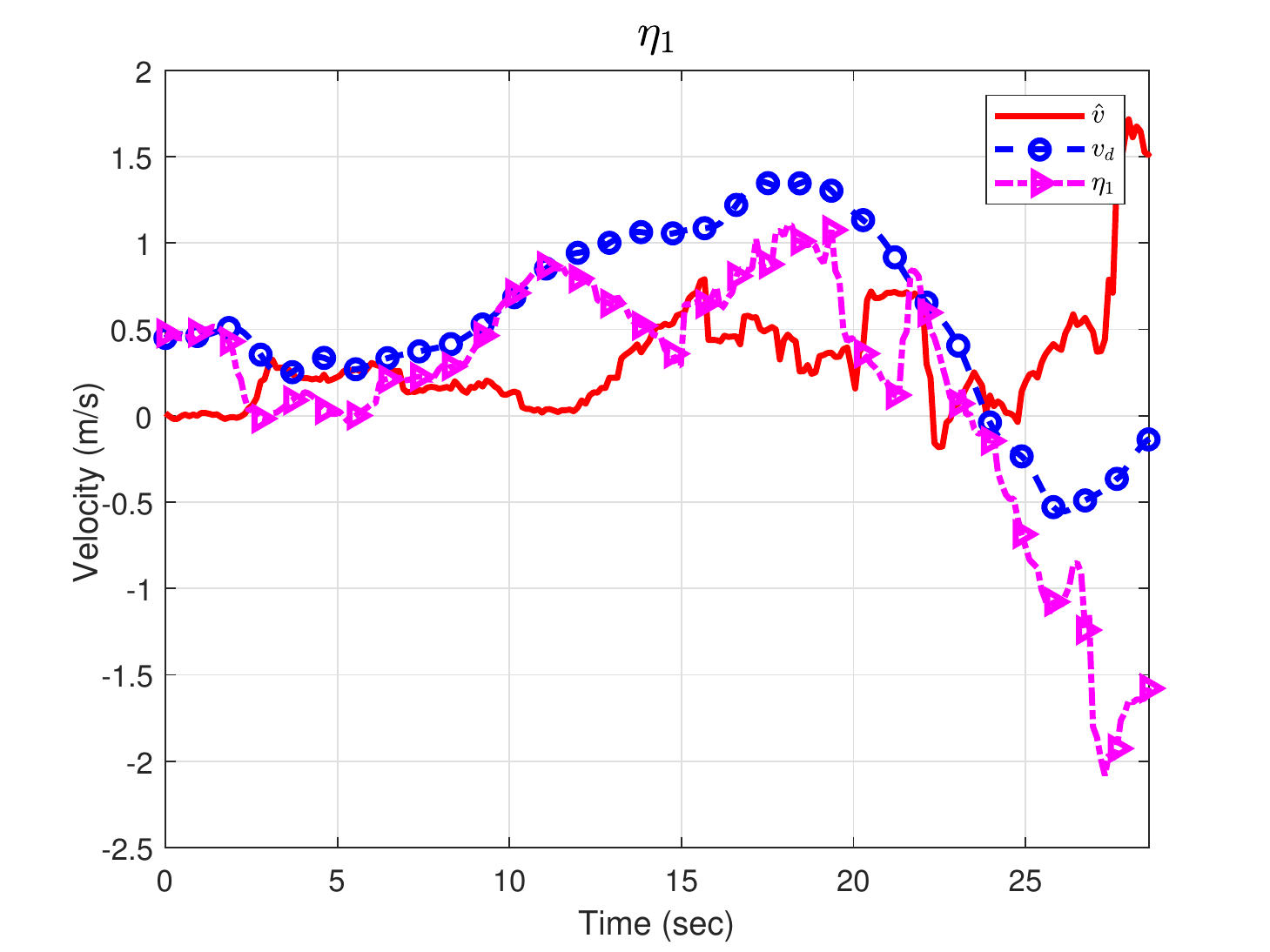}}
\par\end{centering}
\textcolor{black}{\caption{\label{fig:U-eta1}Trajectory U. $\hat{v}$ is estimated by the UKF
described in Section \ref{subsec:The-Second-UKF}, and $v_{d}$ is
the desired velocity according to (\ref{eq:control input of kinematics}).
Signal $\eta_{1}$ indicates the tracking error of the leader controller
during translational motion.}
}
\end{figure}

\begin{figure}[h]
\begin{centering}
\textcolor{black}{\includegraphics[width=0.9\columnwidth]{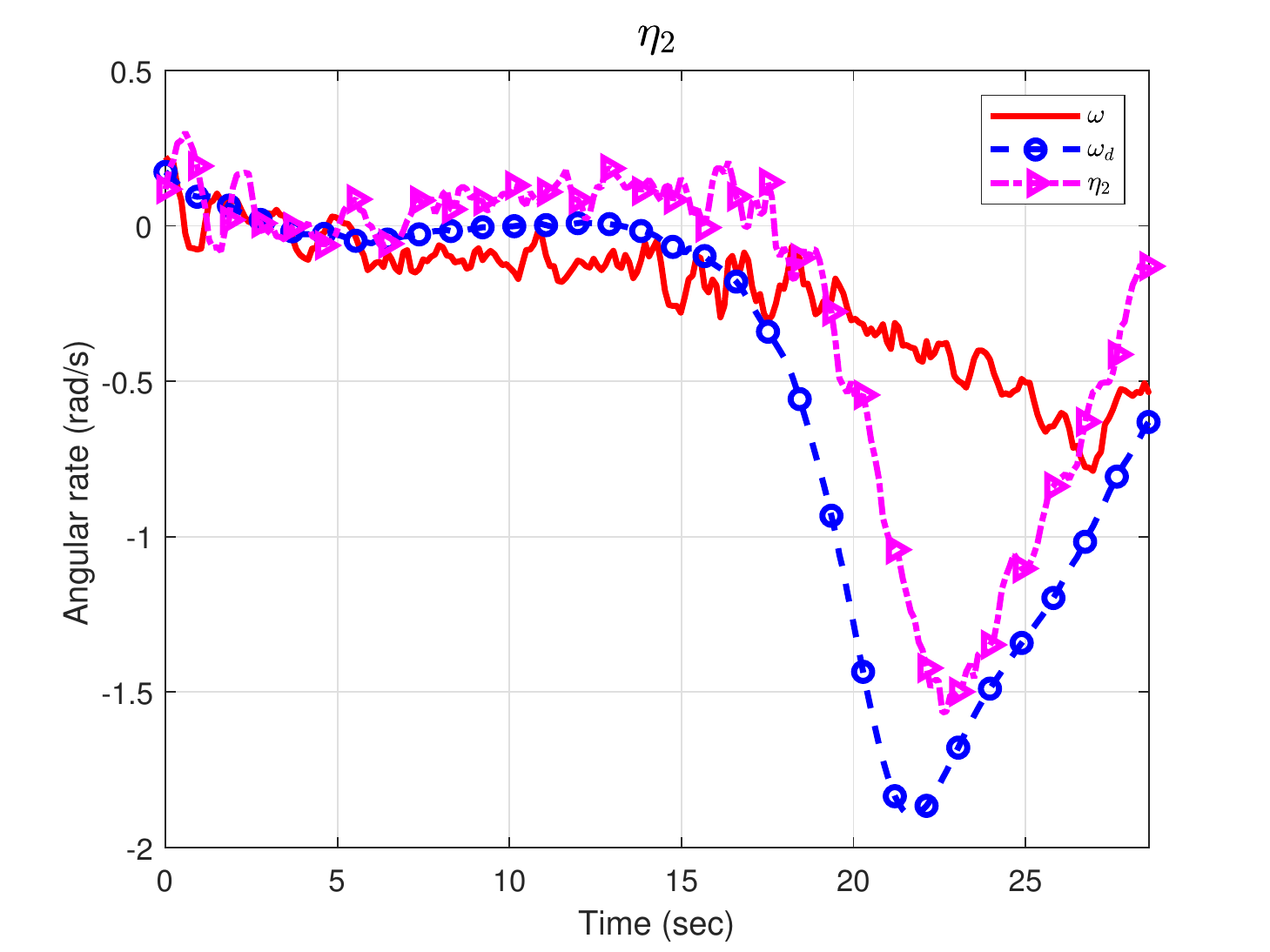}}
\par\end{centering}
\textcolor{black}{\caption{\label{fig:U-eta2}Trajectory U. $\omega$ is measured by the IMU
attached to the payload at $c_{1},$ and $\omega_{d}$ is the desired
angular rate according to (\ref{eq:control input of kinematics}).
Signal $\eta_{2}$ indicates the tracking error of the leader controller
during rotational motion.}
}
\end{figure}

\textcolor{black}{}
\begin{table}[H]
\textcolor{black}{\caption{\label{tab:The-standard-deviations-of-c2}Standard deviations of the
two trajectories.}
}
\centering{}\textcolor{black}{}%
\begin{tabular}{|c|c|c|c|c|}
\hline 
\multirow{2}{*}{\textcolor{black}{Trajectory}} &
\multicolumn{2}{c|}{\textcolor{black}{S}} &
\multicolumn{2}{c|}{\textcolor{black}{U}}\tabularnewline
\cline{2-5} 
 & \textcolor{black}{$\mathit{x}$} &
\textcolor{black}{$\mathit{y}$} &
\textcolor{black}{$\mathit{x}$} &
\textcolor{black}{$\mathit{y}$}\tabularnewline
\hline 
\hline 
\textcolor{black}{Standard deviations (m)} &
\textcolor{black}{0.3063} &
\textcolor{black}{0.1176} &
\textcolor{black}{0.2465} &
\textcolor{black}{0.2479}\tabularnewline
\hline 
\end{tabular}
\end{table}
\textcolor{black}{Figs. \ref{fig:S-eta1}, \ref{fig:S-eta2}, \ref{fig:U-eta1},
and \ref{fig:U-eta2} show the tracking performance of the kinematics
controller designed in (\ref{eq:control input of kinematics}) when
tracking trajectories S and U. The oscillation of signals $\eta_{1}$
and $\eta_{2}$ is due to the payload swinging along the $x^{B}$
direction during the flight, and corresponds to the position errors
that are compensated by the leader controller.}

\textcolor{black}{}

\textcolor{black}{}

\textcolor{black}{}

\begin{figure}[H]
\begin{centering}
\textcolor{black}{\includegraphics[width=0.9\columnwidth]{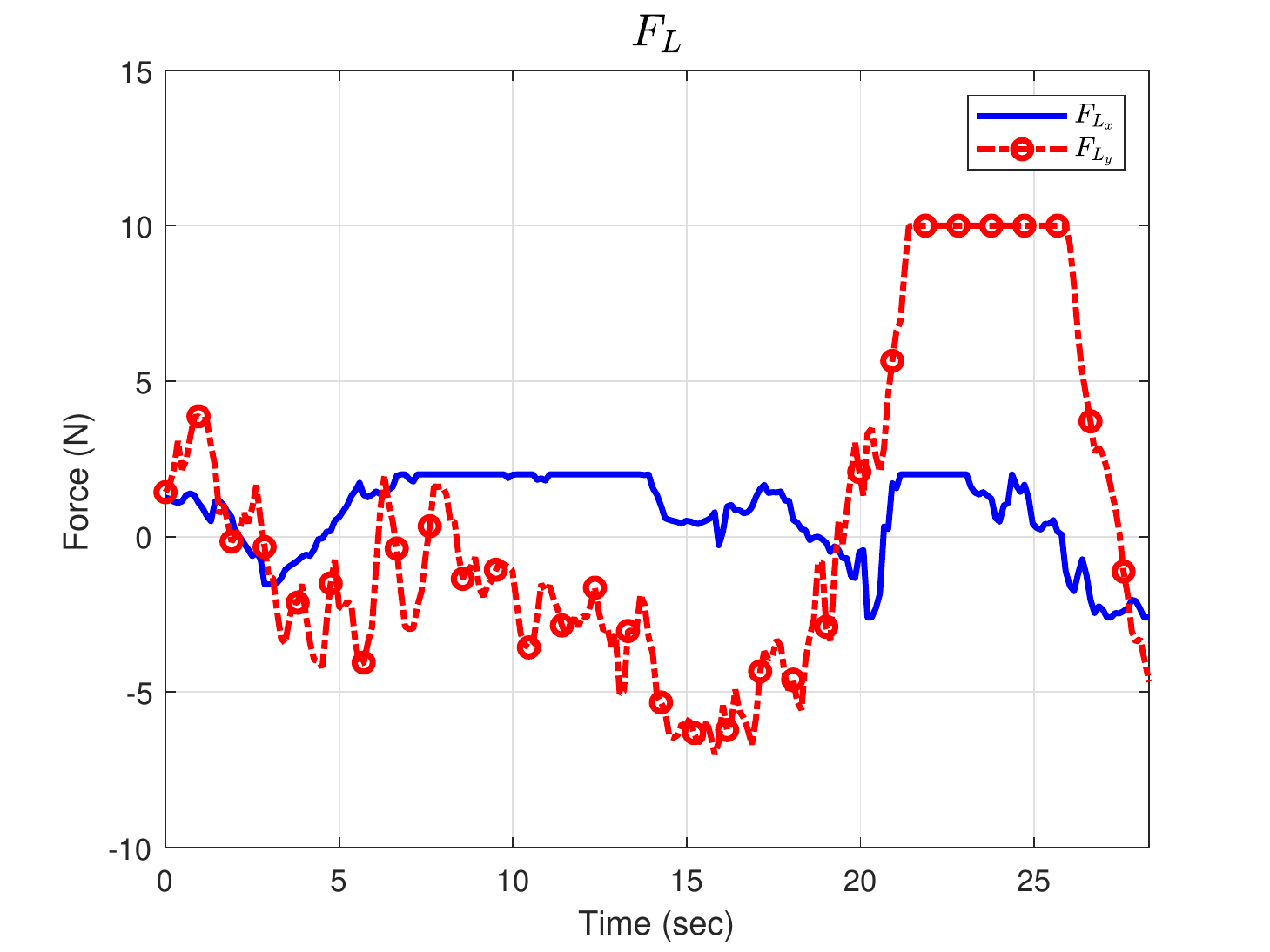}}
\par\end{centering}
\textcolor{black}{\caption{\label{fig:FL-S}Desired forces $F_{L_{x}}$ and $\mathit{F_{L_{y}}}$
for trajectory S computed by the leader controller, both expressed
in the body frame.}
}
\end{figure}

\begin{figure}[H]
\begin{centering}
\textcolor{black}{\includegraphics[width=0.9\columnwidth]{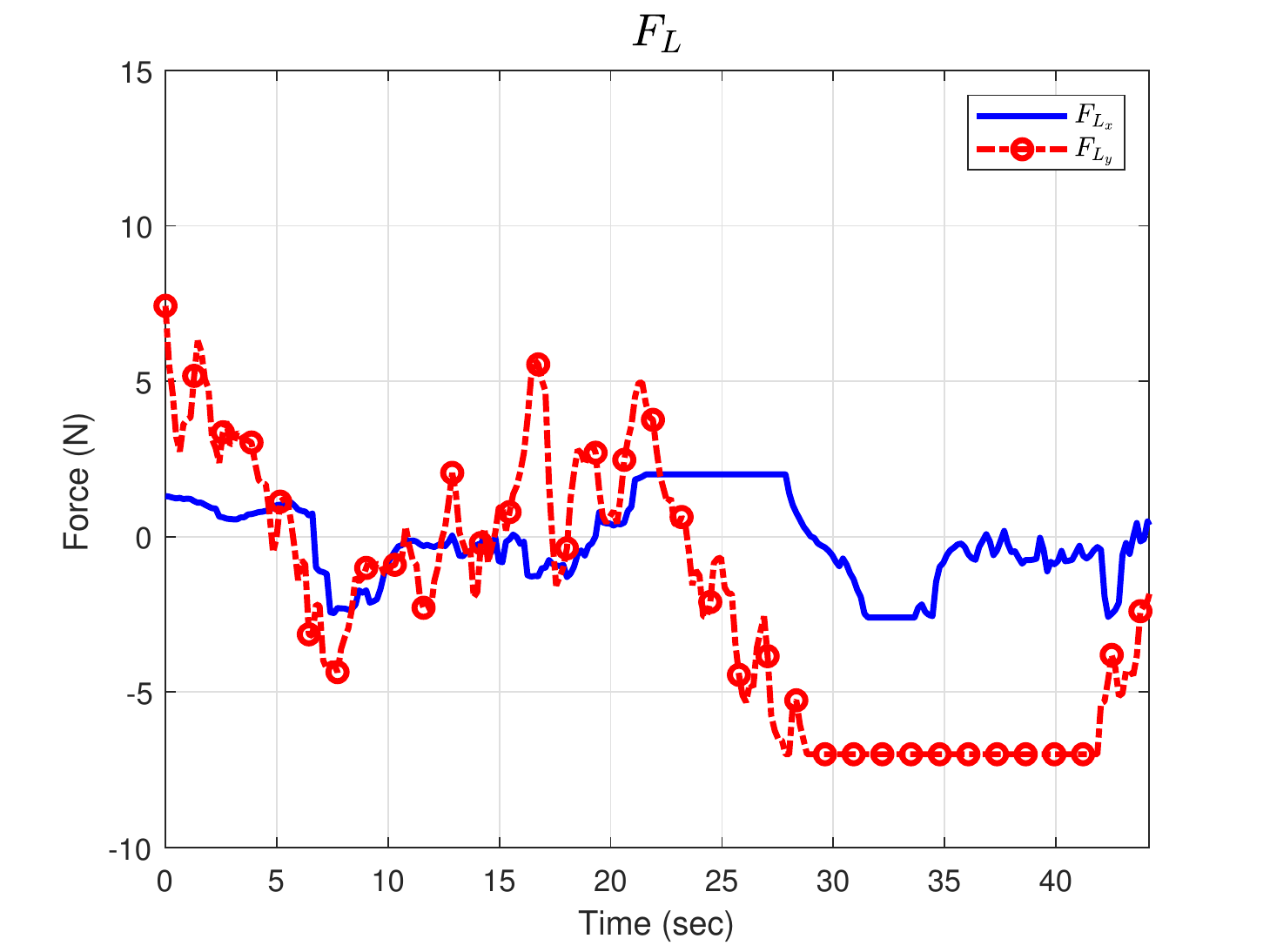}}
\par\end{centering}
\textcolor{black}{\caption{\label{fig:FL-U}Desired forces $F_{L_{x}}$ and $\mathit{F_{L_{y}}}$
for trajectory U computed by the leader controller, both expressed
in the body frame.}
}

\end{figure}

\textcolor{black}{}
\textcolor{black}{Figs. \ref{fig:FL-S} and \ref{fig:FL-U} show the
desired forces applied to point $c_{1}$ by the leader. The upper
and lower bounds are set to prevent aggressive control inputs, since
the rope is not a rigid body. A delay is present because the leader
controller was designed to track the desired $\mathit{c_{2}},$ point
which is not located on the leader itself.}

\textcolor{black}{}

\textcolor{black}{}

\subsubsection{\textcolor{black}{Evaluation of the UKFs}}

\begin{figure}[H]
\begin{centering}
\textcolor{black}{\includegraphics[width=0.9\columnwidth]{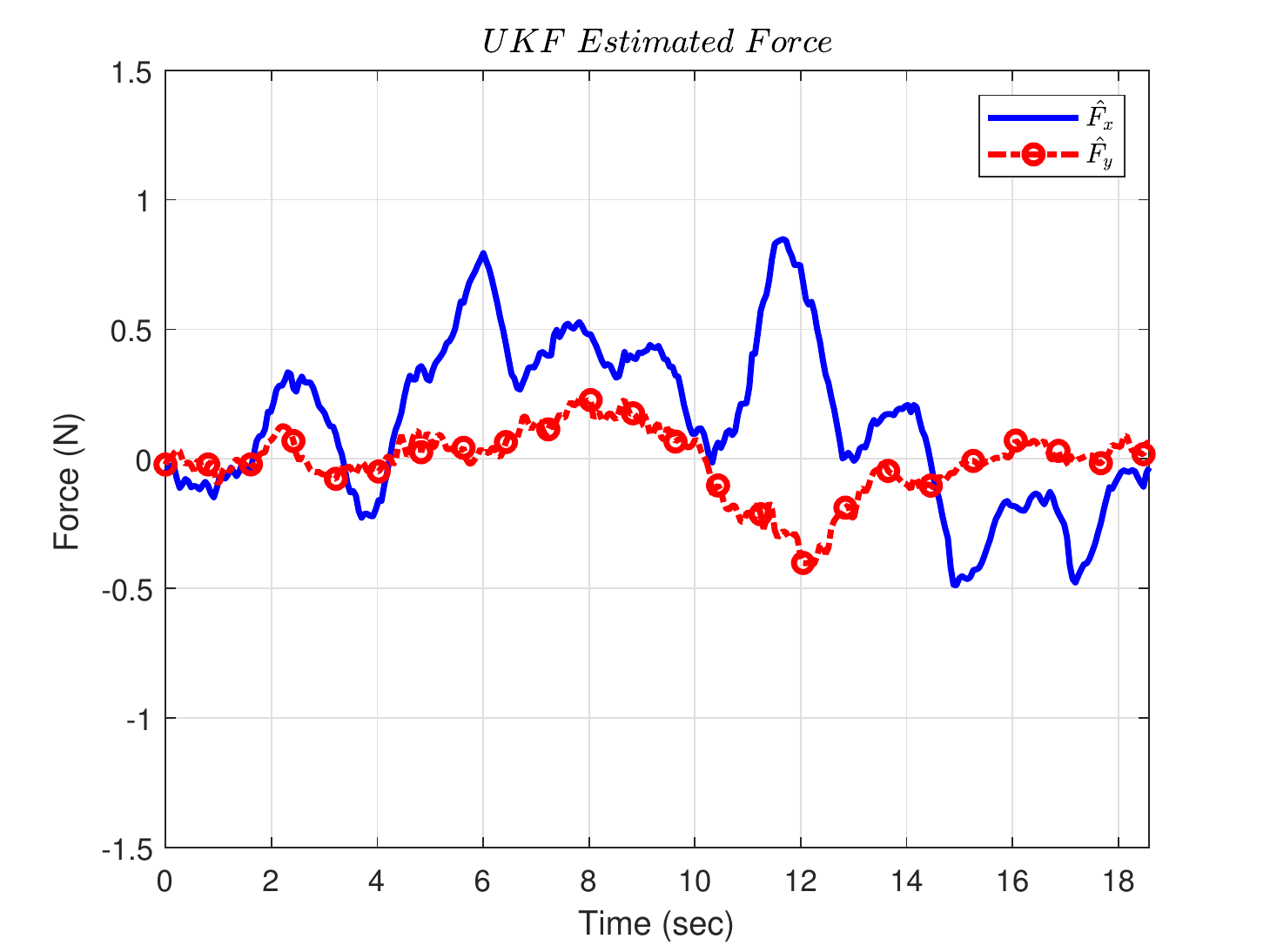}}
\par\end{centering}
\textcolor{black}{\caption{\label{fig:FF-S}Forces estimated by the follower UKF for trajectory
S. }
}
\end{figure}

\begin{figure}[H]
\begin{centering}
\textcolor{black}{\includegraphics[width=0.9\columnwidth]{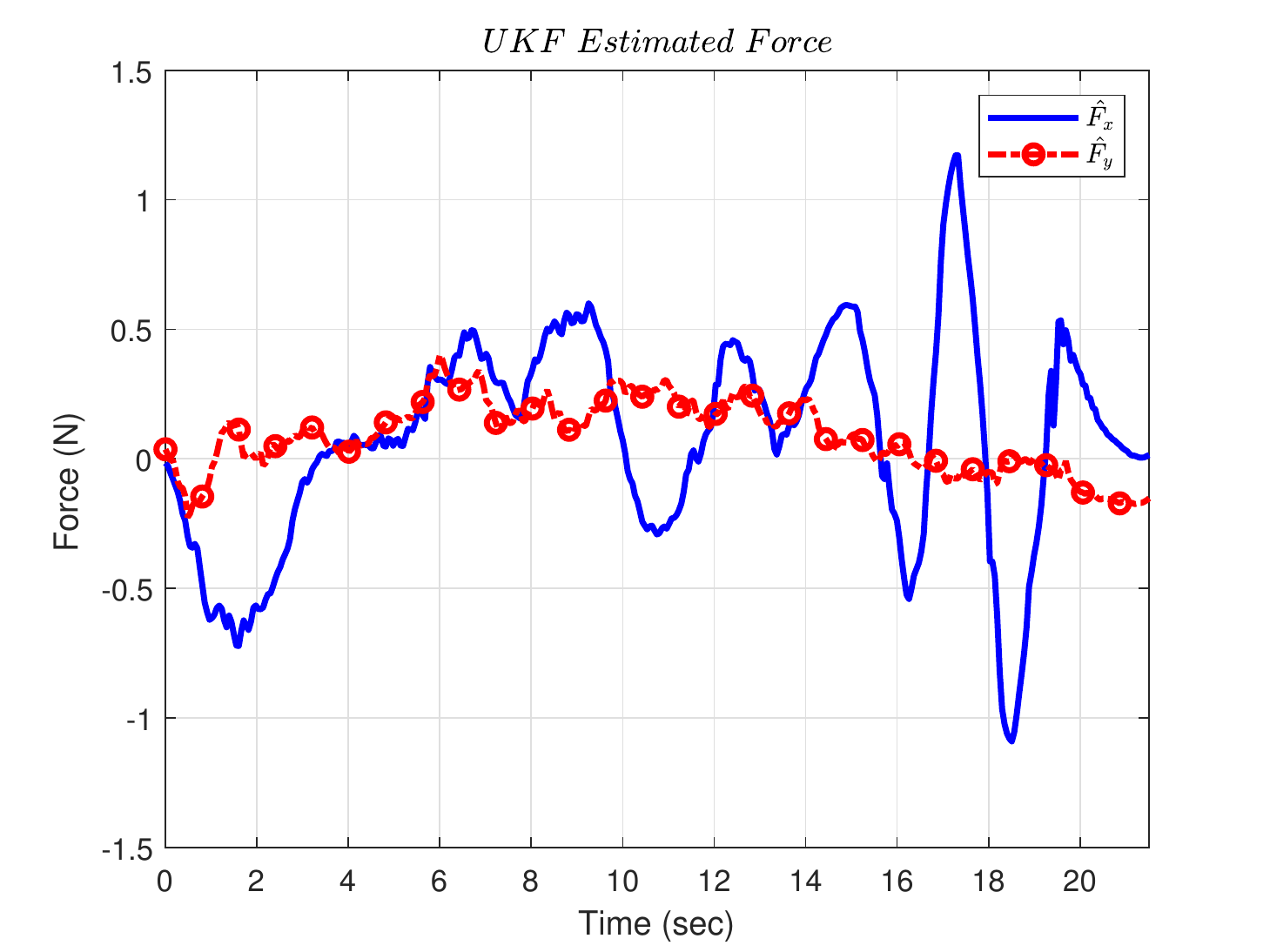}}
\par\end{centering}
\textcolor{black}{\caption{\label{fig:FF-U}Forces estimated by the follower UKF for trajectory
U.}
}
\end{figure}

\textcolor{black}{Figs. \ref{fig:FF-S} and \ref{fig:FF-U} show follower
force estimates $\hat{F}_{F_{x}}$ and $\hat{F}_{F_{y}}$ when tracking
trajectories S and U. Estimate $\mathit{\hat{F}_{F_{y}}}$ is close
to zero due to the nonholonomic constraint, and $\mathit{\hat{F}_{F_{x}}}$
almost remains positive since the leader pulls the follower during
the motion.}

\textcolor{black}{}

\begin{figure}[H]
\begin{centering}
\textcolor{black}{\includegraphics[width=0.9\columnwidth]{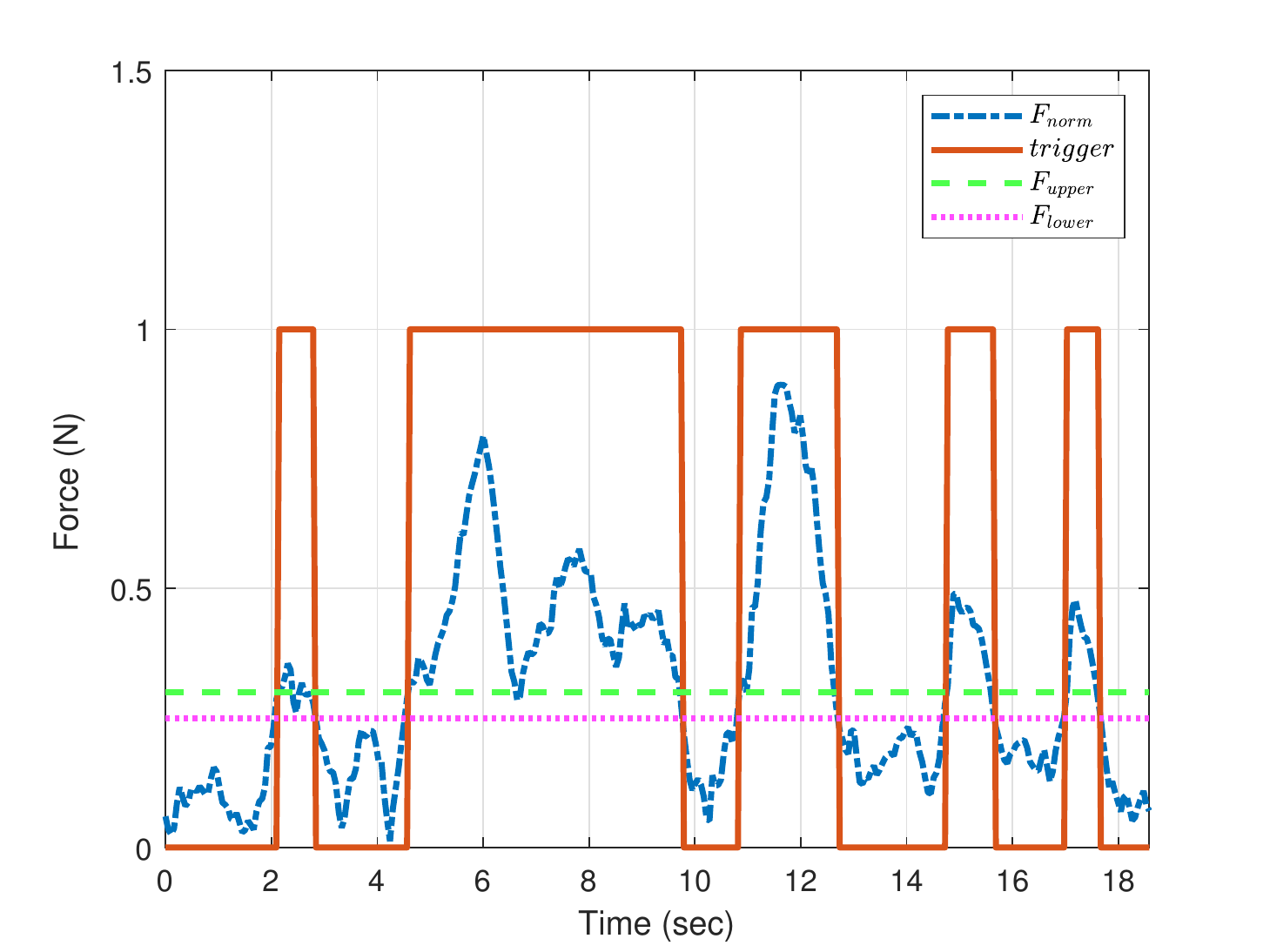}}
\par\end{centering}
\textcolor{black}{\caption{\label{fig:S-trigger}Trajectory S. A triggering signal of $1$ implies
$t\in T^{en},$ while it is $0$ otherwise. }
}
\end{figure}

\begin{figure}[H]
\begin{centering}
\textcolor{black}{\includegraphics[width=0.9\columnwidth]{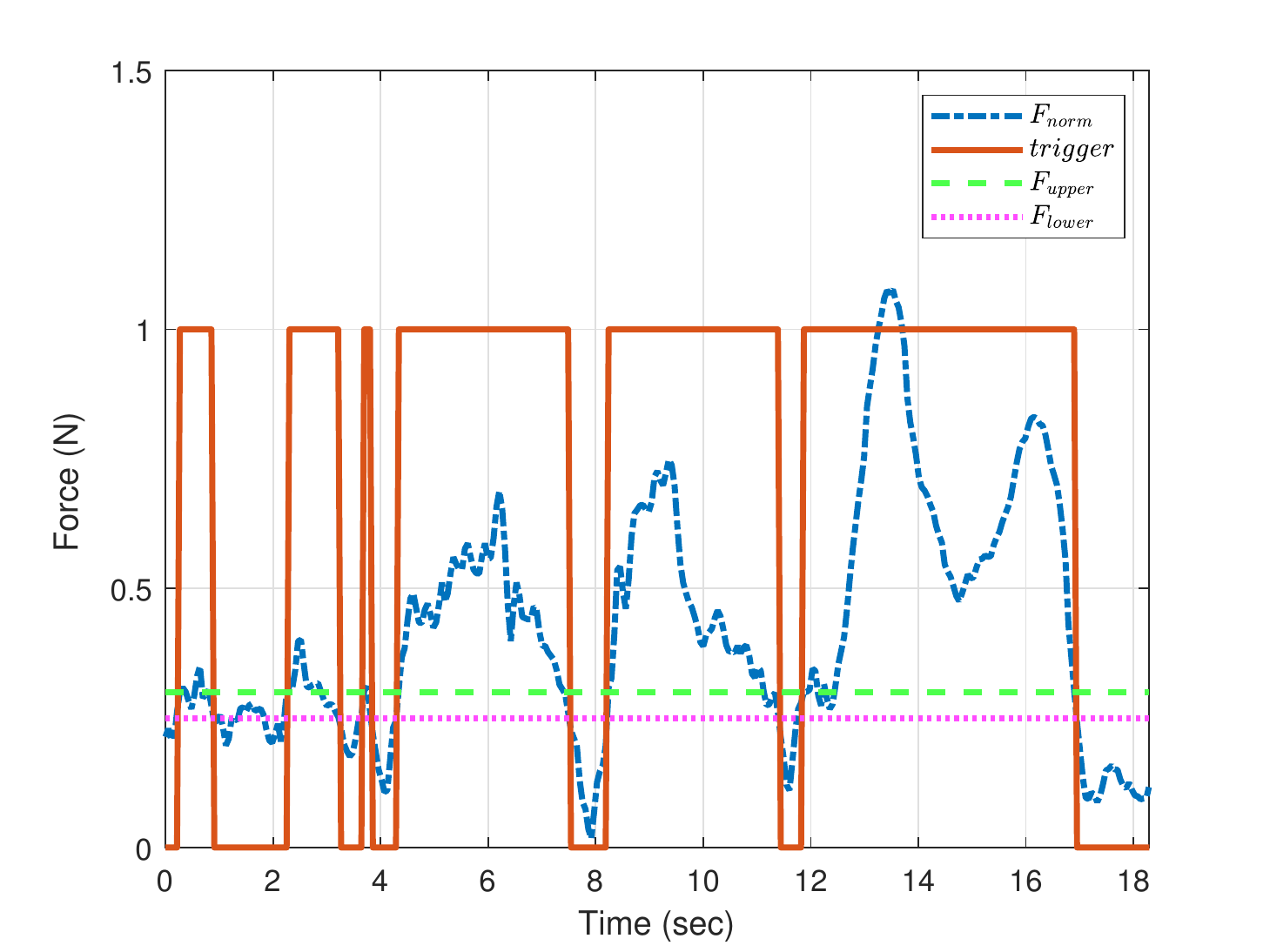}}
\par\end{centering}
\textcolor{black}{\caption{\label{fig:U-trigger}Trajectory U. A triggering signal of $1$ implies
$t\in T^{en},$ while it is $0$ otherwise.}
}
\end{figure}

\textcolor{black}{}
\textcolor{black}{Figs. \ref{fig:S-trigger} and \ref{fig:U-trigger}
show the effects of triggering events, as designed in Section \ref{subsec:Triggering-Mechanism},
when tracking trajectories S and U. Zeno behavior did not occur due
to the presence of only a few triggers.}

\textcolor{black}{}

\textcolor{black}{}

\section{\textcolor{black}{Discussion\label{sec:Discussion}}}

\textcolor{black}{A cooperative transportation using a leader-follower
UAV system has been developed, in which inter-UAV communication is
not required and the reference trajectory of the payload can be modified
in real time. To achieve this, the system is modeled as a nonholonomic
system such that the follower's motion does not affect the tracking
error of the payload; that is, the follower is controlled such that
one end of the payload performs only longitudinal motion, and the
leader is controlled to ensure asymptotic tracking of the $c_{2}$
point on the payload. Robustness of the leader controller is also
proven given the estimate errors in the UKFs and the disturbances.
UKFs are developed to estimate the forces applied by the leader and
follower so that force sensors are not required. Stability analysis
has proven the stability of the closed-loop system. Simulations have
demonstrated the performance of the tracking controller and the feasibility
of changing the desired trajectories online. The experiments show
the controllers of the leader and the follower can work in the real
world, but the tracking errors were greatly affected by the disturbance
of airflow in a restricted space. Future works will consider new control
approaches based on more-practical IMU setups and the instability
caused by the disturbances. }

\textcolor{black}{\bibliographystyle{IEEEtran}
\bibliography{encr,master,ncr}
}
\end{document}